Imperial College London

Department of Computing

**Content Sharing for Mobile Devices**

**by**

**Rudi Ball**

MSc in Advanced Computing

Individual Project Report

Dr Naranker Dulay
Dr Peter Pietzuch

**September 2007**

# Abstract

The miniaturisation of computing devices has seen computing devices become increasingly pervasive in society. With this increased pervasiveness, the technologies of small computing devices have also improved. Indeed, mobile devices are today capable of capturing various forms of multimedia and able to communicate wirelessly using increasing numbers of communication techniques. The owners and creators of local content are motivated to share this content in ever increasing volume; the conclusion has been that social networks sites are seeing a revolution in the sharing of information between communities of people. As load on centralised systems increases, we present a novel decentralised peer-to-peer approach dubbed the Market Contact Protocol (MCP) to achieve cost effective, scalable and efficient content sharing using opportunistic networking (pocket switched networking), incentive, context-awareness, social contact and mobile devices.

Within the report we describe how the MCP is simulated on top of the JiST (Java in Simulation Time) framework to evaluate and measure its capability to share content between massively mobile peers. The MCP simulator uses real geographic data to superimpose simulation on a real world environment using unidirectional, bidirectional and grid pathways, traveled by peers. The paths and data are visualisable using modern tools like GoogleEarth. The MCP is also applied to a collection of real datasets containing contact data. We use this approach to verify the findings and consistency of the MCP simulator on which preliminary assumptions are made. The effectiveness of the MCP is described, discussed and evaluated.

We ultimately put forward two versions of the MCP (Simple and Extended) which can be applied given the considerations of peer mobility, peer density, content size and technologies available to a peer. While expensive, the Simple MCP is useful for the rapid communication of small sized data collections over large collections of peers. Extended MCP is applicable in environments where data file sizes are larger, where communication requires more resources, contact times are longer and battery resources for devices need be reserved. The MCP is shown in conclusion to be a powerful means by which to share content in a massively mobile ad-hoc environment.

# Acknowledgements

A great many thanks go to my supervisor, Dr Naranker Dulay for his time, patience, criticism, motivation and insights. The conversations during his meetings were broad, fascinating and inspirational, fostering an enormous enthusiasm and curiosity to develop and test the ideas of content sharing in mobile devices.

Thanks are also extended to Prof. Jon Crowcroft and his lecture at Imperial College on Opportunistic Networking which provided much of the inspiration for the protocol discussed in this paper.

# Contents

# 1 Introduction

As computing devices have decreased in cost and miniaturised, so computing devices have also become more and more pervasive. At the beginning of the 1980's it was unusual for a single person to own or have access to a single computer. Today, many individuals own more than a single computing device and with the mobile telephonic revolution, many individuals already hold improved mobile devices capable of handling complex and computationally intensive applications. These devices are also increasingly capable of capturing various forms of multimedia in ever improving quality. Improvements in communications techniques (WiFi and Bluetooth), computational power and storage capacity are slowly making wireless applications not previously possible or conceivable on small mobile devices, achievable. Importantly, applications are no longer dependent on the prevalence of a centralised central point or permanent structure for the routing of messages. We expect that the future environment which we inhabit shall be teeming with various pervasive devices and systems providing a multitude of services.

One such opportunity for future application is peer-to-peer (P2P) content sharing within mobile ad-hoc networks. The limited resources of mobile devices, intermittency of contact and the unpredictable behaviour of peers makes content sharing a complex problem to solve. Dynamic systems such as Napster, Kazaa, Skype, Freenet and BitTorrent, provide examples of the differing implementations of the P2P paradigm. These use P2P as a basis by which to communicate and share content however they do not deal with mobility and intermittency of contact between peers. They have however proven that the P2P paradigm has been considered successful in providing a means of scaling and reducing the cost of data communication, utilising a divide and conquer philosophy to information storage and retrieval. P2P could have widespread benefits within the mobile computing environment where ad-hoc networks could consist of a multitude of co-operating and self-organising peers located in geographic proximity to one another. The ad-hoc methodology has already been considered and applied to some extent within military applications. Opportunist networking seeks to provide benefit to its users via the use of ad-hoc networks created through the intermittency of connectivity between roaming or massively mobile peers sharing content or even computation cycles.

Simultaneously, social networking systems have gained major popularity in the past few years. Systems like Facebook, YouTube and Twitter are examples of so-called Web 2.0, where content is generated and consumed by the users using such systems. It is somewhat accepted that similar mobile P2P communities will form and use mobile devices to generated and consume similar content on the move. Some users are likely to form groups which may be interested in only specific content. In considering this explosion of social communication, there are definite benefits within the ad-hoc realm for users, capable of communicating cost effectively in an increasingly physical environment.

While improvements are slowly being realised to make P2P on mobile devices possible, there are many problems which still exist even within present fixed computing P2P applications. When considering the problems still inherent in P2P architecture design for fixed networks, the idea of mobile ad-hoc network content sharing is no small problem. Issues affecting mobile content sharing, which are by no means only common to the mobile ad-hoc environment include unpredictable contact intermittency, search algorithms, peer discovery, message routing, incentives for information storage or computation, communication cost and reliability, scalability, information security, power constraints and dependability. Many of the solutions to these problems do exist in other research, however we should be careful not to make the assumption that mobile computing is just like normal distributed computing. There are clear constraints relating to network connectivity, computational ability and storage limitations. Engineering methods required for the creation of mobile ad-hoc networks are limited. Argument is still ongoing in how to treat mobile devices. They are either simply analogous to large mobile computers or considered as having their own constraints, mostly due to size. Within this report we take a more moderate view on the sources of solutions to mobile computing.

To achieve ad-hoc wireless information sharing it is required that the P2P system be highly available in operation, seamless to use, provide transparent access to services and provide localised services within a secure framework whilst also being timely and cost effective. Achieving these elements within the mobile ad-hoc environment requires the integration of wireless communication, self-organising systems and resource optimisation.

## 1.1 Motivation

At present, proprietary content has been given mass importance due to the profits generated by selling content to users. File sharing communities like Napster, Gnutella and BitTorrent work on the theme of sharing content with other users. As mentioned previously, we should also expect to see content created and shared at a local level within the mobile environment using mobile devices. Many present day mobile devices already carry the capabilities to record videos, audio clips and take pictures. Considering both these content markets of user generated and copyrighted information, it can be argued that a major benefit of sharing content using P2P networks is primarily a spread of load on the system. GSM networks have limited capacities for data transfer. The cost of data transfer using an umbrella mobile service provider is also largely monetary. A truly mobile P2P approach could be decentralised, providing an ever changing subnet or "island" of infrastructures on which to communicate.

Previously overloading of centralised GSM networks (where multiple requests bombard the network's capacity for communication) has been noted many times, especially during disasters. P2P is more easily able to accommodate limited failure. Another benefit of P2P is where umbrella GSM networks fail or no coverage in certain geographical areas. An example of this could be in remote communities where technological change is slow, such as in developing countries. Many users often have mobile devices yet they lack the network to provide a mechanism for cheap content communication. In essence we could consider mobile opportunistic P2P as both an alternative communication coverage tool and a backup system where the centralised communications networks and protocols are less than reliable.

## 1.2  Project Overview

This report describes a new protocol for content sharing utilising opportunistic networking [36] and market based characteristics dubbed the Market Contact Protocol. The protocol uses peers acting as sellers or buyers to share and communicate messages with one another. The MCP is intended for use in dynamic peer-to-peer (P2P) scenarios where networks of peers are unpredictable in their availability to provide a framework to share messages, but where mobile devices seek to share their resources for the betterment of themselves or the systems in which they reside based on incentives. We present the MCP as a protocol relying primarily on peer incentive, contact, resources and geographic mobility.

The approach was influenced by work on opportunistic networking [36], approaches to content sharing and overlay networks [15][21], and social networking web services like Facebook [22] and Youtube [86]. The power of opportunistic networks was considered by research previously but not associated with *motivation* or *incentives* for sharing content. Applying opportunistic networking has real benefits where groups share content hence social networking approaches offer an incentive to share content within a real geographical setting using massively mobile peers. While many approaches describe routing methodologies where message communication is immediate, opportunistic networking could provide a mechanism for decentralised low cost communication between peers, indirectly making peers more social in the process of using the system to retrieve and send content to one another. The same approach could also provide a persistent storage accessible within geographical space.

## 1.3  Report Structure

This document describes the Market Contact Protocol (MCP), beginning with the approach to conceiving the protocol and concluding with the simulations and outcomes of experiments testing the applicability of the protocol.

- Chapter 2 discusses a collection of background concepts pertinent to the development of the MCP, including: peer-to-peer (P2P) methodologies, opportunistic networking, incentives, costs, technical issues, scalability, the limitations of mobile devices, social peer behavior, present and future communications techniques and approaches to content sharing in MANET (Mobile Ad-hoc Network) systems.

- Chapter 3 describes the aims, architecture and scenarios of the MCP.

- Chapter 4 assesses the implementation and modeling of the MCP simulator, built on top of JiST (Java in Simulation Time). The simulator is described and its assumptions argued. The MCP is described in its Simple and Extended form. A brief portion of the chapter describes a quick and dirty implementation of the MCP using J2ME and Bluetooth.

- Chapter 5 discusses the results generated by the various MCP simulators, their expected results and the final results produced. A summary highlights those concepts and knowledge most interesting to the application of the MCP. Most notably this chapter highlights the exhibition of Verhulst's population growth function in increasingly complex systems of sharing peers.

- Chapter 6 considers an alternative to corroborate the generational results and behavior of the MCP in Chapter 5. We use CRAWDAD [18] datasets to highlight the common results of both real and generational simulation data. We show that the MCP behaves similarly for real datasets as for generational datasets, thus presenting that the MCP simulator is realistic in nature and the MCP can be applied within a real world P2P system when such bridging technologies arise.

- Chapter 7 reviews the achievements of the project, identifies deficiencies, discusses the possible broad effect of the MCP and proposes future work and issues to be solved.

# 2 Background

This section describes the background concepts to the generation of the Market Contact Protocol (MCP). It discusses some of the notable concepts which are both beneficial and limiting to content sharing within the P2P realm. We describe the incentives, costs, technical issues, present and near future communication techniques and alternatives to content sharing in the mobile P2P environment. We finally present various schemes for file or information sharing.

## 2.1 Motivation for using the Peer-to-Peer Methodology

In the context of P2P issues of cost arrive for each component of the system as is common with distributed systems. The node computation, communication and storage are all limited resources and hence there must be some sort of motivation in a sharing environment to share resources for the provision of communication. This section considers these three fundamental constituents of the system.

### 2.1.1 Incentives

Sharing is fundamentally the co-operative act of providing some resource, without necessarily requiring a return or payment for the resource. The major advantage of P2P is that peers provide resources including bandwidth, storage capacity and processing cycles to a larger community of similarly interconnected users. Increasing the number of peers within a system also improves the robustness of the system and improves the networks ability to deal with fault. If nodes are lost, there are other routes which communicated data (messages) could be routed through to reach their destination. It is desirable to have large collections of interconnected peers. Where there are too few peers, there is no means of communication. The replication of data on many peers ensures that if a node is lost, data can still be retrieved from an alternative source. The replication of content may however be dependent on the worth associated by a peer for some item of data. Data which is thought worthwhile by a majority of users may be easy to find and retrieve while data thought only worthwhile storing by a minority maybe more difficult to find and retrieve.

### 2.1.2 Resource Cost

We can define resources as those limited commodities required to achieving a measure of quality of service. Resources such as storage or processor cycles are accessible in a shared environment. If we return to the client-server model, clients request solutions from providers (servers). The server would need to have the adequate requirement of resources required to provide a service to requesting client.

Peers act as both clients and servers simultaneously and applications can decentralise resources and control between peers. By harnessing the distributed resources provided by P2P groups, applications can mitigate resource costs. A local or global storage choice either needs to be made by the community of the network or individually by each peer.

An example P2P application which reduces resource costs is BOINC (Berkeley Open Infrastructure for Network Computing) [6]. The authors consider the application "open source desktop grid computing". Internet users donate free processing cycles to the BOINC system for research in the analysis of various scientific topics.

### 2.1.3 Monetary Cost

If finding data associates a monetary cost with that data, then we should consider the robustness, flexibility and dependability of a typical network. Historically the client-server methodology held data in a single location. Points of failure in the system included both the network link and the data storage locations. Clients made requests to servers for their information. A server's only goal was to service requests. Too many requests to a service could result in the failure of the service. Denial of service attacks play on this problem. The monetary cost of increasing scale lies with service providers. P2P seeks to mitigate costs by spreading the cost of computation, storage and communication between co-operating peers. In this sense the monetary cost is lowered because scaling the system is not required. Aspects concerning the loss of control of data could be argued, where information held is associated with value. At present the cost of many mobile devices supporting varying wireless standards like WiFi and Bluetooth is expensive, but as the market adopts specific standards we could expect a reduction in the cost of devices capable of providing ad-hoc P2P capabilities.

## 2.2 Technical Issues and Problems

Various technical issues exist within static and dynamic networks and the various methodologies applied. Within these approaches and paradigms exist advantages and disadvantages attributed to content sharing in mobile ad-hoc networked systems.

### 2.2.1 Static and Dynamic Networks

This document considers static P2P networks as those where access points are statically known. A peer is assumed to be connected to the network during the entire duration of a process' lifecycle or execution (beginning to end). Static peers are capable of dying. A level of fault tolerance is required, however the peers are not considered to be moving from one time instance to another. The network routing topology is constant, with peers either available or not available to route information.

For the purposes of P2P computing a dynamic network is one which changes over a period of time. The network's peers are ever changing and the network is ever changing, due to the mobility of users. Communicating nodes on the network must adapt to the changes in the system to perform their function.

### 2.2.2 Network Methodologies

Within the confines of mobile content sharing we consider two major methodologies, namely mobile ad-hoc networks and overlay networks.

#### 2.2.2.1 Mobile Ad-hoc Networks

Mobile ad-hoc networks are usually dynamic, unpredictable, self configuring networks connected by wireless communication mechanisms. The networks are created without the requirement for a base station (centralisation), by discovering neighbouring devices. This has both advantages and disadvantages.

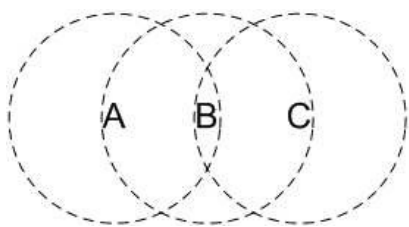

**Figure 1. An ad-hoc network consisting of three devices (A,B and C). Devices within range of one another have knowledge of their immediate neighbours. A does not know the existence of C and vice versa.**

Using multiple devices, the network can be used to connect unseen neighbouring devices, using algorithms such as hop routing algorithms. Mobile ad-hoc networks (MANETs) consist of devices which have the ability to act as both end-points and routers for data packets. Nodes can, without warning, connect to or disconnect from the network, due to mobility. For this reason MANETs need to be rapidly self-organising. Rapidity in organising usually requires much computation and communication.

Devices acting as routers are as mobile as the endpoints making use of them to transport communications. We can think abstractly of devices as possessing vector direction, while also having a limited communication, or sight range. The range in reality is more irregular. This routing assumption is different when compared to normal networking in that the dependability on the network for routing is called into question each time a message is sent along (hopping) the network from a source to destination. Avizienis et al. [1] have proposed more precise definitions relating to dependable and secure computing. The state of the network can be assumed to be truly changing from once instance to the next.

Common MANET route discovery mechanisms include on-demand (reactive) routing, table-based (pro-active), and hybrid (a mix of both) routing mechanisms. Other more unconventional but interesting approaches include hierarchical cluster routing, geographical routing and power-aware routing. Power-aware routing is considered later in this document.

Table-based routing uses periodic updates to provide complete knowledge about the network topology, whilst on-demand routing discovers peers when a node is required to find a route to transmit messages to a destination. Both approaches have their advantages and disadvantages. Table-based routing is costly to bandwidth, as routing tables need to be swapped and updated periodically between neighbouring peers. Nodes within the MANET need to periodically check that the network has not changed to some degree, and that the routing tables are up to date given the state of the network. Routes that no longer work need to be removed. There is also a rather benign cost of communicating tables and routing information for routes that will not be used and depending on the size of the network it may not be feasibly possible to converge on complete knowledge of the network (e.g. Guesswork [57]).

In contrast On-Demand routing protocols often flood (repetitively broadcast) within a network space, sending request messages to all neighbours. This is extremely costly to bandwidth for the entire network. We could also expect a multiplicity of querying messages due to the broadcasting nature of the protocol. Latency also exists before a message can be sent to a targeted recipient (e.g. the Ad-hoc On-demand Distance Vector protocol [59]). Hybrid methods use portions or a blend of both approaches to make logical judgements on where a specific approach is best suited (e.g. Hazy Sighted Link State Routing Protocol [65]).

Node failure is unpredictable. Routing algorithms typically use a backtracking storage algorithm and a hop count (time-to-live count) for initially finding a route between two nodes. A search range (distance from the root) is limited by the hop count, sometimes referred to as a "horizon".

2.2.2.2   Overlay Networks

Overlay networks are defined as those networks residing on or formulated on existing network topologies, usually to achieve improved efficiency, manageability and reliability. Overlay networks exist where all peers provide networking capabilities. A virtual network exists on top of a physical network [15]. A neighbour is described by the content they store. Fundamentally an overlay network methodology breaks down the process of messaging into single distance components considering the network as a local process rather than a graph problem. The overlay used on an existing network can be used to route messages to destinations where the IP address of the receiver is unknown, and subsequently they have been put forward as a solution for the provision of improving quality of service (QoS). To implement an overlay network requires that the routers are modified. In present 'fixed' internet services, this is a problem. As ad-hoc networking is yet to mature, their implementation could still be possible as a standard. The overlay exists where each node has an associated node identifier, generated by hashing some sort of unique associative information, such as IP address or a secure public key. Messages usually utilise key based routing to reach their destination. Querying has a determinable conclusion (i.e. we get a yes or no answer), whereby the querying node initially finds the relation between it and its target node. Once the relation is known, the querying node knows the direction in which it should query. Depending on the protocol being used differing identification methods are used. Some protocols include Chord, Tapestry and Pastry. A common example of a structured overlay is a distributed hash table (DHT), with an example system being FreeNet [17].

Pastry's authors call it a "fault tolerant substrate" for P2P communication [58]. It is an overlay network using a distributed hash approach similar to Chord in functionality and architecture. A collection of value pairs are used for the redundant storage of data. The routing table is dynamically generated based on these pairs. The benefit of Pastry is that it has a number of working applications attributed to it, including a distributed file system named PAST and a publish/subscribe system called SCRIBE.

|  | Broadcast | DHT |
|---|---|---|
| Routing | $O(n^2)$ | $O(\log n)$ |
| Scalability | Bad | Excellent |
| Implementation | Low | Medium |
| Maintenance | Low | Medium |
| Energy Efficiency | Low | Medium |
| Shortest Path | No | No |
| Cross Layer | No | Yes |

**Table 1. Routing approach comparison between Broadcast and DHT [20].**

A summary of P2P file sharing approaches to ad-hoc networks was conducted by Ding and Bhargava [20]. They concluded that distributed hash tables were the best means by which to formulate such networks. There is a desire for the network to achieve decentralisation, scalability and fault tolerance (reliability).

As previously remarked, within the context of on-demand networks, the overlay network has the ability to route data to a particular location without prior knowledge of the address of the destination of a specific message. An overlay network by definition guarantees data retrieval (convergence), a set *O(log n)* look-up time for *n* nodes and the exhibition of load-balancing and dynamic self organisation [20]. Fewer resources are used if compared with the pure form of P2P as the pure form utilises time-to-live packets to search nodes a particular distance from the root. Overlay networks use a method for finding and routing data based on divide-and-conquer techniques. The system is also reasonably dependable, as there is a guarantee of information retrieval using iterative means, and non-graph traversal.

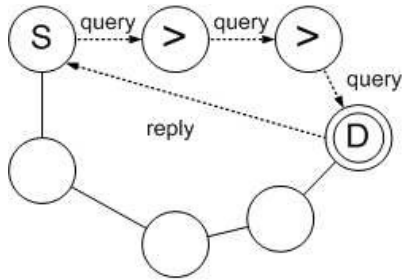

**Figure 2. Overlay network routing of a query from source S to destination D. Intermediate nodes calculate their relation to the destination. The destination node replies along an intended route [20].**

Overlays provide a means by which to concentrate on scalability, however they require increased bandwidth and logic processing (computational) requirements to measure all inter-node paths (edges) and exchange routing information. Probing neighbouring nodes has a cost of $O(N^2)$. The natural multi-hop alternative routing of overlay networks could be used to aid the spread the communication energy cost within ad-hoc networks.

### 2.2.2.3   Opportunistic Networking

Opportunistic networking [36][45] as a paradigm considers the communication of information between peers based on peer mobility and availability. The paradigm is considered a type of delay-tolerant network. Peers exist in an environment where there are intermittent periods of contact between peers and where the network topology is highly dynamic in nature. Peers can take advantage of the process of contact between peers to share information. Opportunistic networks usually exhibit short paths from peer to peer. The short hopping paths exist for brief moments during a peer's lifetime of movement. Connecting these short paths can facilitate the sharing of information. Long end-to-end paths are absent from such extremely dynamic systems. The ad-hoc nature of unpredictable mobile peers makes the topology of the network highly dynamic.

Devices are required to make control and management decisions based often on their context, hence context-awareness is an enviable characteristic of devices operating in this environment. By understanding the environment in which a device exists, that device with the provision of some enhanced intelligence has the opportunity to improve the efficiency and effectiveness of content sharing, given limited resources. The dependability of the system is associated with its connectivity. Peers not connected at some time to the network have no way of receiving information addressed to them or a grouping of peers.

### 2.2.3 Scalability

Scalability is termed as a "desirable" [8] property of system, where by increasing the number of requests for resources on a system, the system can increase the availability of resources to counter the increased demand. If the performance of a system improves after adding these resources proportionally to the capacity added, the system is considered scalable. The system is assumed to be able to deal with resource sharing efficiently. It is in effect a model of balance between two or more properties of a system, where the desire is to observe a controlled and proportional improvement to a system. Termed alternatively we could consider scalability an economic problem based on supply and demand. As demand for a particular system increases, the system should supply more resources, to maintain a certain level of performance.

$$speedup = \frac{1}{\sum_{k=0}^{n} \left( \frac{P_k}{S_k} \right)}$$

**Figure 3. Amdahl's Law [2], where k is a label for each $S_k$ and $P_k$, $S_k$ is the speed-up multiplier, $P_k$ is the percentage of instructions which could be enhanced, n is the number of speedups resulting.**

Indeed as the size or scale of the system increases, the system should be able to cope with increasing load "gracefully" [8]. Defining scalability is challenging because the context in which a system resides often determines the factors attributing to scalability.

In the context of improving scale in processing ability we can apply Amdahl's law. The law stipulates that for a given network and parallel performance, a greater benefit to performance is achieved if a new node is added to the network, with the alternative being that each node could be tuned or optimised for further performance. The law is used to calculate the possible maximum speedup achievable for a multiple of processors. The method is not however perfect, as speedups greater than those predicted by the formula have been achieved in certain systems. For the purposes of parallelism, the fastest performance is dependent on the slowest processing node.

It is argued by Zeinalipour-Yazti et al. [87] that existing searching techniques found in P2P systems do not scale well, because they either require message flooding (on-demand routing) or a global knowledge (table based routing). To solve this problem more intelligent and complex algorithms are proposed by several papers.

Peer-to-peer structured overlay systems such as Oceanstore [61] and FeedTree [64] are examples of scalable systems, capable of maintaining performance while using low maintenance traffic. Oceanstore seeks to provide a dependable "global persistent information store" using untrusted storage locations. FeedTree has been produced as a possible low bandwidth cost P2P alternative to single RSS server querying. Proving scalability with these systems is however more theoretical than practical.

### 2.2.4 Mobile Devices and their limited capabilities

Providing various strategies to achieve improved energy-efficiency would without doubt be greatly beneficial. Mobile devices are limited predominantly, within the three areas of communication link, computation and storage, by their size. Most problems however arise due to power limitations which have a knock-on effect on the computational and communication abilities. Alternatively, if an implementation of an application is complex, the power required to run that application is heightened affecting the power usage of the device. Signal strength is dependent on the antennas characteristics and the amount of power used to broadcast a signal. In certain cases mobile devices may be connected to a network on the downlink (able to see the network), but unable to communicate on the uplink.

#### 2.2.4.1 Computation Costs

Computation incurs an energy cost, and it is therefore desirable to make applications efficient. Whether an application is compiled prior to runtime or at runtime (Just-In-Time compilation), has an effect on the life of the battery. J2ME applications are such an example of costly JIT compilation. Chen et al. [13] found that the point of compilation has a major effect on the battery life of a mobile device. Compiling an application on a remote server is a strategy used to reduce the cost of local execution, where and if the cost of communication is less than the cost of local compilation. An energy cost choice needs to be made between the remote compiled size of an application and the alternative local compilation cost. This information is helpful if we consider the opportunity of using roving agents within mobile P2P.

#### 2.2.4.2 Communication Costs

Communication, depending on range factors, can have a large resource cost on the battery of a device. Added to range issues, the desired concept of constant "always-on communication" connectivity anywhere is limited because to maintain constant connectivity with a network, requires that the network be using energy at all times. Greater than 64% of energy consumed by a device may be in the waiting-for-transmission state [29]. Gurun et al. [29] argue that the power problem is still the major constraint to application development for mobile devices, but that it is feasible to deploy protocols on devices. Batching messages with periodic sending could have significant benefits to power usage. Within ad-hoc systems, many strategies considered also include "power aware routing mechanisms".

There are several approaches proposed [70]. DQRMA, POCSAG and FLEX are some of the single hop or reservation paging protocols which operate on the MAC layer. Nodes transmit periodic messages with the target user ID. When a receiver node is alerted that a message is waiting for it, it waits to receive the message, while other nodes ignore the message, defaulting themselves to a battery saving state. Work with single hop systems suggests that message contention produces higher power consumption. In contrast scheduling messaging reduces costs, however increases latency.

Singh et al. [70] extends the paging idea, describing the use of mechanisms like power aware metrics to determine the routing of messages. Their work suggests a reduction in power cost per packet of between 5-30% over the shortest hop routing distance. By combining their metrics with their PAMAS (Power Aware Multi-Access protocol with Signaling for ad-hoc radio networks) it is possible to further reduce energy costs. Similar protocols are used in sensor networks.

### 2.2.4.3   Information Locality and Load Balancing

The decentralised nature of the P2P methodology seeks to distribute load on a network (load scalability). The more redundant data within the network the better or faster data can be queried and downloaded. Also, information locality provides a method by which data can be used according to context. Information is nearby.

By having multiple fragments downloaded from various sources, the demand and usage of bandwidth is spread, affecting each data fragment provider less than it would if they were the sole provider of a specific resource. This is the basis of load distribution (load balancing) [17].

### 2.2.4.4   Communication congestion and bottlenecking

If there is increased message or packet load on a subnet a single node or parts of the network may have difficulty spreading this communication load. A subnet may be congested and a single node's ability to route packets may be hampered by the volume of packets to be routed.

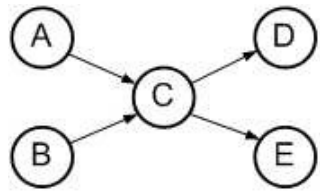

**Figure 4. Congestion generated by A and B produces a communication bottleneck at C.**

If traffic becomes too congestive the node (router) may fail or lose packets. Too much traffic therefore produces congestion which in turn degrades the performance of the network. Other bottlenecks may exist as characteristics of the server. Where there are limited resources available there is the implication of possible bottlenecking.

A test for congestion within a system is possible as described by Tanenbaum [74]. If a subnet is considered to have maximum capacity called $\sigma$, then we could expect the outcome described below.

An example could be that C has limited memory, to deal with concurrent requests for message forwarding. This is a reasonable example because the "trivial" answer would be to give C vast collections of memory, perhaps even assuming that C had infinite memory storage capacity. Nagle [54] argues that infinite storage may indeed cause further problems to routing. In this case there is the possibility that packets which spend too much time queued in the memory of a routing node, may time-out. Their time-to-live may expire. Packets are dropped thereby requiring further resolution, incurring cost within the system.

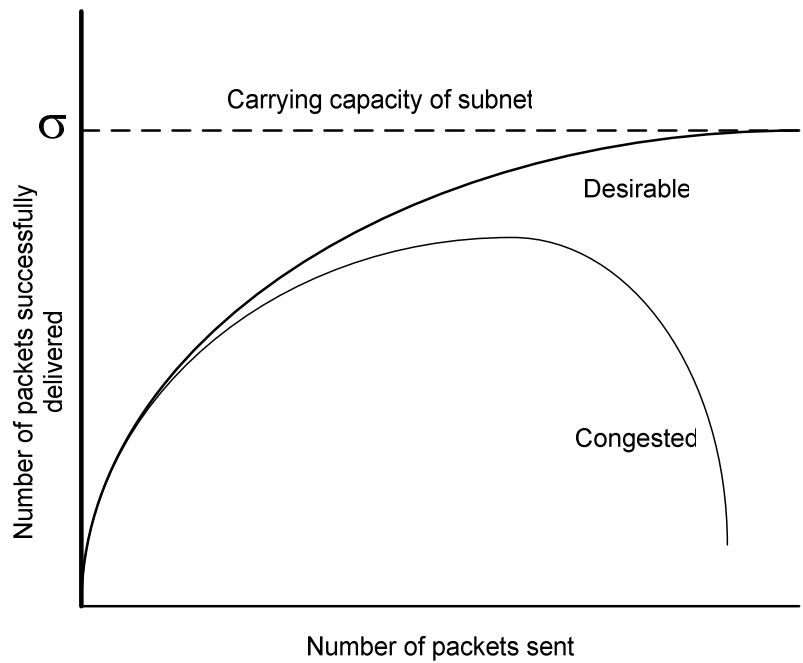

**Figure 5. Expected congested and non-congested systems.**

Methods of resolving congestion in a network include open and closed loop solutions. Open-loop seeks to solve congestion through good system design, avoiding congestion occurring. Closed-loop solutions apply feedback loop control, which entails detecting points of congestion and acting upon this knowledge of the system. As previously described another method to solve load is to use an overlay network.

## 2.2.5 Quality of service (QoS) and Reliability

Quality of service is characterised by "information flow reliability, delay, bandwidth and jitter". Requirements of the characteristics are dependent on the application which is being used or provided. File transfer may only require high levels of reliability and reasonable amounts of bandwidth. Techniques used to achieve QoS include over-provisioning (providing excess routers and network infrastructure), buffering (removes jitter with a trade-off of delay), traffic shaping (controlling or regulating the flow of packets), resource reservation (regulating resources available to nodes, these include bandwidth, buffer capacity and processor cycles), scheduling (providing a fair allocation of routing capacity for different nodes) and proportional routing (using the best rated routing path to a receiver).

QoS is closely associated with overlay networks, where the concept of a Resilient Overlay Network (RON) is considered. RONs aim to provide a reliable network on top of a physical network. The reliability and robustness of the network is improved by nodes watching the condition of connections (edges) between themselves and their neighbours. When a packet needs to be sent to a receiving peer, the sending node makes a logical decision about whether the packet should travel directly or indirectly to their packet's intended recipient. Metrics define the state of the network between neighbours. RON networks suggest that reactive routing best applied to nodes which have multiple connections to the network, i.e. there are multiple neighbouring nodes. Using "observed routing", messages could be used to predict failure within the network, reducing the reaction time required to deal with network node failure [24].

## 2.2.6  Social peer behaviour, structures and opportunity

While this report is not predominantly interested in the social behaviour exhibited by P2P systems, the behaviour of peers within a system due to policies may have effect on our assumptions and how best to model a mobile peer to peer system. This behaviour could be translated into how the peers organise themselves within the context of sharing.

Many of the incentives associated with P2P content sharing can be associated with ideas in game theory. Game theory provides a formal approach to modelling social situations and interactions between peers. Participation in a P2P network is voluntary and to be considered reliable the network must provide a particular level of QoS (reliability) to its users. Buragohain et al. [11] consider a simplified interaction of heterogeneous peers of a size $N \times N$ matrix. With this they use a function to describe the benefit between peers within the P2P network, producing qualitative properties with the Nash equilibrium algorithm. From this, incentives policies can be derived. Alternative approaches use metrics such as the participation level (ratio of upload to download) to conclude on a given policy within the P2P network. Huang et al. [35] argue that a policy framework may not be needed at all for incentive in early adoption of a P2P system and that by adopting a policy you may only hurt the deployment of a system in its early stages.

A contradiction within P2P behaviour is the principle of "free riding" [27]. Free riders are those individuals or entities within a system who consume more resources than they provide to a community. Free-riders are sometimes referred to as "leechers" within the file sharing context. In the static peer-to-peer context, there exists the paradox that while there is a great deal of free riding by the majority of users, the file sharing networks still seem to thrive. Jian and MacKie-Mason [39] argue that the phenomenon may be party because of altruistic effects, because of the requirement of the file-sharing programs used for downloading. File sharing applications are kept permanently sharing by default.

Hui et al. [36] take the behaviour of users of devices one step further, by looking at the social network and the groupings created through user contact and communication purely via the ad-hoc network. Pocket switched networks use human mobility and local forwarding to their advantage. This process of contact between peers creates the opportunity for "opportunistic networking". Using this contact network the information could be distributed (perhaps very slowly) through a socially created network. This approach has advantages and disadvantages, most notably that information travels slowly and only within a specific set of users, however it could be improved with further redundancy of data seeding within a system. Grouping in this manner is a good way of dispersing information which is relevant to a specific set of users. Privacy is another issue within this approach. Applications using grouping could include news feeds only relevant to people working together. Disaster information is also considered, because in the event of a disaster umbrella GSM networks may cease to operate and information could still flow via ad-hoc networks, granted that there most users still have access to electrical power.

Grouping people together into communities of peers or clusters is likely given that there has been an increasing interest by online communities in content sharing. Systems like Facebook [22], YouTube [86] and Twitter [77] are all examples of Web 2.0, i.e. websites where content is generated by the users who use such systems as a local point of exchange. The same could be true in the mobile P2P context, where particular individuals could be interested in specific topics, ideas or subjects. This would make routing information common to specific subsets of the total set of available mobile community.

Tuple spaces could be used as an approach to storing data within a P2P system. The model used is considered a generative communication model. Tuple spaces provide an associative memory structure where there are a collection of set tuples. The tuple space references a tuple without relevance to where the tuple is located. The tuples can be accessed or modified by any node or peer within a network. Tuple spaces provide a simplistic method of placing data into a shared pool. No knowledge of the network is required by individual peers, leading to improved decentralisation. Some tuple systems include Comet [48], Java Spaces [26] and Linda Spaces [62].

## 2.3 Physical Communications Techniques

Various communications mechanisms have been developed and adopted for mobile communication. Standards within the context of radio communication between mobile devices commonly use Bluetooth and Wi-Fi (IEEE 802.11x), however there are other standards being developed as alternatives including PAN (Personal Area Network) systems like ZigBee which could hold some promise as good low-power and low-cost peer-to-peer alternatives for communication between peers. The fundamental factors for a communication technique used in mobile ad-hoc content sharing are such that the latency to discover and communicate with one another must be low, coupled with low power usage.

### 2.3.1 Bluetooth (IEEE 802.15.1)

Bluetooth provides a means of communicating between devices within a short-range (1 to 90 meters) using the 2.4GHz range (an unlicensed radio band which includes devices like cordless phones and microwave ovens). Its aim was to provide a seamless method of connecting various devices. Bluetooth radio components are required to be small and consume little battery power if they are to be used on mobile devices. Users can connect, in an ad-hoc fashion, up to eight devices in a local short range network dubbed a piconet. The piconet however requires a main control device (master). All other devices connected in the network connect to the master as slaves. If a slave wishes to communicate with another slave it needs to route its packet data through the master. This is a disadvantage if considering a pure peer-to-peer communication environment where all devices can see one another and are within range of communicating with one another. Depending on the range between the devices communicating, data can be exchanged at speeds up to 723kbps. As with most radio signals, attenuation and signal loss (signal degradation) affects this bandwidth the further devices are from one another. Frequency hopping is used to avoid interference between other Bluetooth piconets within particular vicinity. Devices communicating with one another do not need to be within line of sight. The user of Bluetooth does not need to have specialist knowledge on how to connect devices.

Within the initial specification of Bluetooth [30], the major aim was to provide a short range mechanism to connect varying local devices with one another without cables, thereby replacing infrared and cables as means to transfer data between devices. In this sense the protocol lacks the ability to quickly discover and resolve other devices automatically. This is mostly because the protocol was designed to connect local devices together for simple data transfer. It is argued that many applications of Bluetooth seek to use the protocol for purposes it was initially designed for.

Major flaws were found to exist within the protocol [9][72][75][83]. Issues with authentication and data transfer mechanisms provided several means by which personal data attacks could occur [9]. It was found that confidential information stored within a phone could be accessed (calendar and phonebook data) through a process called "snarfing". Trifinite using improved antennas with Bluetooth devices found snarfing possible even at a range of 1.78 kilometers [75]. Individual phones could be identified using their IMEI (International Mobile Equipment Identity) and associated with a specific user, providing location and activity information about the phone's owner or user. Previously paired Bluetooth devices were found to be capable of reading the complete memory of a device. Access to higher level commands was possible using the AT command (BlueBug attack). Methods of intercepting voice conversations between devices and headsets were also discovered. Worms are also capable of propagating using the protocol [72], however application installation still requires user intervention, hence viruses have not yet been a major problem. Wong et al. [83] provided a measured investigation of the security of the Bluetooth PIN pairing protocol, showing its shortcomings while providing recommendations on how to improve the protocol. Their recommendation is to bolster security using asymmetric cryptographic methods.

Bluejacking is a recent phenomenon, where hackers seek to access or "bluejack" an unsuspecting user making use of the Bluetooth protocol [72]. The hackers sometimes leave a message proving that a device has been compromised using the protocol. Bluejacking has also been used on occasion as a means of communicating or posting information in public locations, such as in shop windows.

The implications of these concerns within the protocol make it difficult to implement content sharing in on the Bluetooth protocol. Security issues hamper the speed and efficiency content sharing models.

### 2.3.2  ZigBee (IEEE 802.15.4)

ZigBee [88] is a communication standard specifically aimed at providing embedded applications which require low-power communications and low data rates. ZigBee is particularly interested in mass automation. Intended applications for its use include embedded devices in buildings and medical data collection tools. ZigBee supports three different device modes: Co-ordinator (ZC), Router (ZR) and End-device (ZED). The Co-ordinator mode is necessary to provide a root of a network, storing information about the network and security keys. Routers, act as intermediary nodes transporting messages. End-devices are limited in functionality that they can only communicate with a parent node, being unable to relay information like the Router. A decreasing amount of memory is needed between the ZC, ZR and ZED, this can influence the cost of the device.

ZigBee has the ability to utilise protocol algorithms like ADOV and neuRFon. ZigBee examples have used the Sun Sunspot devices.

To use ZigBee commercially requires a purchasable licence; however a licence for its use non-commercially is available through request. An open variant of ZigBee is also available, however Wibree exists as an open competitor to both ZigBee and Bluetooth. The disadvantage of ZigBee is that its design is structured towards mesh networks, not mobile ad-hoc networks.

### 2.3.3 Wibree

Wibree [80] is an open-standard put forth by Nokia which exists within the 2.4GHz range, providing a bit-rate of 1 megabit per second. The aim is to produce an "ultra-low" power consumption radio standard for button batteries, which could work at ranges of between 10 and 30 meters. The standard would supposedly complement Bluetooth. Wibree was merged with the Bluetooth specification in June 2007 [80].

If Wibree is indeed similar to Bluetooth in its usage, it may be a good alternative for future mobile P2P content sharing applications. A 1 megabit per second bit-rate could be legitimately acceptable depending on the intended P2P content sharing application. Unfortunately little information has been release on Wibree and the many holes in Bluetooth warrant scepticism on the part of developers.

### 2.3.4 WiFi (IEEE802.11x)

WiFi [81] was initially produced as a means by which to network larger devices together into a so-called Wireless Local Area Network (WLAN). The more recent versions of the standard are compatible with one another. Initially 802.11a and 802.11b were not compatible. This standards issue affected WiFi deployment in the computer market. 802.11g sought to combine the most advantageous characteristics of 802.11a and 802.11b together.

More commonly WiFi is used to provide an easy means by which to provide devices connection to the internet. Devices connect to access points within so-called "hotspots" provided in a client-server fashion by a WiFi router, which in turn could be connected to the fixed internet (WiFi access points broadcast their availability for connection, by broadcasting their identification known as the Service Set Identifier). These broadcast packets are known as beacons. A hotspot's size is variable and overlaying hotspots can provide a mesh on which to communicate messages, between WiFi peers (acting as routers and end-points simultaneously). In this regard WiFi routers also have the ability to form ad-hoc networks. WiFi is still prone to collisions, but cannot detect collisions; therefore collision avoidance techniques are used.

In the context of P2P mobile devices, the advantages of WiFi are such that it provides a widely used means for ad-hoc LANs to be created reasonably easily without the necessity for cabling. The protocol is increasingly found embedded in many devices and form factors. WiFi Protect Access (WPA) looks to be a good replacement for WEP (Wireless Equivalent Privacy), providing improved security to users. The standard has started to include improved QoS mechanisms to improve the power efficiency of devices.

WiFi's disadvantages stem from a number of technical issues based on spectrum availability and power consumption. Power usage is extremely high for WiFi, limiting battery life for the device. As previously described security is an issue with WEP's breakability however this has been repaired to some degree through the protocol's replacement with WPA. WiFi like all wireless radio communication techniques has a limited range of between 45 and 90 meters. Other problems also stem from the increasing numbers of devices embedded with WiFi connectivity. The spectrum band is becoming noisy, sometimes referred to as "spectrum pollution". Long-range WiFi is presently being researched to extend WiFi to range many kilometres. Within WiFi at the P2P level this is still affected by power issues.

### 2.3.5 Alternatives

While Bluetooth and WiFi are the most common techniques presently employed to create ad-hoc networks, they are not the only techniques available for this purpose. Other interesting alternative systems include: xMax [84], PicoRadio [60] and the more futuristic SkinPlex [37].

xMax has been developed as a combination pairing to WiMax (IEEE 802.16), providing low-powered usage requirements with extended range to WiMax. The target applications for xMax are streaming media and voice-over-internet protocol (VOIP) applications. While WiMax requires a dedicated frequency band, xMax does not. xMax's market is however the broadband connection market.

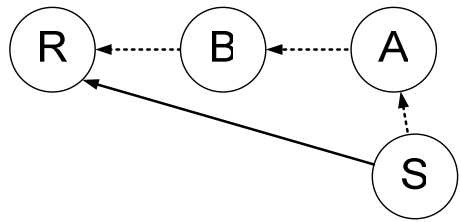

**Figure 6. PicoRadio's routing mechanism, where the sender (S) seeks to send a message to receiver (R). Instead of S broadcasting with increased power to R, the peer routing is intelligently distributed over neighbours A and B.**

PicoRadio was aimed at providing piconets for sensor network, to provide true ubiquity for applications. Rabaey et al. [60] believe that there the experienced improvements of using their PicoNode system would be two orders of magnitude more power efficient than existing communication techniques. SkinPlex transmits at the skin level of the human body, using the skin as the ether for communication. In generally its application is originally authentication and security applications. The communication range is up to one meter from the embedded user's body. How viable such a close range PAN could be for P2P content sharing remains to be seen. It could be argued that SkinPlex should remain termed as a Body Area Network (BAN) communications standard. While none of these standards are absolutely perfect for fast paced mobile ad-hoc networks, further standards like IEEE 802.15.3 (Wireless USB) are expected in the future which seek to provide high bandwidth connectivity with low-power requirements in the facilitation of mobile ad-hoc network connectivity.

## 2.4 P2P Implementations and Platforms

Various implementations for information communication have been developed. This section outlines some approaches to message communication in ad-hoc networks. Most involve the setting up of overlays, routing sets of peers or range based communication techniques.

### 2.4.1 Mobile Service Overlays (MSO)

SOLAR [14] and JEcho [15] are two middleware approaches to achieving efficient and effective P2P communication using the publish and subscribe methodology.

SOLAR [14] is an open platform, implementing a middleware layer, utilising a publish and subscribe mechanism for peer-to-peer communication. The approach of placing contextual information into the system is known as dynamic injection. It was originally created to reduce the amount of complexity inherent in context-aware systems development, providing an easier means by which to locate and use resources within the network. Context information is encapsulated within the infrastructure, allowing applications running on mobile devices to execute more efficiently. Solar shares a collection of context modules across the network, thus reducing costs within the system.

Contextual information is considered published information. Context information is customised for those peers who are subscribed to a specific kind of information. Processing contextual information is sharable between peers, reducing computational costs between groups interested in similar information. Described alternatively, if a node in the multicast path interprets a specific published data collection in a specific way, then this interpretation can be pushed onwards to other interested nodes. Because context information is common to multiple parties, this specific approach works very well. SOLAR generates a graph of events to traverse, to reach a conclusion on the state of the environment. For mobile P2P this could be useful, because state information from multiple sources could be similarly grouped and interpreted by mobile peers.

JEcho [15] is similar to SOLAR, however it is a Java implementation and it utilises "opportunistic overlay networking" to achieve its goal. JEcho uses an opportunistic overlay with the publish and subscribe method using special objects including Modulators, Producers, Consumers and Broker overlays, to "dynamically construct" broker network topologies to match the physical network.

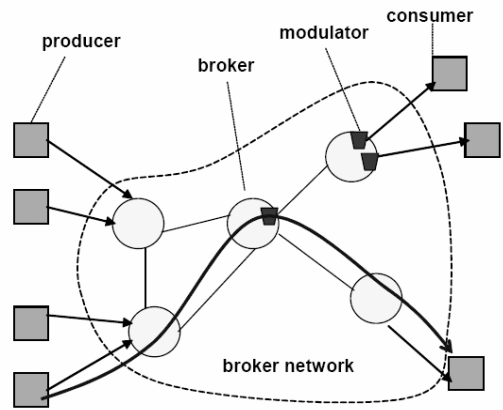

**Figure 7. JEcho's specialised objects where the Broker acts as a router [15].**

The broker effectively acts as a router to provide base by which information can be routed communicated through the network. Each time the brokers change the overlay routing paths are recalculated and a new overlay routing path collection is generated. Event Brokers provide event information to event Consumers who are subscribed to specific information. JEcho was tested using the AODV protocol. Overheads for the opportunistic overlay were considered "moderate" whereby the average bandwidth used between brokers. The topology of the physical network is periodically checked with the overlay. Topology tables are kept by brokers about the state of the network at a specific instance in time. Each Broker updates its neighbouring nodes by using an Expanding Ring Search algorithm. After this first stage is completed the table of each Broker is broadcast to other Broker nodes (known as Broker Propagation). Information is only sent about changes within the topology. A sequence number confirms the state of a network at a specific update. Tests on JEcho suggested that opportunistic networking were better than static methods for event distribution, even in cases where the peers were significantly mobile.

Juxtapose (JXTA) [43] is an open source peer-to-peer platform from Sun Microsystems, which aims to provide interoperability and platform independence. Being an open standard makes JXTA easily portable. The system consists of so-called super and edge peers, and each peer has a specific role within the system. Message passing uses XML. This has the benefit of providing a commonly understood language for messages allowing multiple devices to use the platform.

```
<?xml version="1.0"?>
<!DOCTYPE jxta:PipeAdvertisement>
<jxta:PipeAdvertisement xmlns:jxta="http://jxta.org">
<Id>urn:jxta:uuid-59616261646162614E50472050</Id>
<Type>JxtaUnicast</Type>
<Name>TestPipe.end1</Name>
</jxta:PipeAdvertisement>
```

**Figure 8. An example Advertisement message in XML.**

JXTA uses an overlay network created by the peers to communicate with other peers. This provides direct communication overlay. Each peer has a unique ID which remains constant even if changing its location. Peers have the option to form groups, thereby providing a logical method by which to propagate messages through a network.

Various protocols are required for the JXTA to provide the P2P network; these include a Resolver, Information, "Rendezvous", Membership, Binding and Endpoint protocols. Advertisements are made in XML stating where resources are and what types of resources are available. JXTA provides a high level yet simplistic approach to providing a method of finding peers and resources across a network, sharing content and communicating securely. Example applications using JXTA include myJXTA [53] and WiredReach [82].

An alternative to JXTA is P2PS, which aims to provide a simplified layer of abstraction over JXTA. A lightweight approach is beneficial if we consider that the JVM running on various form factors can have detrimental impact on the computational speed of a specific device. A major benefit of P2PS over JXTA is that the P2PS protocol is more customisable [78]. Expeerience is a proposed middleware for JXTA [7]. The middleware provides self management services transparently.

The LightPeers [16] platform seeks to provide a so-called "lightweight" approach to P2P using a specialised architecture, data model and set of protocols. LightPeers is motivated by mobile learning opportunities. The approach is applicable to a multitude of form factors. The software stack of LightPeers consists of four components. At the lowest level the P2P layer has direct access to the Network Layer, thereby providing direct access to various network standards like Bluetooth, WiFi and Zigbee. The approach is holistic in that it builds the entire peer-to-peer approach from basics. The approach is still however under development. The stack is unique because it does not use TCP/IP, rather introducing its own P2P Layer. In effect the LightPeer system improves performance for mobile P2P by building a basic low level interface between the application and the Physical Layer.

## 2.4.2 Implementations

### 2.4.2.1 Friend Relay

The Friend Relay [12] is a generic "resource sharing framework" for wireless devices which provides automatic service discovery, configuration and resource management. The motivation is to pool resources within a mobile context and make those resources available to a set of mobile devices. Resources in the context of Friend Relay refer not only to storage available, but also hardware resources (e.g. an internet connection, printer etc). The target device is that which has limited power resources; hence the approach is a lightweight. Computational load is spread or balanced between peers to improve the performance of sharing peers.

The authors of Friend Relay produced a prototype system dubbed Friend Relay NAT, which sought to prove the feasibility of the system. Their findings determined the benefits and feasibility of sharing mobile peer resources. While Friend Relay holds great merit, the system's feasibility has only been tested and simulated on a static configuration. This is probably because of limited time. The system is in its early development, yet the idea holds promise. The authors are taking a holistic approach to the systems development. This could be argued against given their constraints. Friend Relay highlights difficulties both in design methods available for mobile ad-hoc systems and tools available for the development of mobile resource sharing applications.

### 2.4.2.2   Presence and Proximity Systems

Shan and Shriram [68] propose and implement a P2P system for so-called Enterprise mobile applications based on the presence of devices and the logical proximity of devices. A backend of enterprise servers serve mobile ad-hoc devices using a common gateway comprised of many possible communication interfaces. They assume that the availability and connectivity of mobile works is questionable, hence enterprise mobile applications could help groups or sets of users share common (logically relevant) information.

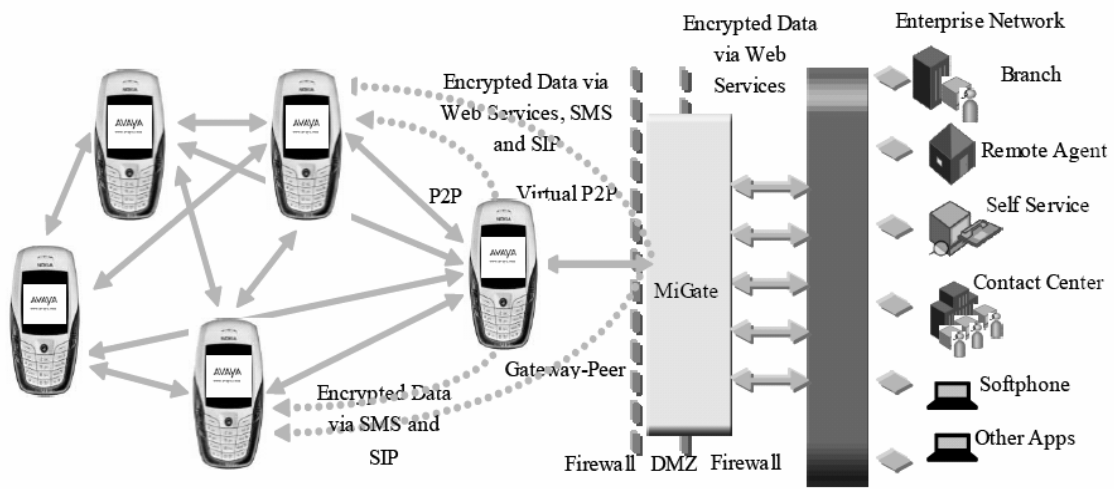

**Figure 9. The system architecture and arrows defining the flow of information.**

The system seeks to provide a way of lowering the amount of data required between enterprise servers and applications in the field and resource use could be predicted for users of services given their proximity to other users. This would mean effective utilisation of resource for mobile users. Highly distributed enterprise applications are argued to be better applied using the P2P methodology. Two models of decentralised and semi-centralised P2P for enterprise applications are considered. Deciding which approach to use is dependent on the applications being provided to enterprise application users.

Their method groups users together into sets. The groups thereby determine the logical distance between two peers. Peers within the same group often need the same services and same information. Communication is assumed to be achieved using various short or long range wireless methods. Changes in the presence of a peer trigger the event that peers should communicate.

The most interesting characteristics of this approach include the logical association of information to mobile users and the flow of information in and out of the MANET.

BlueTorrent [42] is a file swarming protocol which seeks to provide P2P file sharing between mobile Bluetooth enabled devices. Swarming is implemented due to limited bandwidth and short connection time constraints. Users can share content between one another while moving within range of one another. If both peers have data to share symmetric communication is used while where only a single peer needs to transfer data, the asymmetric mode is used. Factors such as signal range, user speed, user contact and density affect the ability of users to share content. The pervasiveness of the Bluetooth protocol makes it a good protocol to use, however there are problems with the protocol in relation to service discovery, power usage and security. Service discovery is slow and power usage can sap the battery if Bluetooth is constantly being used.

Prior to connecting to one another, devices must discover one another. Bluetooth's piconet limits the effectiveness and time taken to achieve this step. To solve this problem the BlueTorrent protocol randomly switches between master and slave states using the so-called "periodic enquiry mode". Discovery latency is defined as the recorded elapsed time between a two devices discovering one another. To reduce the discovery latency, BlueTorrents authors have used analysis data to find the optimal settings. The change in settings included reducing the channels for discovery from the normal 79 to 32 channels. Reducing channels results in a smaller range of channels to search for other devices, while there is a tradeoff in the possible channels used for communication. This could also have security ramifications.

Jung et al. [42] express the feasibility of the protocol within dense moving dynamic environments, where there are a large number of sharing peers. Within the test domain BlueTorrent was prone to error where influenced by wireless interference.

### 2.4.2.3   Vehicular P2P Systems

The Smart Web Project [71] has been successful in producing a vehicle warning system using ad-hoc networks. Using a network created between cars on the road information can be shared about accidents and other road information. Cars waiting in a specific lane could be alerted to the problems ahead of them. The information collection systems used could be advanced enough to intelligently decide what information should be collected and reported to neighbouring peers.

The benefit of a vehicle P2P system is that power is available in large quantities to the devices required to provide the P2P connectivity. The computers can also be larger, thereby improving computational capacity. Vehicles travelling within range of one another are more likely than not travelling in the same direction (or are stuck in the same ranged location) and their broadcast range is larger than a small mobile device. They serve as a verifiable base from which to state that mobile ad-hoc content sharing is possible and that it could have many beneficial outcomes. Some of the ideas of Super Peers could be ported to the Smart Web system.

P2Ping is a mobile system proposed by Seet [67] implemented on a collection of city buses. P2Ping makes super-peers more relevant within a P2P network. Within this architecture a fleet of city buses provide a backbone for communication. Using the backbone content sharing and distributed computing applications can be provided to users within range of the buses. The buses in this case are the super-peers, nodes of greatest traffic and load carrying capacity.

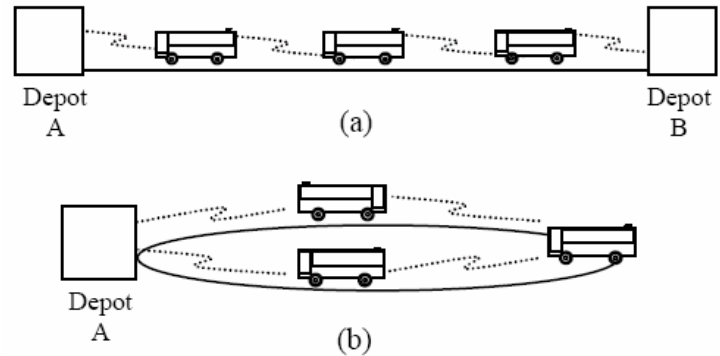

**Figure 10. Buses act as Super-nodes, routing through other buses to reach other depots. (a) considers a trunk service between depots, while (b) considers a loop service [67].**

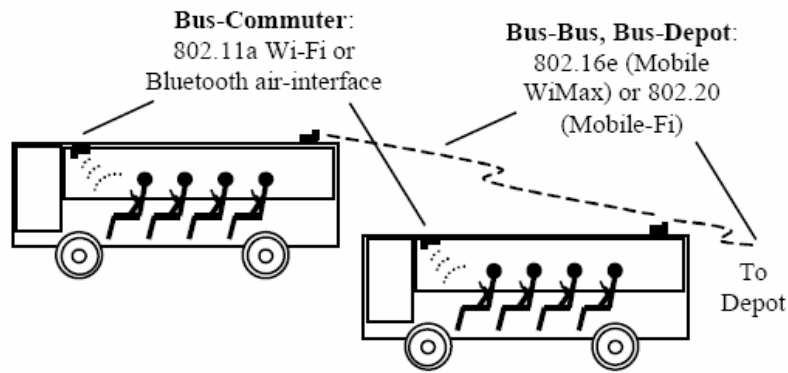

**Figure 11. Commuters use over-the-air bus services and buses are interconnected [67].**

Seet [67] argues that buses have many advantages as they are frequent, have power resource capacity and they come into contact with a large proportion of a city's population each day. Their coverage determined by their route taken each day also maximises their coverage (contact) and therefore their ability to act as carriers for MANET services. Using Singapore as an example, there are 2500 bus services, travelling 190 routes. This coverage is unparalleled except for planes and perhaps MANET peering embedded in all motor vehicles. Services to provide to customers include mobile content sharing and mobile gaming. Super-peers would keep an index of their users.

Seet [67] proposes that such systems would be wasteful in using P2P architectures like Kazaa and instead should opt for systems based on Chord or Pastry, which utilise distributed hash tables (DHT). A structured look-up service is also considered for the purposes of locating information objects relevant to specific users. P2Ping is an extremely interesting approach because it comprises many elements from many ideas, including MANETs, overlays, opportunistic networking [36][45] and social networking, into a single P2P sharing scheme.

NOM [21] is a resource location and discovery system, which is able to maintain a lookup of resources within the dynamic environment. Each peer runs a copy of the Nom client. The client monitors local message traffic and the neighbours it is connected to.

When queries are detected it uses a standard set of messages to resolve the queries. The algorithm acts on the types of messages received. If a message has already been processed it is ignored. The neighbourhood is periodically checked. If a query message is detected, then the Nom client will send the query onward to other neighbouring nodes. Every message has a TTL (time-to-live) counter associated with it. In this respect, Nom is similar to FreeNet [17] in that it uses a controlled flooding approach to resource discovery and data collection. Flooding is however extremely expensive and wasteful of resources which could be used for another purpose. NOM could be useful in single hop instances of contact.

### 2.4.3  Summary

Given the present pace of systems of P2P within the mobile realm, we can expect an increasing number of mobile devices to exist within everyday life. With this in consideration we could expect individual mobile devices to increasingly merge towards a single communicator carrying many capabilities (phone, camera, GPS, PDA). We can also expect a multitude of different communication techniques available to users of this specific device (GSM, WiFi, Bluetooth and other wireless standards). Each technique is required because no single approach is a perfect solution, where implementation is dependent largely on the application. From this basis, content sharing is a logical outcome as wired communities begin to share self-produced content in a style synonymous with the coined terminology of Web 2.0. Web 2.0 could be argued as major motivational factor in the generation of ad-hoc content sharing network infrastructures. API-wise, there is still no standard providing direct and easy implementation of ad-hoc services.

There are a multitude of exciting possibilities for applications operating within mobile ad-hoc environment. The more common ad-hoc content sharing systems already include military applications and vehicular information systems, with P2Ping being one of the most interesting P2P applications. Message passing could entail applications which have the ability to find professionals to solve problems, within a specific range or proximity of a user. Logical proximity applications are feasible. Disaster information provision is the common example with opportunistic networking. Applications could exist which provide communications capabilities to poorly developed communities, who lack a base station umbrella system. As communication standards and batteries improve it is likely that users will use audio and then video communication tools with mobile ad-hoc networks for multi-hop (Voice-Over Internet Protocol type applications). A future of resource pooling ad-hoc mobile devices as put forward by the authors of Friend Replay is feasible.

As previously described in this report, P2P content sharing applications will require the ability to conduct peer discovery, transparent P2P communication, multi-hop communication, multicast communication and content sharing, coupled with node adaptation techniques based on the types of applications and data being shared. Users should have the flexibility to use services on both fixed and mobile peers, i.e. a single platform which supports multiple services.

There are varying arguments to what mobile computing problems entail. Some argue that approaches to application development for mobile computing are no different to other computing paradigms. In effect they state that mobile devices are just small versions of larger computers, so the design models and methods to produce these applications should be the same. The other extreme alternative approach is that mobile devices are a new area of computing and the resource constraints which they are subjected to provide a basis to argue that they need completely different approaches to application resolution. A more moderate view is likely. Mobile device application design is dogged with problems relating to the reliability of services. Both new and old approaches need to be merged to provide an improved basis on which to develop applications. In some ways the design problems for mobile devices are similar to space science applications (less the detrimental effect of severe failure), where there are fixed and limited resources, which need to be efficiently harnessed to achieve a goal. The constraints with respect to dependability are very different from larger computing paradigms.

At the lowest level, the fundamental requirements of a system providing ad-hoc connectivity for P2P file sharing applications requires hardware which is low in power consumption, mobile and low in cost. Intelligence is required either locally at each device or globally for the system to adapt to changes in the dynamic environment. Another challenge in providing P2P services in ad-hoc networks is firstly finding data and secondly to route that data to an intended recipient. Latency needs to be reduced when connecting devices, as contact time between devices may be very short. Issues arise where peer-discovery and handshaking is slow, especially for Bluetooth and WiFi.

At a higher level, some applications need a dependable network on which to support content sharing applications. To achieve this may warrant the use of overlay networks (Chord and Pastry), middleware alternatives like SOLAR and JEcho and architectures such as LightPeers, JXTA and P2PS. DHT exhibits a promising approach to content routing in real-time, but the nature of content sharing makes sharing non-immediate. Hence methods like opportunistic networking are an extremely useful alternative approach within this context. Utilising existing approaches used with static P2P networks, we can devise approaches for mobile ad-hoc P2P, which are more efficient given limited resources.

Content sharing using mobile devices has strong and clear motivation and feasibility. The creation of such applications is still however difficult. Some of the problems previously encountered at the software level, in systems which are static P2P, serve as a basis upon which to build and improve mobile ad-hoc systems. Technical improvements increasingly improve the feasibility of content sharing systems, while Vehicle P2P implementations serve as a precedent for the possibilities of content sharing applications for mobile devices in dense structured mobility settings.

# 3  The Market Contact Protocol

This section presents a description of the market contact protocol (*MCP*) in both its Simple and Extended forms. The MCP is concerned with the controlled *infectious* spread of messages within a dynamic system of collaborating peers, which utilise intelligent policies and actions to influence the course messages travel. It bases it's usage on the *opportunistic networking* paradigm. The power of MCP relies largely on a peer's ability to infect other collaborating peers with messages when they come into geographical range of one another based on their sharing policy attributed to *incentive*. As the MCP is a broadcasting protocol, it needs to find an optimal point at which *infection* is maximised and *message quantity* is minimised.

This section considers the aims, assumptions, incentives and architecture of the MCP discussing the environment in which peers live and the effect this has on the choices made in the protocol's design and application.

## 3.1  Aims, Concepts and Assumptions

The MCP seeks to provide a generic method for content sharing within a dynamic system of communicating peer-to-peer devices using local forwarding, social interaction and human peer mobility. A dynamic peer-to-peer (P2P) system is one where the distributed system changes its topology (structure) and peer properties unpredictably for successive time steps. Peers are unpredictable in their availability to collaborating peers. The P2P methodology seeks to exploit collaborating peer resources and connectivity to provide services which are easily scalable. Load within the network is not unduly confined to one point in the system. As an extension of the client-server methodology, P2P peers can act as both clients and servers simultaneously. As more and more peers join the system they provide opportunity for more resource capacity. Hence, a major benefit of the P2P paradigm is the opportunity for peers to provide and share local resources which include computation capacity, storage capacity and communication bandwidth. The distributed nature of peers adds robustness and service availability within bounds. Where services or information is redundant in the system we can achieve robustness and fault-tolerance.

*Redundancy* refers to the replication of data or services in the system. In other words, where a servicing peer is lost, another peer can replace that lost peer if available in the provision of data or services. In a multi-peer environment this removes a single point of failure from existing within the system. The issue of redundancy is considerable in the context of the MCP. A lack of information or service redundancy would reduce the robustness and fault-tolerance of the P2P system.

Within this report, peers are considered dynamic mobile user devices or computing devices which are capable of moving freely within geographical space and communicating using various communications techniques locally available (e.g. Bluetooth, WiFi or future techniques). The notion of geographic space is important. Geographical spaces influence the performance of communications techniques. To achieve the aim of message sharing, the protocol makes some assumptions about the allocation, movement, positioning and distribution of peers, the reliability of communication methods and the dynamic environment in which the peers exist. MCP requires an environment consisting of mobile peers and ether to facilitate communication. We assume that each peer travels towards intended goal-locations using *paths*. A path is defined as an intermediary route traveled between two points. In reality peers travel from a start point, to an end point using intermediary points. We define a path as a collection of so-called waypoints. Peers are assumed to cross paths or come within close range of other peers. Hence we assume that peers could exist within the same local space at a specific time. Periodically peers broadcast their stored message collections. When peers intercept messages they apply policies to decide on how to treat the messages. A policy might be that when a specific message is received the peer should store it and in the future forward that message to other peers. In basic terms, policies describe the rules attributed to messages when they are intercepted by a receiving peer. Peers have the option to share messages that they have received with other peers and so a recursive process shares messages repetitively. In all systems of MCP peers we would assume that given a set of peers, enough time and all peers sharing, a single message would ultimately be received by all peers within a system.

The MCP seeks to capitalize on the procedures of a market system. MCP peers act as either sellers and/or buyers. Peers can take on either of these roles simultaneously. We use the terminology of an originator to describe a peer which a message originates from. Sellers (content creators or originators) are located at a specific time point in a specific location. Sellers in general, announce or advertise the items they wish to sell to a collection of MCP buyers, whom may or may not be interested in the goods or items they are selling. The sellers achieve communication between listening (MCP buyers) parties by broadcasting a message. Simply, sellers broadcast messages and buyers interpret those messages. *Contact* is defined to be true if two MCP peers are within range of one another's broadcasts. A message is interpreted by the MCP buyers and subsequently they have the choice of replying if they are interested in a specific sale or message. The reply by an MCP peer can be direct, use the MCP further (a more statistical means of reply) or choose another method of communication.

With some *incentive* a subset of the buyers who may have heard a specific sale announcement may decide to carry that announced message to other peers and act as a *proxy* for the original seller. Incentive is required for the usage of an MCP peer's resources. Within the MCP, incentive descriptions travel within messages. In essence, the provision of incentive is fundamental to MCP message sharing. Without an incentive to share there is no means by which messages can travel to other peers, there is no added value to the system by peers. The free-rider problem exists [27], where leeching peers simply benefit from the use of other peer resources and do not provide benefit to other peers. Incentive could be achieved by compensation or altruistic group behavior. MCP associates incentives with peer policies. We discuss approaches and alternatives to incentives more completely when discussing the implementation of the MCP.

The *contact* portion of the protocol is such that the protocol makes use of the social contact between roving peers, maximising the relevance of period social interaction of groups. The protocol exploits both more predictable and less predictable peers to share content. The geographical social network serves as a method by which to communicate specific information to specific types of users. Messages have the opportunity of being carried by social groups interested in that type of message or by subsets of peers who have an incentive to carry a message. The group association implies an altruistic incentive on the part of collaborating peers to carry a message, which uses precious power, communication, computational and storage resources.

As mentioned previously, the protocol's ability to share content is grounded fundamentally by these three device properties: a) power consumption or usage, b) communication cost and c) computational costs.

Many mobile devices are up until now, dependent on the batteries or cells used. Without power, the devices are rendered useless for communication. There are many methods which seek to make efficient use of cells. The knock on effect of power consumption is that it affects the communication and computational abilities of a device. Communication costs are exacerbated due to environmental changes (effects such as radio signal attenuation, refraction, reflection, diffusion exist) and the relative positioning of communicating peers and any other objects within the effected range of a radio signal. Radio propagation affects the characteristics of both signal transmission and reception. A greater range between two communicating peers requires more power from a transmitting device. The MCP assumes a reasonably short range contact, most commonly between zero and 150 meters although the MCP approach could be used for greater ranges as technologies improve. Computationally, the more computation or processing that a device is required to do also requires more power to be consumed. If peers are highly intelligent then we can be reasonably certain that the device will use more battery power to compute those extra decisions from policies. Storage capacity is required for peers to hold information for other peers. The necessity of storage by peers is dependent largely on the manner in which the MCP is being used. Large content collections would require longer contact times for communication (dependant on bandwidth). Large storage capacities are more likely to allow a peer become involved in the MCP for sharing purposes. We assume near unlimited storage capacities. The larger the content being shared the more effective the MCP must be in context-awareness, so as to intelligently examine the cost-benefit of broadcasting a message and its failure to receive. Described alternatively, each time a Seller peer broadcasts a message it needs to invest time, energy and computation in the sending of that message. It needs to be reasonable sure that the message will be received and not dropped before it can broadcast.

Within the MCP, it is assumed that peers have a selection of properties, which include vectored motion (speed), a geographical position in time (navigational component) and an irregular transmission and receiving range (communication range limit or communication coverage area). The radio coverage area is considered irregular because of the previously described radio propagation effects.

Peers are unfortunately as predictable as their users in their *mobility*. The term mobility is used as a broader term to describe the movements of peers throughout the world over a period of time. An awareness of mobility and its affects are important to the MCP and the protocol's ability to share content between neighbouring peers within the P2P environment. *Predictability* refers to the ability to predict a peer's future mobility. Predictability has two extremes, on one extreme peers could have *goal-orientated movements* and this could be predicted within variable certainty, while in the other extreme, peers could exhibit *random movements* making their mobility unpredictable. Goal-orientated movement judges that peers have specific goal locations which they seek to reach. From this concept we can identify the intended end point which a peer should find, however intermediary waypoints may be more difficult to confirm. Predictable goal-orientated movement provides for the possibility that some intelligent logic could determine with reasonable accuracy the whereabouts of a node within future time steps. This could have beneficial consequences for future communications. Random movements are assumed by MCP peers. The protocol is not in any control of peer mobility. Hence MCP needs to handle this environment and cannot be sure that its messages will be received by any peers. Indeed much of human nature is more random. In essence the MCP at present is unable to use much of this information. Content sharing is more probability based and predictability is attributed to contact durations and times. The MCP has the possibility of logging and learning during a daily trip when it comes into contact with the most number of neighbouring peers. In this way it could intelligently tailor its usage to specific hours of most certain contact in Extended MCP. This is however not possible in Simple MCP as there is no provision for a peer to know or learn anything about its environment.

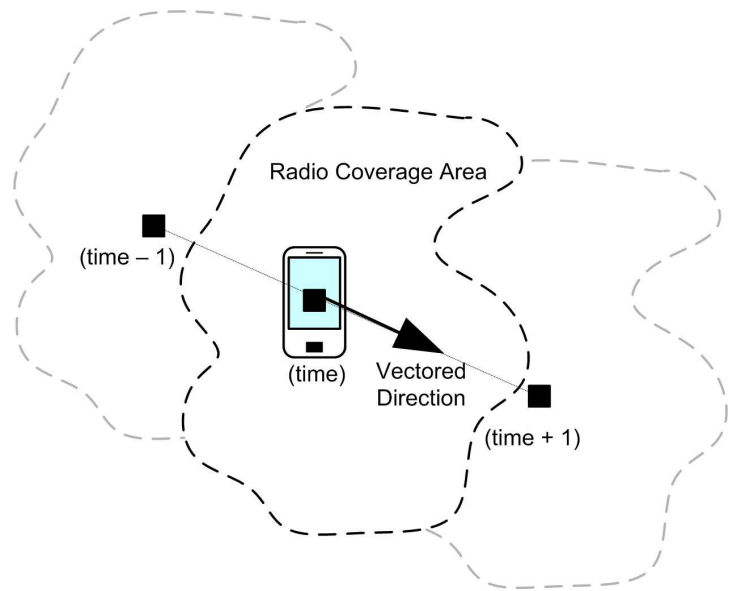

**Figure 12. Peers have properties which include a changing radio coverage area and a vectored direction which changes the present location of a peer to a new location in a following timestep (time+1). The waypoints are expressed as filled boxes. When the waypoints are connected they describe a path expressed over a period of time. We assume that the path follows a straight line intermediary processes.**

Before discussing the optimal environment for the MCP let us discuss *contact* and its definition. Hui et al. [36] highlight in their research, what they term an opportunistic network of contact points, to provide messages a mechanism by which they travel over an extended period of time. In their consideration they consider a perfect system where all peers are willing to carry messages for other peers. The messages of these systems act similarly to viral worms, infecting grouping sets of peers with information. The ideas expressed by Hui et al. [36] are fundamental to the MCP. However the MCP also assesses the importance of ideas put forward by systems such as JEcho [15] with its specialised intermediary objects and the message XML approach of JXTA [43]. The MCP does not however use an overlay network to determine a routing for messages.

An MCP peer in both its forms (Simple and Extended) is uninterested in the movements of collaborating peers (other peers moving around it). Peers are more interested in the contact availability of neighbouring peers and the durations of contact. Within Simple MCP, peers acting as sellers repetitively broadcast their messages at a constant *period* (they do not actively sample the environment to check other peers in contact). Peers acting as buyers repetitively listen for messages (passive implication). We can think of the *contact* approach as Boolean in nature. We could formalise that a set of peers {a,b,c} are in contact in truth at successive instances of *time*.

$$\text{contactA}_{\text{time}} = \{\text{b,c}\}_{\text{time}}$$
$$\text{contactB}_{\text{time}} = \{\text{a,c}\}_{\text{time}}$$
$$\text{contactC}_{\text{time}} = \{\text{a,b}\}_{\text{time}}$$

So if without warning b is removed from the network from the network in a successive time step (time + 1), then the interaction of connectivity between peers could be considered as:

$$\text{contactA}_{\text{time}+1} = \{\text{c}\}_{\text{time}+1}$$
$$\text{contactC}_{\text{time}+1} = \{\text{a}\}_{\text{time}+1}$$

We associate ownership with contact. In other words within the environment although peers do not know they are within contact they have the opportunity of communicating their message with a subset of other peers. In Simple MCP the seller peer does not know anything about its environment; there is no need to confirm that a message has been received. However, extensions to the MCP make allowances for *incentive* and *peer value* to be negotiated between peers, prior to a peer carrying a message for an originating peer. Hence in the extended form of the MCP peers have some idea of how many peers are around them at a specific moment in time.

*Contact counting* also termed *threshold* in the MCP is essentially the record of social networking. The notion of contact counting attributes contact to counters. When a peer receives a message broadcast from any other peer it increments the counter associated with that sending peer and records the time. As an example in a local storage on peer A, the peer has two entries: B = 1, C = 1. In this way peers can improve their worth or value in the system. This is *peer value*. Peers who see many other peers over a period of time may have a higher worth for content sharing than those that don't. Contact counting comes in two flavors, a simplistic one-way counting (counting only the sender) and a more complete two-way counting (counting sender and receiver, requiring a reply). Peers who have many statistical contacts during a daily path are good carriers of messages. These more sociable or gregarious peers provide a stronger method by which to share content or messages. As well as counting, extending MCP could associate counters with last known positions of peers using GPS (Global Position System) co-ordinates, however this is more contentious an issue due to privacy. This is not included in the MCP, but could be a possible extension. Similarly peers who carry more messages for other peers could also have increased value in the system. They are themselves knowledge bases and seeders, and may know more about a given system than other peers. This is analogous to looking for a person who may have many answers.

We consider that possible extensions to the ideas of peer value are a possibility. In the MCP we make the provision for the protocol to value peers on average distances traveled by a peer during a day, contacts seen, individual peer ratings (fellow peers rating each other) and transactions done. Fundamentally this is content-awareness which affects policies.

The MCP as with most P2P systems is cornered, in that it requires a base of users to make the system operational. Peers form the infrastructure on which messages travel. No active peers means no system. The MCP in this way is tailored for best results to dense collections of users and we highlight the settings in which the MCP is most beneficial. As we can expect messages to last longer than usual in the MCP, we can assume that the messages live and travel for days, perhaps even months until they frizzle out and die. The city setting is a commonly good dense arena for the MCP to be supported. We expect fluctuations of peer density surrounding a user over a period of a day. A commuter traveling to work may start his or her day in the sparse setting of a home (Figure 13). If traveling to and from work using a train or mass transport, then we expect that user to see many people (high density) for interludes during the day. A train would be a perfect place where the MCP could switch on and share content, before peers move off once again. In this way we exploit human mobility which is often made up of short bursts of action and then reasonably long intervals of rest. The intervals of rest are those opportunities when the MCP can take best advantage of peer to peer content sharing.

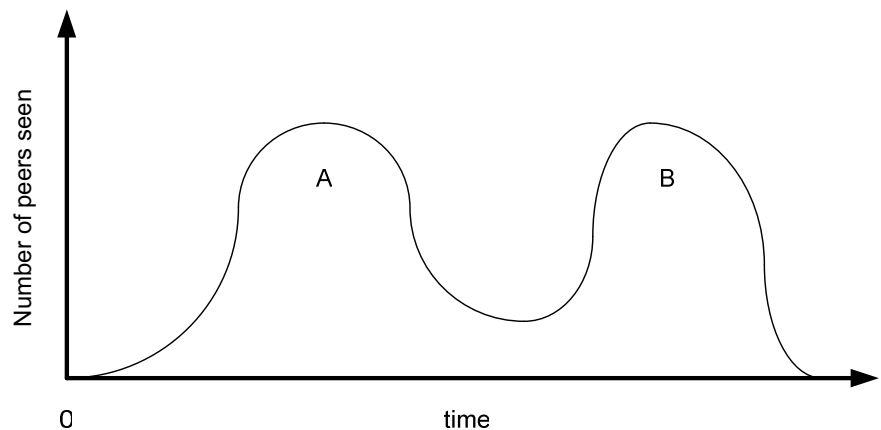

**Figure 13. An expected peer contact count for a person commuting to work each day in a dense collection of peers graphed according to time. A represents the time of greatest peer density during a commute to work, B represents a time of greatest peer density returning from work (peak points are most the most desirable points at which to broadcast messages to maximise message coverage versus communication cost using limited resources).**

We can make the analogy of *density breathing*. There would be a pattern of increase and decreasing density over time. The density of available peers increase, then decrease, then increase as users go about their daily activities. The points in a day where the density is greatest may exist for brief periods of greatest movement or activity followed by rest. As alluded to previously, we expect that when changing density settings (environments), there is a period of massive peer activity in a short space of time, followed by rest, when a peer reaches its goal for a period of time. We could expect increased communication success, rather than erroneous incomplete message communications. Also the best time to share content or messages would be that time in the MCP where there is the largest number of contacts seen. This more intelligent approach to content sharing is discussed in Extended MCP. The peak points of density provide the most desirable (cost effective) points in time at which to send messages to achieve optimal message coverage.

The protocol makes further assumptions about the technologies available for communication between devices. Devices are expected to use either present day or future radio communication specifications like WiFi and Bluetooth, but the protocol lends itself to also using standards like GSM and other similar standards for reply purposes. Some advances like master/slave switching are assumed to be possible (limited numbers of present day mobile devices have this ability). It should be noted that present protocols are in reality not perfect for the implementation of the MCP. The MCP would best be applied in the case where communications channels are controlled and managed as full-duplex rather than half-duplex, however the broadness of one way communication in MCP also makes half-duplex a consideration for more simple radio communication techniques. The MCP would benefit from a simple low energy and short range broadcasting technique. This is due to the simplicity of the protocol. Context switching is also fundamental if peers are to act and behave as both Buyers and sellers simultaneously.

Communications standards have their relevant limits to bandwidth and range usage. The protocol makes allowance for future content sharing provision allowing content of varying type to be distributed and shared between peers as long as the future communications standards provides the ability as such. The payload of a message could hold anything. We could describe this alternatively by stating that as the methods of communications improve, so more bandwidth intensive content can be shared. With the mobile device converging into a single "communicator", we could expect self-created content like videos to be shared using this protocol. In the Extended MCP we have the opportunity of using *context-awareness* using *sampling* to determine the best usage of mobile peer resources. This approach is motivated by data locality. The Extended MCP is put forward as a method by which to analyse the cost-benefit of communication using contact counting. Before sending a large message a smaller message tests the receptiveness of the system to a possible broadcast and determines if a message should be broadcast at that time.

The "contact time" of the protocol refers to the assumption that given the location and movement of devices, there is enough time for devices with range of one another to establish a connection and communicate their shared data. In this regard we reiterated that the MCP is based fundamentally on broadcasting shared content to other peers. It is assumed that some peers may be more or less mobile than other peers.

The content which is shared is assumed to be small enough for the consistent operation of content sharing. In Simple MCP packets need to be small, such that the periods of contact between devices may be short. Extended MCP considers that the protocol operates where large data content is shared, however the periods of contact would need to be long enough to allow for the complete transfer of data. Simple MCP could be used in mobile settings while Extended MCP could be used in cases of a device being at rest for periods.

## 3.2 Architecture

This section describes the basic architecture required for the provision of the MCP in its Simple and Extended forms, discussing the generation and reasoning of the architecture. The Simple form of the MCP does not use an approach to investigate its environment dubbed "sampling". The Extended MCP makes use of preemptive sampling before continuing with Simple MCP content sharing, to improve the effectiveness of communication and reduce the resource costs associated with broadcasting at regular periods. The section also considers some of the extensions of the basic model to accomplish possible enhancements for applications and the improvement of the protocol, to optimise the sharing of generic messages and content. We discuss the differences between Simple and Extended MCP more extensively in section 3.3.

### 3.2.1 Structure

The approach taken for the Simple MCP architecture is grounded on the ideas discussed previously. Fundamentally the peer has limited intelligences determined by the policies associated with the device. Policies are rules which influence control logic following a simple input output regime. Each peer consists of four major logic components: a) a *peer information*, b) *seller* (broadcaster) *policy*, c) *buyer* (receiver) *policy* and d) *contact database* components.

The MCP *peer information* component houses the local identifiers of the device and any specific local contact information. We could think of this component as a kind of business card stating the unique identity of the peer and how that peer can be found or communicated with alternatively. The unique identity of a device has benefits in those cases where MCP is used to authorise transactions and where *peer value* is implemented. If a transaction takes place between two parties then the buyer has a description of the seller's unique device International Mobile Equipment Identity (IMEI) number and contact information, such as email address, telephone number, digital certificates and other information. When dealing with monetary transactions we need to ensure non-repudiation and authorisation. At present, the IMEI number is a unique code of fifteen digits which allows a mobile service provider to identify individual devices on the network. The IMEI number could however be replaced with another unique identifier. In this way devices could be validated and a web-of-trust framework could be used to provide a trustworthiness rating for selling and buying peers. Trading reputation and peer value would help to reduce indecent trading. Each time a buying peer makes a transaction they can rate the seller peer on that transaction and thereby attribute worth to that Seller peer. This approach to peer value is useful to both Seller and Buyer. The Buyer leaves behind their details and their rating making themselves available for future questioning by other peers, improving their worth as a source of information (perhaps selling this information). The Buyer improves their reputation by having more favourable and referencing ratings.

The MCP selling and buying policies are based on the choices made by a specific peer (user) or they may be provided in a default form. By extending the Simple MCP idea we provide the MCP peer with the opportunity of making its own intelligent decision on what to do with a particular message. This approach is fundamentally *context-awareness*.

We provide the peer with the opportunity of learning to interpret and identify messages which a peer's user may be interested in. The benefit of such an intelligent approach is that is provides a foundation to improve the MCP; however the added intelligence could have drawbacks on the computational abilities of a peer device. More time, computation and finally power is needed to add policy intelligence to a peer. This may only be a short-lived limitation of present technology.

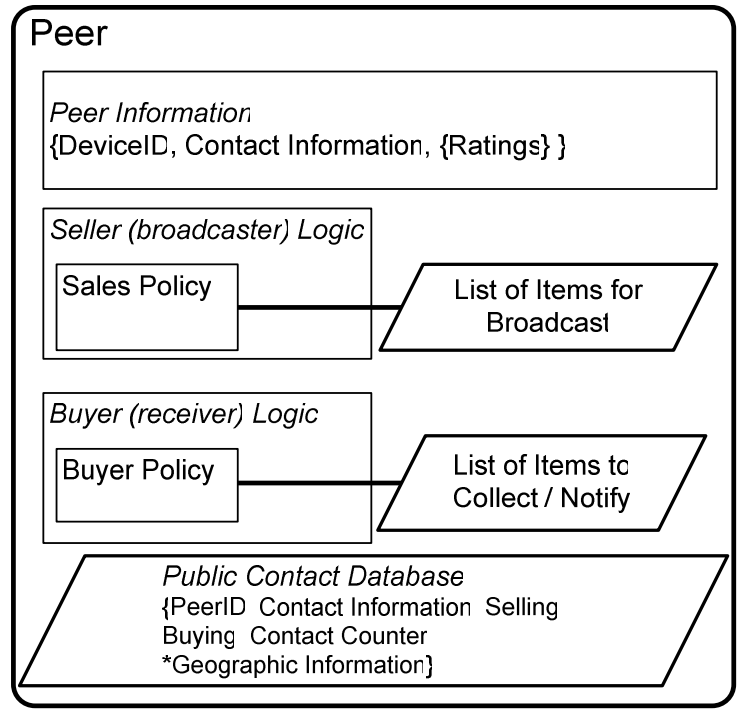

**Figure 14. The components required for the protocol to operate. Seller logic accesses a list of items which the local device aims to sell. Buyer logic contains a list of items the local device aims to buy. The public contact database holds proxy buyers and sellers (actions which the local device is communicating to other peers on behalf of those peers in the database).**

Policies describe what action to take when a specific message or request arrives. Policies describe the process which a peer should follow in action or inaction. The MCP sales policy states the *period* at which a message should be broadcast. We should expect that the lower the period, the more messages broadcast, the more likely a message is received by traveling peers. At each period point the MCP examines its policies and messages stored. Hence there is a relationship between the MCP sales policy and the list of items for sale (messages to be broadcast).

The public contact database is concerned with the storage of proxy requests. Each peer, that is not the originator, can decide on whether to store a message for future announcement. The contact database records the an announcement's originator peer, the originator's contact details, the originator's items for sale, the items the originator requests, a contact counter (referencing how often the local peer has seen the previous announcing peer and any geographical coordinate information (if it is available). Where the contact database is used to store the announcements of originator peers, we should expect that there is some incentive for the intermediary proxy peer to announce sale or buying requests on the behalf of another peer.

### 3.2.2 Messages

As previously reiterated, Messages (Figure 14) dictate not only the information being communicated between peers (data), but also the logic of the system. In this regard logic and policy are not directly or simply attributed locally to each peer, but also to the message originators requests. There is combined intelligence in both a peer and a message. Described alternatively, the message has some input into how it is handled by the peers carrying it within the distributed system.

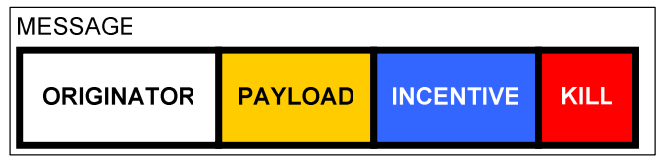

**Figure 15. A message breakup into its relevant data spaces.**

While each peer has the possibility of over-riding any recommendations associated with a message, messages provide some intelligence on how they should be communicated. We could consider the tags representing message policy to be recommendations to the peers whom handle them.

We can better describe Messages such that they follow a simplistic four tag format using XML (Figure 16). The originator, payload incentive and kill tags are the only fundamental tags but the message base is flexible enough to provide a generic means by which to communicate, depending on how the protocol is used.

While XML is a useful means by which to conceptualise Message format the MCP experimented with in reality does not use XML messages due to the redundancy of information found in XML data and the aim to minimise the required durations of contact between communicating peers. The major goal is to communicate information as compressed and quickly as possible to improve the effectiveness of the MCP to receive communicate messages. In cases where this approach is not fast enough to communicate information, MCP peers would receive fragmented Messages. With increased logic we could argue that fragmented messages could be logically repaired with other fragments, however the MCP does not make any provision for this fragmentation regeneration of a dropped Message it could do in successive versions perhaps collecting fragments from varying peers to constitute a whole message.

The originator tag is usually placed at the front of the message followed by the payload and the kill holders. The originator tag space should hold the IMEI number or a unique identifier containing the original seller, but it can hold contact information about who originally sent the message.

```
<message>
        <originator>
                <IMEI number>ABCDEF123456789</IMEI number>
                <phone number>+440123456789</phone number>
                <email>student@doc.ic.ac.uk</email>
        </originator>
        <payload>
                <selling>
                        <item>
                                <descript>ifony</descript>
                                <price>300.00</price>
                                <expires>14/06/07</expires>
                        </item>
                </selling>
        <payload>
        <incentive>
                <payment>0.10</payment>
                <currency>GBP</currency>
        </incentive>
        <kill>
                <hopcount>2</hopcount>
                <hopsexpire>10</hopsexpire>
                <date></date>
        </kill>
</message>
```

**Figure 16. The XML message fundamentally contains originator, payload and kill holders.**

The payload is concerned with the message content. In the seller-buyer scenario, we could expect that the protocol would be used to list items being sold, along with descriptions and prices associated with each item. Alternatively the payload could hold data which is specific to a data sale. We could expect an entire music, video or text file to reside inside the payload portion of the message. In this way, the protocol is generic enough to provide a basis by which content can be shared as messages are broadcast.

The incentive tag in the example used is empty; however the protocol makes provision for the possibility of providing an incentive for messages to be distributed. The kill space attributes a lifetime to a message. Beyond the lifetime the message is destroyed. Messages in the dynamic social environment could live for extreme periods. In this case there is a danger that old and useless data could live indefinitely within the system, using up bandwidth inefficiently. Killing messages is beneficial to the bandwidth and loading of MCP peers. However, this characteristic of messages also provides a space for which the MCP could provide *persistent data storage* to peers (unlimited message lifetime). The kill space will usually contain a default time-to-live based either on hop counts or time. We can expect multiple messages to be sent between peers. Those messages may or may not belong to that specific intermediary or proxy peer.

### 3.2.3 Policies

Policies are applied by MCP peers to handle either messages or context change when sampling in Extended MCP. The MCP policies are rather the reactions of stimulus from the environment in which a peer exists and travels given the local parameters of the peer. For the purposes of Simple and Extended MCP we chose a simplistic set of policies for peers to utilise. Policies are applied predominantly due to either Messages or external environmental inputs.

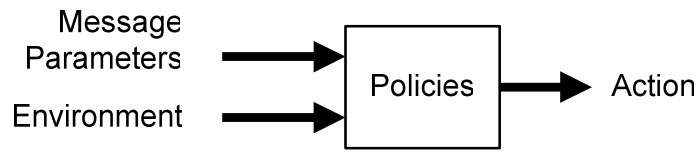

**Figure 17. MCP peer policies act on messages and environment inputs.**

We firstly consider Message policies, those policies applied due to changes in the parameters of Messages. We have previously discussed the originator, payload, incentive and kill parameters. By manipulating these parameters an MCP peer has the opportunity to decide whether to view and share content on the behalf of other peers. In general policies seek to limit resource usage, however over control by policies stifles the carriage of Messages by cooperating peers. Policies may apply such that a peer may only act as a proxy for a peer specific originator peer, social group, priced item, incentive payment or period of time. The MCP peer default policy is applied as altruistic. MCP peers act as intermediary peers no matter the conditions or the Message which was being shared. In any case that the Message parameters are changed, the resources of the MCP are thereby reduced. Fewer peers are available for exchange of content and communication. We term these modified policies and in those cases the MCP is dependent on the end-user to provide input into what Messages are passed on and what Messages are filtered out of circulation. As an example, to limit the number of messages in the system a policy would be employed stating how many times and at what interval a message should be broadcast.

Secondly the MCP considers policies relating to the effect of an MCP peer's environment. This is approached as *context-awareness*. Within Extended MCP sampling is applied to improve the performance, efficiency and effectiveness of an MCP peer investing limited resources in the communication of information. The policies applied follow such that peers who are surrounded by a specific density of peers within a periodic sampling state decide on a threshold number of neighbouring peers that must surround it, before in broadcasts its messages. The MCP peer samples its environment, counts the number of receptive peers and then broadcasts its messages dependent on this count called the *threshold*. The application of this Extended approach has limitations. Highly mobile peers may be out of range by the time an MCP peer has finished sampling and decided to broadcast. The threshold also may be too high to warrant a broadcast, stifling opportunities to share content. The usage of energy for sampling may in fact also be more than the cost of simple broadcasting because the initialisation of checking an MCP peer's environment requires the same resources as for Simple MCP broadcasting. Within the project we propose both approaches and evaluate the advantages and disadvantages of both approaches.

### 3.2.4 Process and Actions

Within the MCP we can define each peer as a finite state machine. Each peer executes a collection of steps depending on whether it is acting as either a Seller or Buyer or both. Messages are broadcast containing content. The peer to peer nature of the system provides that peers could act as both message receivers (sellers) and message transmitters (buyers). It is useful to think of the originator peers as the broadcasters and any intermediary peers as receivers (buyers) of messages.

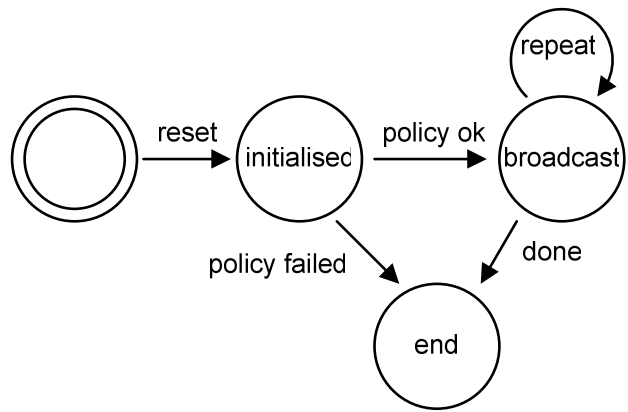

**Figure 18. Simple MCP: State machine for a Seller.**

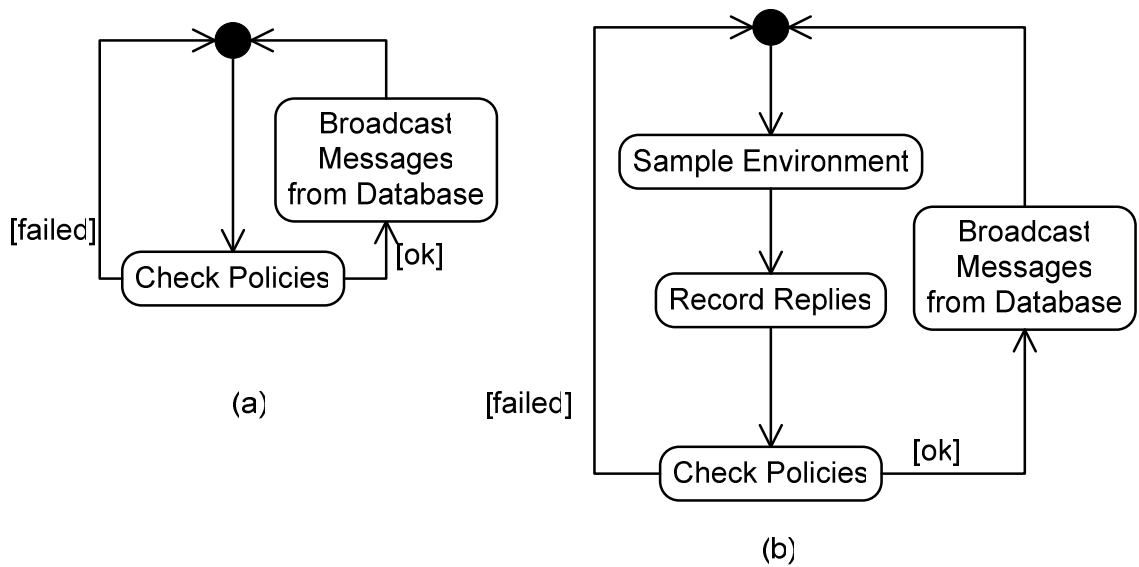

**Figure 19. Activity Diagrams representing Simple MCP Seller process (a) and Extended MCP Seller process (b).**

The MCP Originator peers or the sellers are intent on broadcasting or announcing a message (content) to any peer which is within range to hear a specific announcement. We should reiterate that an MCP peer acting as a receiver of messages is in effect simply listening for communications. Announcements are made periodically depending on the period settings of the peer. Reducing the *period* between announcements increases the probability of a message reaching a roving audience. The issue of period is important because it improves the chance that messages are received by fellow MCP peers. The system is dynamic, so we cannot be sure when and if peers are within contact. If the peers are entering and exiting the radio coverage area rapidly, then we have more chance of communicating a message to the majority of peers by repeating it rapidly. However if the environment is changing slowly, then there is limited benefit for repeating the announcement to those peers who may have already received the announcement.

In the simplest scenario of the protocol, sellers construct a message which is then announced on to potential buyers, who are within range of receiving the announcement. In this manner messages are passed from sellers to buyers in single hop jumps. Figure 18 describes these simple actions. The process is repeated until the device user decides that the announcements be terminated. A simple policy is that if a peer has an announcement (content) to share, then it should broadcast periodically. The alternative is such that the policy once consulted returns that no broadcasting should take place.

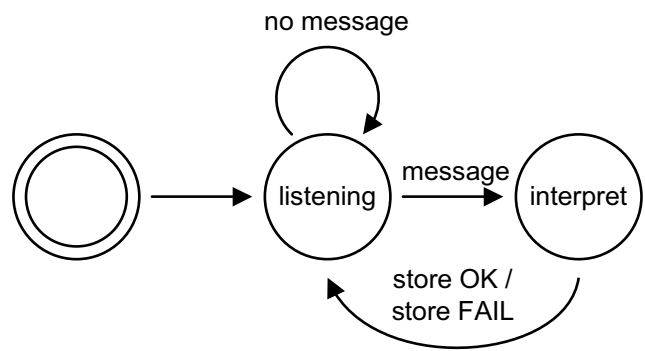

**Figure 20. Simple and Extended MCP: Buyer state machine.**

A Buyer works in a similar manner to a Seller; however its action logic could be more complicated. In a simple case a Seller receives a message describing a specific item, information about the item and the Seller's information tagged with the message. Just as in a normal market, if a buyer is interested in a specific item and they decide they would like to purchase that item, they can do so, by immediately contacting the seller using the calling card information attached to the message. Figure 20 and Figure 21 seek to clarify the process.

Fundamentally a buyer needs to make a decision on whether to purchase an item and then decide whether it will carry and broadcast the message for another peer (store-and-forward). Using this method, messages make single hops between peers, where there is some type of incentive provided for the mobile device to use its limited resources for the transport of a message.

As previously mentioned, the protocol makes provision for peers *incentive* to spread messages on the behalf of others for the benefit of the social network.

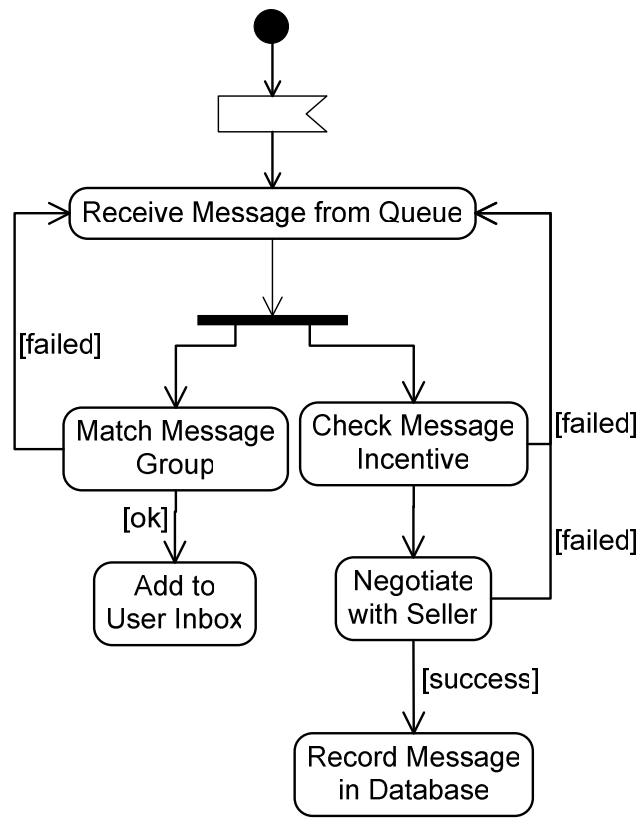

**Figure 21. Activity Diagram representing the Simple MCP and Extended MCP Buyer process.**

## 3.3 Scenarios

Within this section, we consider a collection of scenarios where buyers (broadcasters) and sellers (receivers) exist. We show how the MCP is formulated from these basic scenarios and argue the effectiveness of the protocol in its simplicity. Later we make extensions to the protocol to include *contact counting, peer value* and *incentive*.

For the purposes of explanation, Alice is a seller, while Bob and Charlie are both buyers. Simple MCP refers to the simplistic form of the MCP while extended MCP refers to MCP with extensions to improve the protocol's message sharing capacity. In each scenario case we first describe Simple MCP and then Extended MCP, arguing the necessity for improvement.

### 3.3.1 Single Buyer Single Seller

The most basic interaction between peers considers a single buyer and a single seller coming into *contact* with one another. A seller broadcasts it's announce message and a seller's message is received by the buyer. The buyer makes a choice based on policy to act on the message and then whether to store or ignore the message received, based on the content of the message. Simple MCP does not require a reply (point-to-point, non-broadcasting). There is however the opportunity to do so. In the Simple MCP, a reply would only provide a true interest and action by the buyer (Bob) upon the seller's item. The broadcasted message could be received by multiple sellers within range. While the broadcast is assumed to use a radio communication mechanism, the reply in contrast could include other single point communications solutions e-mail, SMS, physical acknowledgement or a voice connection. In essence this provides the buyer the opportunity of making decisions at a later time about the purchase of items from a seller. It also improves the privacy of listening buyers.

A reply action can be instantaneous if the buyer has registered a search for a specific item or gradual. In this scenario, *contact counting* is however one-way, unless a buyer decides to act on a seller's message. The seller has no knowledge of the buyer's existence because it is not always provided with a reply for its broadcasts. Only the contact counter of the buyer is incremented. In this case an entry is made by the buyer (Bob) that it saw Alice at some particular time in its travels. We should remember that each message is sent with some incentive for the receiving peer to pass it on.

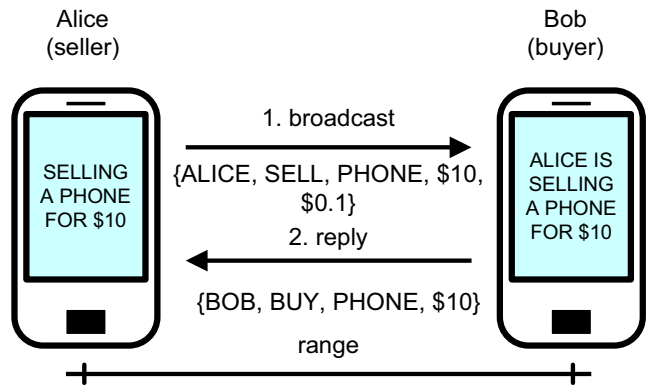

**Figure 22. Simple MCP, A single buyer and a single seller.**

We extend Simple MCP to Extended MCP by providing the sellers and buyers the opportunity of *sampling* their local space (active context-awareness), broadcasting their intent to get involved and testing the context in which they reside. A reply from sampling provides a wealth of information on the Buyer peer to the Seller, such as the Buyer's peer value, peer count and social groups it is affiliated with. The sampling would be executed prior to broadcasting a message. In this extension a sampling approach is used. Prior to any sale broadcast by a selling peer, peers can simply announce that they exist. In this way peers have a better opportunity to build social contacts. Peers have a choice to conduct this socialising based on their policy choices. The more sociable a peer the more it reduces it privacy, but at the same time, its worth within the network rises. Sampling is almost akin to making friends or meeting people and then determining their status or willingness to help share content. We should note that the payload of a message could provide incentives to replying, such as grouping. For example, a message could be sampling, but it could also say for which particular grouping. Peers could be sampling for film enthusiasts or music choices. A peer receiving a broadcast still has the option to not reply. Indeed this brings up an issue of benefit. The peer initiating the sampling gains nothing if no other peer replies. However by not replying the receiving peer still gains because it knows that it has seen another peer. We could however argue that Simple MCP also has this problem and that in any case a seller needs to go out on a limb to gain something.

While this *doubles* the number of messages broadcast it provides two-way counting. The buyer and seller can record each others contact adding increased knowledge to the system. sellers can decide whether broadcasting a message will reach an optimal number of peers. Replies can be via the previous reply methods mentioned earlier (e-mail etc). Where replies are broadcast it may be preferential to reply in this alternative manner for privacy implications. However if we assume that a reply could add benefit to the peer replying then the very act of replying adds benefit to all nodes surrounding a peer replying. All nodes within the receiving range of a no-reply broadcast can increment their counters to the new peer detected.

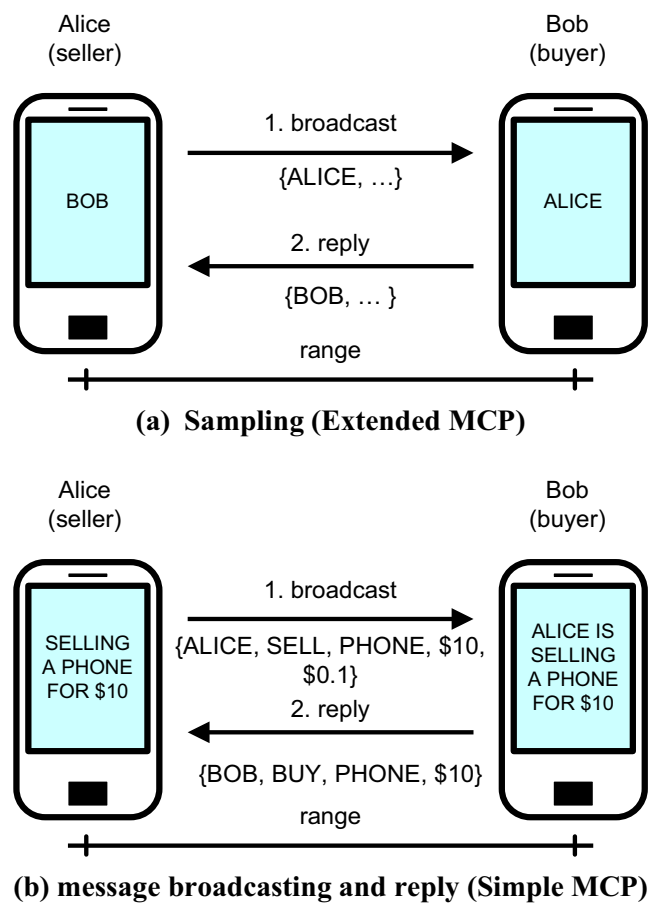

**(a) Sampling (Extended MCP)**

**(b) message broadcasting and reply (Simple MCP)**

**Figure 23. Extended MCP, peers sample their environment (similar to handshaking). The receiving peer decides based on policy to reply.**

Opposition to sampling is that it requires the sending of an extra message utilising limited resources. There is no payload of data shared, only circumstantial evidence of the environment at a period in time, where the next moment or iteration in time may have changed due to the dynamic nature of the P2P environment. Peers may be moving so quickly that sampling is useless. In this respect it is obvious that we may need to implement an intelligent approach to sampling in real systems. It may be that future mobile devices have the ability to provide accelerometer readings (via GPS or other means) to the overall operating system of a device. These readings could be used to determine the best opportunity in which to apply sampling.

If we extend the simple seller-buyer relationship we would expect that after a reply is made by an interested seller negotiation may take place between the seller and buyer about details relevant to executing the transaction. The MCP assumes that this problem can be solved using relevant negotiation and authorisation protocols and point-to-point communication.

### 3.3.2  Multiple Buyers Single Seller

In this scenario the MCP assumes that buyers are found in different coverage areas. The seller cannot see the end buyers directly. There is an associated provision for incentive on the part of other potential buyers to carry the messages broadcast by sellers. An incentive might include a commission for advertising a specific item for a seller.

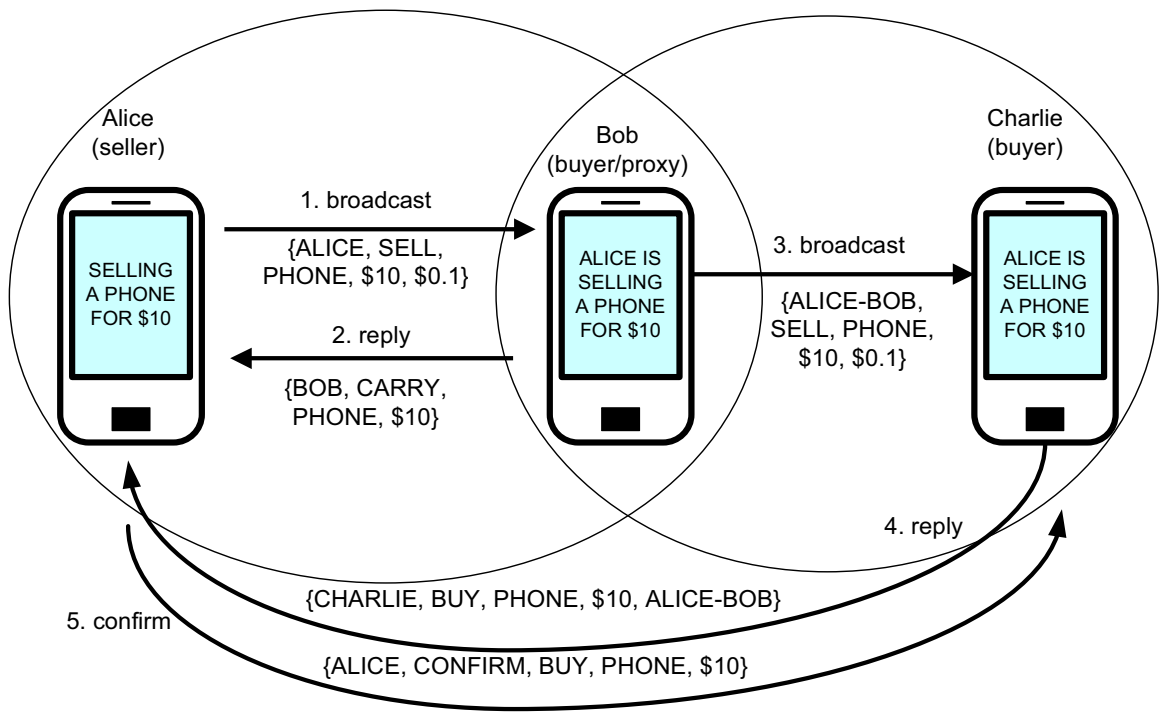

**Figure 24. Illustration of buyers acting as proxies for sellers.**

The scenario follows a similar progression as that of the simple buyer-seller case (1, 2), however due to the incentive for intermediary peers to act on behalf of the seller, the intermediary peer (Bob) becomes a proxy for the seller (Alice). After amending the message, Bob promotes the item for Alice, broadcasting to peers it comes into contact with over a period of time (3). We should note that any peer acting as a proxy should decide to notify the seller that they will be promoting their message. In this case we could expect a subset of the possible peers to act as proxies.

Finally a peer might act upon the message originally sent by the seller peer (Alice). A reply would then be sent via a long range means to the originator (Alice). Long range means might include an e-mail, SMS or voice call. If the seller has already disappeared from the proxy's space then we assume that at sometime in the future both seller and proxy will eventually be able to communicate be required. The message modified as it travels from peer to peer with the proxy peers information. A confirmation of the sale would then need to be made back to the buyer (5). In this context security protocols could be applied to secure the reply and confirmation aspects of the protocol. Once again Extended MCP *sampling* gives peers the ability to make more complicated and intelligent decisions on when and in what context to share content.

If we consider that a message might eventually arrive at a specific peer and we assume that the common daily tasks of a user force intermediate peers to cross paths, then we can extend the MCP one step further for the reply of messages. As illustrated a proxy peer has the opportunity to look for routes during a day or interval of time, which would allow the originator peer to read a reply. There may be huge time delays based on the contact of peers, however Extended MCP makes allowance for this routing of messages to specific receiving peers (in this case an originator). This reply approach to use the MCP for based more on probability that precision.

### 3.3.3  Single Buyer Multiple Sellers

In the case where a single buyer considers multiple seller's offers, we expect that the single buyer would make a choice upon which seller is best to purchase from within the local market. The Buyer Policy would consider the broadcasted messages received from the subset of sellers.

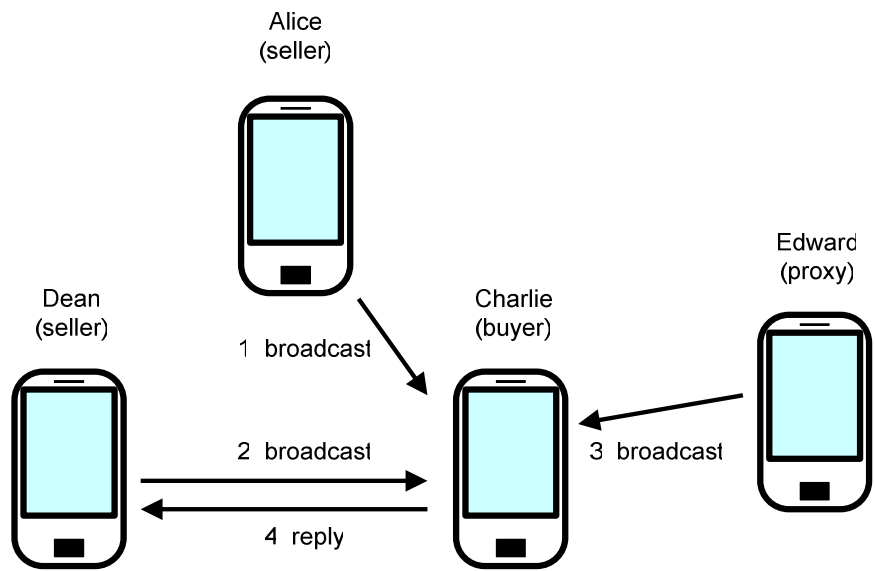

**Figure 25. A possible single buyer, multiple sellers scenario. Buyer (Charlie) concludes by choosing to reply to specific seller's offer.**

The scenario would be similar to the single buyer single seller approach; however there would be multiple messages to consider. The offers could be displayed or acted upon automatically by the buyer peer. In this regard we could expect that the buyer who considers its options for the longest would have the best possibility of considering more collected prices. In this respect, the MCP leaves designers the choice on how to handle purchases and in essence when and in what context to reply to messages.

### 3.3.4 Multiple Buyers Single Seller

Where multiple buyers exist a seller could consider the best price comparison or best incentive option, assuming that the content is equal. Indeed once again the MCP leaves the application implementation the opportunity of acting within this scenario. It may be beneficial for a seller peer to consider all the replies it gets from buying peers. Once again the logic and application of the market is associated with policy and in this case the seller policy used. We could expect a multitude of negotiations to take place between peers and policies are expected to follow market protocols intended for use. As an example an auction market could be created or a first come first serve policy could be implemented. The MCP seeks to provide flexibility in these cases. Indeed we could say the MCP is interested more in the sharing of messages and the provision of a flexible geographic market based system.

While we consider this scenario a possibility it would be a further technical policy enforce on the part of the peer. There is provision for it occurring however its use would be defined by MCP peer policy.

## 3.4  Contact Time, Environmental and Technical Effects

The model of the Market Contact Protocol revolves on the idea of a dependable system of interconnecting peers who maximise their usage of local social networks for the infectious spread of information. Issues pertaining to contact times, peer mobility, peer density, information carrying incentives, bandwidth, handshaking and the period between broadcasts need to be considered.

Contact times can be defined as the time period by which a buyer is within a seller's broadcast range. Stated alternatively, the time interval in which two peers have the opportunity of communicating with one another. Within this report the period of a peer is considered the time between successive message broadcasts. In the case where the contact time is short, we should expect that the period would need to be shortened. Messages would need to be broadcast at a higher rate to ensure that they are received by the maximum number of peers. Seller peers cannot determine this without sampling the active environment. With this in mind, we should expect that a tradeoff is associated between message communication and power usage. The lower the period, the higher the power cost and vice versa.

Peer mobility can be defined as the path which a peer takes over a period of time. Within mobility we can expect to see daily patterns arising. People traveling to and from work may follow a specific path each day at roughly the same time. It is fairly unlikely that people would act in a random manner, with a person's behavior being more likely goal-orientated where movement is concerned. Hence we should expect to consider credible human behavior when considering the protocol's use.

Peer density is a fundamental requirement. In peer-to-peer networks a lack of peers renders the network of little use. A dense collection of peers would benefit the delivery of content to multiple peers. A sparse collection of peers would only be limited in use, dependent more on the paths followed by relevant peers. We should expect that message should travel more quickly in dense collections of peers rather than in sparse ones.

Incentives to carry data may be required in circumstances where information has value. As previously mentioned the protocol makes provision for the description of an incentive to carry information, however those incentives may be varying depending on the information being shared. Incentives would be tied mostly to the resources of peer devices (bandwidth, power and storage constraints). We should anticipate that mobile devices of the future will have mass storage capacities larger than those presently available. Present technology for mobile devices provides the user with the ability to storage gigabytes of information. It is not impossible to assume larger capacities in the future. Storage may become an unimportant characteristic of a phone, where there may be seemingly endless storage capacity available to peers.

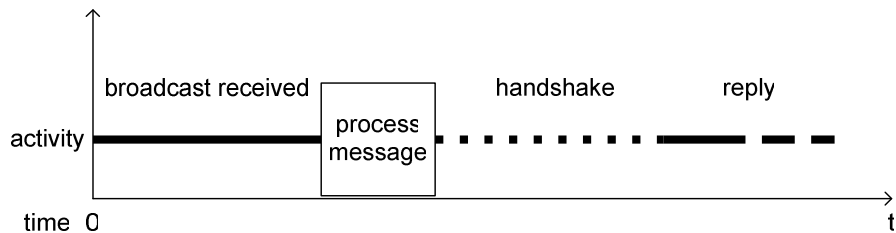

**Figure 26. The possible activities and their times for a device, receiving a message, and then replying to that message. Handshaking and negotiation could take extended periods of time, dependent on their extension to negotiate or not to do so at all.**

Handshaking is associated with the communication technique employed by a specific device. In the context of a dynamic mobile peer-to-peer network we need to optimally reduce the time taken for devices to discover one another and to communicate information. The protocol assumes a broadcasting capability for these devices. Broadcasting information has a benefit such that any listening device should simply have to listen to communications. Communicating with an originator peer is not always required, however the protocol makes the provision for incentives and to act on incentives, peers need to communicate with one another to negotiate terms. Handshaking is hence a secondary priority for the protocol to provide benefit to proxies. Where a fixed network exists the peer may be able to handshake and negotiate using the fixed networks communication methods such as GPRS or SMS.

Bandwidth constraints depend on the communication method (e.g. WiFi, Bluetooth or future methods). The larger the available bandwidth to a peer the faster it can communicate its message. Bandwidth requirements could be varying on the type of data being shared. Within the context of the MCP we expect small data payloads due to maximising communication during short contact, however the payload could hold large data collections such a text, video or audio files.

## 3.5 Summary

Predominantly the aim of the MCP is to provide a system on which content sharing can take place within collections of ad-hoc networks of peers, where all peers available possess various resource capabilities. The protocol seeks to achieve this by exploiting the opportunistic networks created by socialising peers and groups. We have also assumed that mobile devices which are massively mobile in this setting have limited resources pertaining to storage capacity, computational ability, communication range and battery power.

The MCP uses a structure which places emphasis on local data stores and logic to interpret and share content between peers. Messages while possessing payload also have the possibility of defining their movement and lifespan by groups of peers, through actions. The actions of peers are outcomes of policy and input injection from a peer's environment and messages received. We describe the process which both Simple and Extended MCP use to achieve the goal of content sharing, where Simple MCP is concerned with the rapid sharing of content without context-awareness and Extended MCP requires context-awareness to determine the benefits of broadcasting its messages. The various scenarios highlighted peer interactions relating to the number of seller and buyer peers within range of one another. We concluded by briefly reiterating the effects of contact time on the protocol's ability to perform. Contact time is an important factor for consideration in Extended MCP due to the application of sampling and power usage.

# 4 Implementation

This chapter considers the process used to test and evaluate the MCP. It considers the approaches taken and those alternatives which were possible, arguing and discussing the merits and pitfalls of the various approaches to implementation and testing.

## 4.1 Market Contact Protocol Simulator

To test and experiment with mobile distributed systems we considered the choice of both *actual physical* or practical experiment and *simulation* (modeling). Actual testing would involve the development and implementation of the MCP using real mobile devices. The benefits of this approach should not be discounted. Testing the MCP in a real world setting would without doubt provide a more rigorous and uncontrollable framework in which to experiment with the effectiveness of the protocol. More unpredictable factors such as signal propagation, peer density, power consumption and peer mobility could be more intensely scrutinised however the monetary and time overheads would have made this approach very expensive and time consuming. Where the MCP is sought to provide a mechanism for message communication and market provision we could expect to require the protocol to run amongst hundreds of thousands of peers. Time constraints made this real approach impractical.

The best, and perhaps the only, alternative is plain simulation or a hybridised version of real data with the execution of the MCP in a simulated environment. Simulating the MCP required a simulation framework on which to conduct tests. A model is however limited as well. Models are generally more predictable than real life tests. Chaotic factors would be more beneficial especially to wireless mobile device tests. Realistic assumptions were necessary and if necessary random fault injection. It is possible to overlook important effects which should be included in the model. To implement a realistic simulation framework is hence not a trivial matter as the model is only as good as the implementation and application of its assumptions.

Various mobile ad-hoc networking simulators were considered for the testbed on which the MCP could be implemented. Their choice was based on previous benchmarking research by Kurkowski et al. [47]. These included Network Simulator 2 (NS-2) [55] and the Scalable Wireless Ad hoc Network Simulator (SWANS) [4].

NS-2 was initially used to build the a basic MCP model, however it soon became apparent that NS-2 was extremely complex to not only code, but also to test. These problems arose such that NS-2 coding is separate from the scripting of the language (there are in fact two languages C++ and TCL). There must be a full understanding of both coding language and the scripting language to fully utilise NS-2. In this case it required time to learn and become fluent in TCL which was not feasible. A benefit of NS-2 was its track record and extensive knowledge base. NS-2 already had a reputation created through its usage and it could supply a benchmark upon which to compare various other protocols, however at the time, no known implemented content sharing protocols were known to have been simulated using NS-2.

SWANS had a powerful scalability provision due to its use of JiST (Java in Simulation Time). Massive systems could easily be built and tested within the SWANS platform and Java was used in its implementation and scripting, where JiST provided the basis on which instructions were converted. SWANS however was lacking in the respect that it did not have a large knowledge base, it had limited mobility models and the process of following data and procedures was difficult. SWANS did however prove that JiST was a powerful base on which to simulate systems.

With these considerations in mind it was decided to build a fresh simpler, more manageable and adaptable simulator on top of the JiST system utilising those benefits exhibited in SWANS. The concluding result was that the MCP simulator borrowed some initial ideas for implementation from SWANS - most importantly JiST. The MCP simulator aimed to provide a robust simulation environment to test the feasibility of the MCP, single hop messages, their movements and fully verify the actions and locations of peers in their mobility. The simulator also aimed to be extensible enough to at a future point provide a method by which to use real data mobility sets to test the usability of the MCP. Modeled simulations and assumptions could thereby be verified and checked with real mobility sets, a half way point in the confirmation of real and generated results.

### 4.1.1 Java in Simulation Time and the MCP Simulator

JiST (Java in Simulation Time) [4] is a "discrete event" platform developed for the scalable and efficient simulation of objects. A fundamental benefit of JiST is that it provides for the efficient execution of simulation programs using a popular and well known language (Java). The JiST architecture is comprised of four components: a compiler, a bytecode rewriter, a simulation kernel and a virtual machine [4]. JiST interprets JiST code, written in normal Java syntax on top of the Java Virtual Machine (JVM). Java Virtual Machine 1.6 was used for this simulator. A benefit of using Java is its ease of use. There was no need to learn a new way of describing or scripting objects within the JiST environment. Figure 27 describes the process of simulation on top of the JiST layer which begins with the compilation of JiST code using the Java compiler. The compiler classes are rewritten by a bytecode rewriter and the final product is then run on the Java Virtual machine.

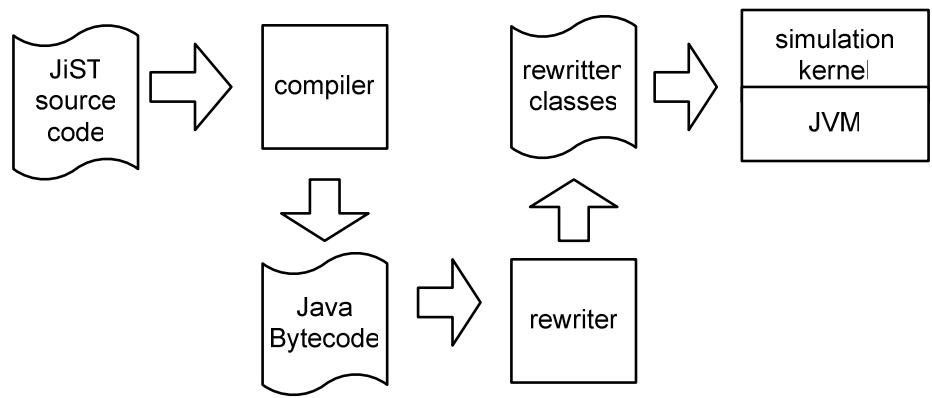

**Figure 27. JiST architecture comprises four components: compiler, bytecode rewriter, simulation kernel and virtual machine.**

The benefits of using this discrete event simulation model are such that much code written in Java can be reused. Many of the benefits of JiST are directly inherited from Java such as type-safety, reflection and garbage collection. The JVM provides a portable base on which JiST objects and simulations can be repeated. There is the opportunity of cross-layer optimisation between the JiST kernel and an executing simulation [4].

JiST provides a surprisingly good method on which to build and execute simulations due to the scaling, assumptions and parallel nature of the environment in which the protocol would be expected to perform.

The MCP simulator was successfully simulated with up to 125 000 peers. The limit to simulation is memory available on the JVM heap. Each experiment ran for a specific period of time concerned more commonly with the rate at which content messages could be shared. The MCP simulator itself used global geometry comprising the latitudinal and longitudinal coordinate system and vectoring. The benefit of this approach is that mobility data can be incorporated from the real-world or generated in an overlaying manner within the simulator on real world coordinates. Node (peers) objects ran in parallel on the JiST layer. Each Node was modeled along the MCP architecture discussed previously in this report. The primary attributes of nodes were such that they held information such as identity, storage capacity, location, vector direction, goal, incentive, a quantified sharing incentive, a content database and a contact database.

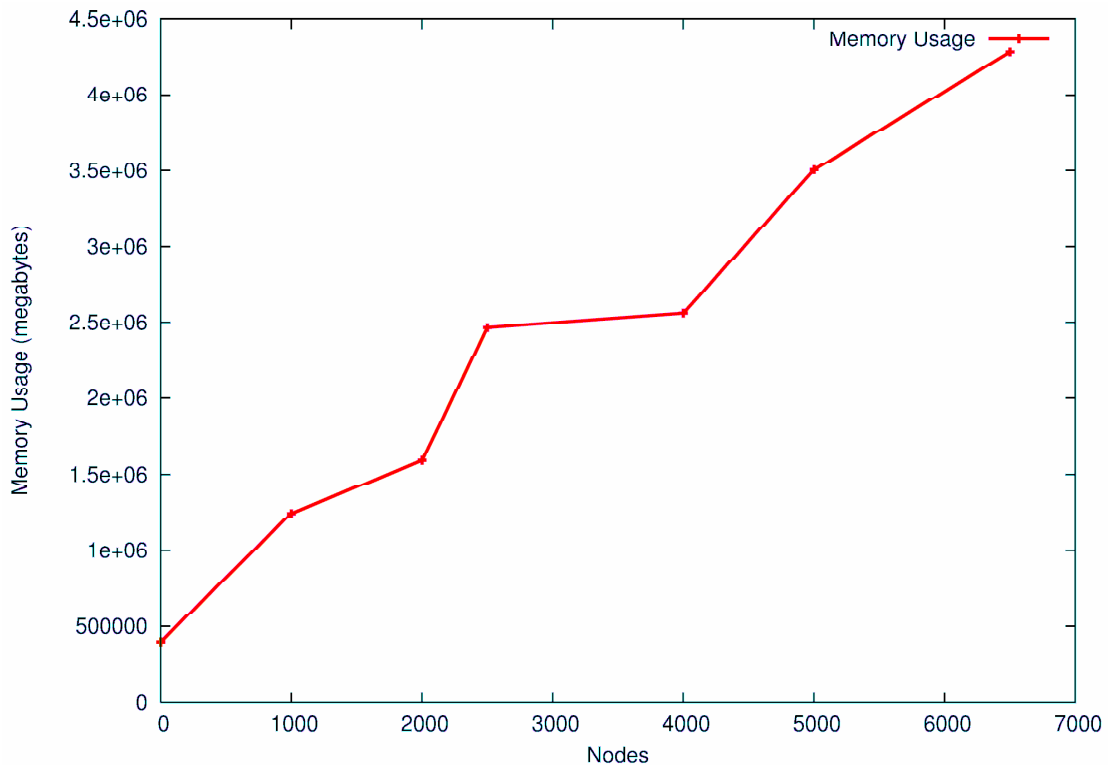

**Figure 28. Scalability: Memory usage increases as the number of nodes increase.**

Following an object orientated approach the Nodes existed inside a World, which provided the ether by which nodes could communicate. The World also dictated what was possible for a Node to achieve and what was impossible. Such an example would include the sending of messages between peers. The World object would consider Messages traveling and determine if that Message could be correctly received by various peers within range of the broadcasting peer. Realistic effects were easily added, where a statistical object recorded the statistics of the simulations during experiments.

## 4.1.2 The Earth and Objects

The simulator is bound by the geographic constraints of the Earth using the geographic coordinate system. In reality, each location is specified in the "degree, minute, seconds" coordinate system format using three parameters: a) latitude, b) longitude and c) altitude above mean sea level. The simulator converts between the degree, minute and seconds format to provide decimal degree coordinates.

As briefly described previously the simulator consists of three fundamental objects, namely the Node, World and Message objects, however various other objects provided extend means by which to manipulate the simulation environment and thereby provide more realistic conditions in which the peers might reside. These include the Tracer, Waypoint, Navigation and Statistics objects.

The most basic data structure used for mobility, location storage and computation is the Waypoint object. Each waypoint has a latitudinal and longitudinal position. Waypoints are interpreted by the World object for the determination of environmental effects, such as radio attenuation, reflection and message verification. Depending on the intended effects of the environment Messages use waypoints to verify the point at which they were sent. Using this model we can simulate many affects on the messages.

To measure the distance between any two geographical points we used the Great Circle Formula (GCF). This was implemented to improve the realism of the surface of the Earth which is slightly angular rather than truly Euclidian. Where we have two geographic points $P$ and $Q$, their latitude is represented as $\alpha$, and their longitude as $\beta$. The formula (4.1) follows such that $\Delta\sigma$ represents the angular difference or distance and $\Delta\beta$ represents the longitudinal difference. We can consider the points as $(\alpha_P, \beta_P)$ and $(\alpha_Q, \beta_Q)$. We should note that the Earth is more of a flattened sphere, due to its spin (geodic). When applying the GCF (4.1) we however achieve an error of up to 0.5% in the angular difference evaluated.

$$\Delta\sigma = \arccos(\sin\alpha_P \sin\alpha_Q + \cos\alpha_P \cos\alpha_Q \cos\Delta\beta) \ \dots (4.1)$$

Messages use the GCF however we should note that this distance is the maximum distance possible. We add limiting factors such as simulating radio propagation and limiting this maximum distance to add realistic effects to the simulator.

The Tracer object is the initial object to either interpret apply real mobility data from read data collections or used to generate paths using a collection of start and end waypoints. The Tracer object is important because it specifies the parameters of the environment and the experiment in which the peers seek to be tested. The Tracer also adds extra flexibility to the experiment. An experiment consisting of multiple peers can be superimposed on real road systems or on ground paths which can be collected or specified using points (waypoints). In effect we have the opportunity of intertwining what exists in reality with a possible conceived path of peers.

To calculate navigational results and statistics the simulator has a Navigation object. The navigation object is primarily a collection of functions provided to calculate peer or message distances, speeds and times, while remaining true to the effects of the environment. The Statistics object is concerned with the metrics of the simulator. Initial parameters such as the number of nodes, range, mobility, period and sharing probability of nodes are recorded here, as are the effects of the simulator which include experiment time counters, message counters, peer contact counters, content shared lists and erroneous message counters.

### 4.1.3 Peers

Peers have local logic which they require to interpret and decide on actions to take when messages are received. The local logic presented in simulation is modeled on the protocol directly. Nodes check the World object for messages which may have been broadcast within their zone and which the World object deems to have correctly reached them. The Node object receives the Message and interprets the payload and depending on the parameters of the simulation and its local settings it can decide whether to act as a proxy for the original Seller Node. The local logic is flexible for the simulation. A lag time is associated with receiving or computing and it is at this point that we can directly influence the realistic effects of each device.

### 4.1.4 Simulation Process

The simulation process comprises two basic sets of inputs and two outputs. The simulation parameters include the simulation duration, number of buyers, number of sellers, radio range, sharing probability, period and max speed of peers. The map data specifies the routes which peers will follow be them generational or real. The intended output is statistical data coupled with mobility traces which can be viewed and replayed. The simulator generates KML (Keyhole Markup Language) traces describing geospatial data and tracks of nodes. The data can then be fed into 3D geographic tools like Google Earth [28] to view the paths and activity of peers over time.

Figure 30 seeks to describe the simulation process. Initially the simulator would initialise the experiment parameters using the Tracer object. The Tracer object is used to set up the simulation for execution defining peer paths and simulation parameters. Waypoints are read and collected to describe the mapping which is sought within the simulation. The MCP simulator inputs for paths were designed to be easy enough for a user with little coding experience to define the simulation parameters.

```
U (51.501427,-0.180414) | (51.492243,-0.178214) S 1 B 100
```

**Figure 29. Simulation parameters held within the map file. The inputs state, uniform speed per peer, start and end points, number of sellers and number of buyers for the path.**

After this process has been completed the Tracer starts successive sellers and buyers as stipulated by the input parameters. Each node runs in parallel, executing its local logic in each time step (every second). Within the context of the simulator peers act as Servers if they provide content or are sending messages and act as Clients if they are receiving broadcasted messages for interpretation. It should be reiterated that peers have the ability to function as both Servers and Clients within the context of the protocol. In this way local behavior is switched depending on whether a peer is acting as a Seller, Proxy Buyer or a simple Buyer.

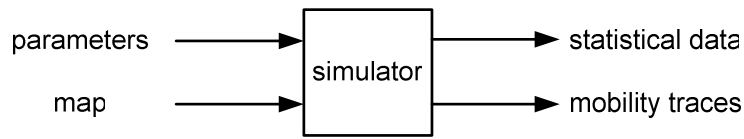

**Figure 30. Parameters and Map (path) inputs produced statistical data and mobility traces.**

As peers are executed they exist within a common World object. When considering the World we use the analogy of Nodes attaching themselves to the World and updating the common World information. Each node sees the changes other Nodes are making in the world, if they accomplish the contact requirements of the initial input parameters. The World provides the ether and affects the ether to provide a realistic means by which to test the credibility and performance of the protocol under specific conditions.

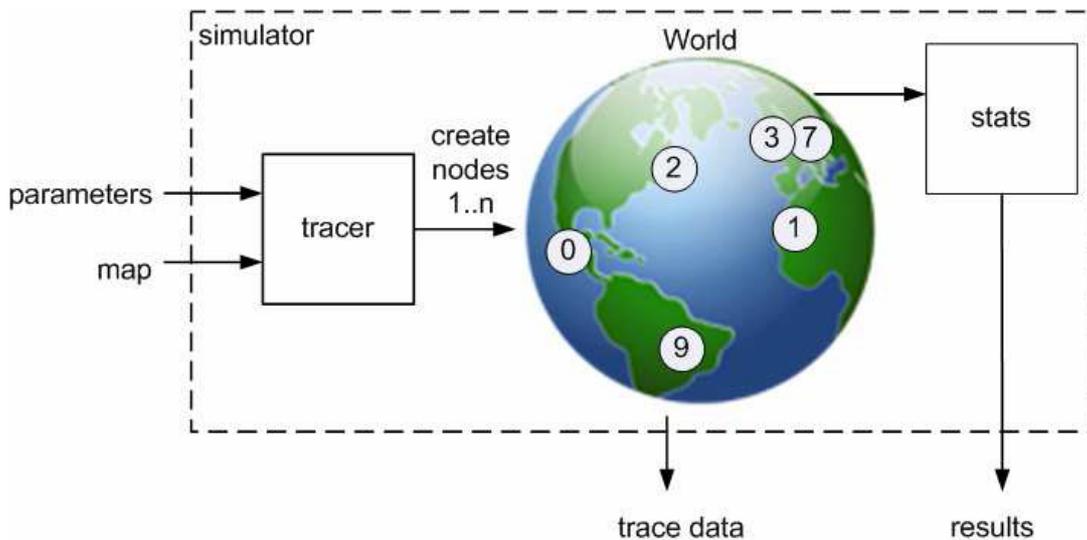

**Figure 31. Simulator and the major objects used including inputs and outputs. A few numbered Nodes exist and travel on the surface of the World.**

During the simulation Messages are generated and shared between buyers and sellers (peers). The Messages had a window period of 1 millisecond to be intercepted by peers. We should remember that the Buyer and Seller aspects within the simulator are best described as behaviors which peers can turn on or turn off. Messages created have a realistic element such that they are built similarly to the message description of the protocol; however they lack the size of the XML layout. In the case of simulation messages contain just simple String content. The interpretation of messages is not necessary for the spreading of testing. We are more interested in how messages pervade through the network, at what speeds, in what times and in what scenarios. XML layout is more beneficial a guideline where the protocol is fully implemented in real systems. This is subsequently achieved in a test J2ME (Java Platform Micro Edition) described later.

The experiment terminates where the simulation time lapses. Nodes are killed and the Statistics object is then used to collect and output the relevant statistics. Information collected with reference to each Node's position in time is written to a mobility file in KML format. The file output can be viewed in geographic visualisation tools to illustrate the interactions of peers and their state at specific moments in time. This is a similar trace output as GPS trace data. The trace data can be used in the future to corroborate either simulation inconsistencies or confirm likeness to real trace data collected on users.

## 4.1.5  Mobility Models

Mobility models capture the location of nodes as they move through space at successive time instances. In essence we create and use directed graphs of movement. Fundamentally two mobility choices exist for the experimentation of the simulator: a) generational mobility and b) real path mobility. In generational mobility we generate the paths used by nodes over a period of time. Real path mobility requires the accurate and if possible highly resolute collection of data points.

We expect that in a real-world environment we need to deal with scenarios which include various types of mobility. The simplest mobility can be assumed to be peers moving towards a specific geographic goal; however we should also expect that the protocol provide a service in a more realistic and richer mobile environment. Mobility modeling can be defined into two major groupings, those generated and those mobility models created from real data collected. The more realistic the mobility data, the more sure we can be that the protocol will behave as predicted in the real world.

### 4.1.5.1  Generational Models

For some time there has been an aim by some researchers to define mobility using algorithms and there is significant importance in providing rational and realistic generational mobility data. For the purposes of this simulator we consider a basic approach and then iteratively build up the complexity of the model while taking many of the algorithm mobility recommendations into account. The basic generational goal-orientated path consists of a single node traveling in a straight line (described as a pipe) at an average speed towards a single location (very stochastic in nature). For the purposes of this text and the simulator, we consider this a "unidirectional pipe" This kind of mobility is obviously extended further. Generational mobility has been grouped from this base case into successively complex cases, which are easy to setup and test. From this simple case (unidirectional), three derived cases exist: i) bidirectional pipes, ii) grids and iii) various irregular constructions.

A path (pipe) exists from a start waypoint and ends at an end waypoint. Nodes travel from one side of the pipe to the other. We can split nodes to enter and exit at specific points, while overlaying various areas with constructions using either unidirectional or bidirectional constructions. These constructions are useful for creating landscapes for the representation of city grids and other irregular road systems.

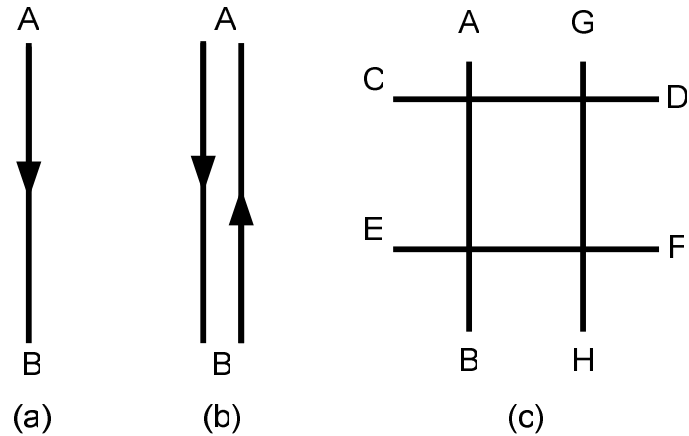

**Figure 32. The unidirectional pipe (a) is used to construct bidirectional pipes (b) and successively more complicated paths where nodes can travel such as a grid structure (c).**

The Node can also be controlled as it moves along the pipe. Simplistic motion would consider an average speed from A to B. This isn't particularly true to human or peer movement in reality, hence we can state more details in the simulator about more complicated Node movements, such as a random speedup slow down mobility or add further complexity to the model by affecting the pipes. Indeed irregular (random) mobility was added.

4.1.5.2   Real Paths Models

Various real data stores exist for GPS (Global Positioning System) mobility and contact data such as CRAWDAD [18]. MCP's use in dense collections of peers was difficult to investigate given the limited datasets, however we approach the MCP's application using such data, to justify its ability to operate even in small sets of inter-relating peers in Chapter 6.

## 4.2  Mobile Bluetooth Implementation

The MCP was further implemented using the J2ME (Java Mobile Edition) development kit for mobile devices with Bluetooth connectivity. This implementation sought to investigate the feasibility of MCP on Bluetooth and the highlight the barriers to MCP's implementation using presently available communication techniques and devices.

Following the manual initialisation of communication between peers we could achieve content sharing at a very basic level. Messages were received by Buyer peers. The Buyer peer was then required to make a manual decision on the acceptance of the first message between devices. After initial contact the peers could interact automatically. Figure 33 illustrates this initial contact procedure and the sharing of content. The Seller MCP peer discovers a receiving Buyer MCP peer (left). The content message is sent by the Seller MCP peer and once the Buyer MCP peer has accepted the connection, the message is transferred. Both peers count one another and record each other's identifiers.

We can see this in the line 0000BB1A077E = 1.The message content is printed on the Buyer MCP peer's screen and stored locally.

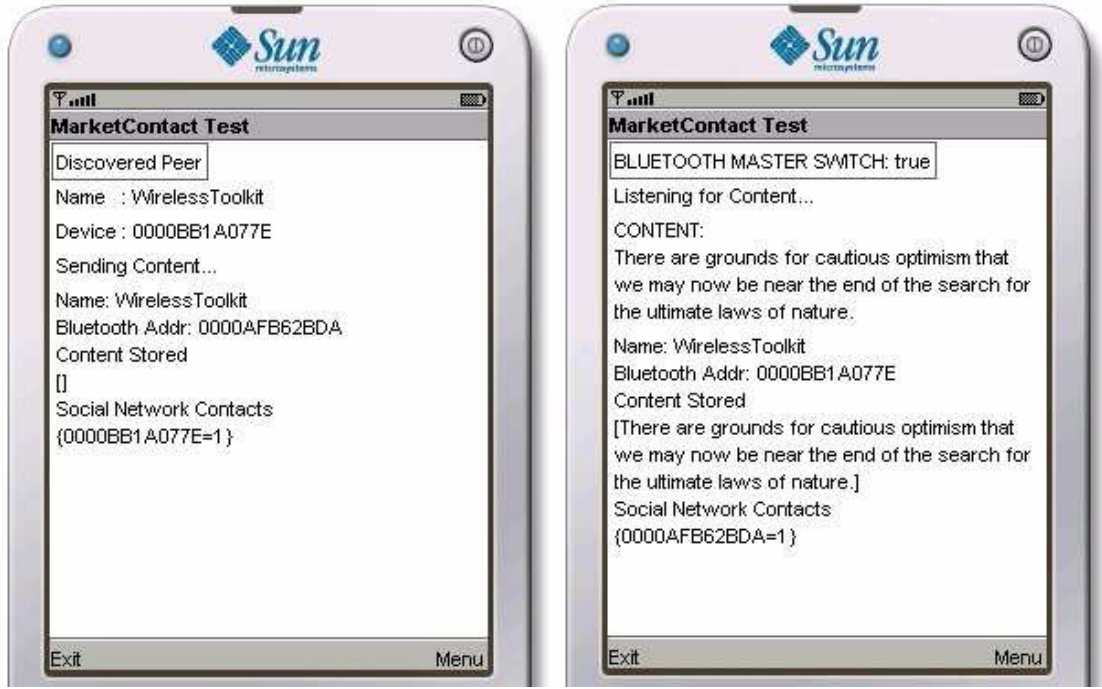

**Figure 33. J2ME application implementing simple MCP, emulated on Sun Wireless Toolkit. Seller MCP peer (left) and Buyer MCP peer (right).**

Bluetooth's many limitations made the experimentation and use of the midlet application troublesome; however the application provides positive impetus for implementing the protocol within wireless P2P networks. Within Bluetooth however, the Client and Server activities were swapped between buyer and seller due to the nature of the Bluetooth specification and the two way requirement for handshaking between communicating Bluetooth devices. buyers acted like Clients and sellers like Servers. Bluetooth *master switching* was also required as Bluetooth is limited to a specific several devices connecting to a master device. This is fundamentally one of the largest limitations of the present Bluetooth implementation and the MCP protocol implementation. Amending the Bluetooth standard to deal with broadcasting or improved context switching would provide a possibility for the implementation of the MCP to achieve success.

## 4.3  Summary

This section has discussed the implementation of the MCP given the constraints of testability in P2P systems. Two major simulators (a preliminary simulator and MCP simulator) were written using JiST (Java in Simulation Time) where the MCP simulator was adapted to deal with Extended MCP and CRAWDAD contact datasets. A single basic J2ME application was written to consider the feasibility of the MCP using the present day Bluetooth provision found on many common mobile devices.

JiST was chosen as the base on which to build the scalable and adaptable MCP simulators. JiST had been proven in previous research to provide a scalable and flexible foundation on which to implement discrete time based simulators. A realistic model was the major aim of the MCP simulator, which sought to test the MCP in its ability to share messages of content using period, range, incentive and mobility paths as the major parameters for input.

The components of the simulator were object orientated. Peers are described so as to live and interacted with one another in a World environment using Messages. A Tracer object was used to initialise the World environment. All objects which existed within the World object had real geographical and environmental constraints applied to them defining when actions by MCP Peers were successful. The MCP simulators used generational unidirectional paths with increasing complexity to create more complicated grid systems defining both regular and irregular (random) mobility.

The J2ME implementation of the MCP sought to define some of the dilemmas which present day technologies would have if the MCP were to be implemented using Bluetooth. The implementation proved that the MCP is feasible using J2ME however the Bluetooth protocol was seen to be lacking with respect to the provision of context switching. This demonstrated that the software capability was presently available the technological capability was not. Using Bluetooth for extended periods would ultimately drain the battery and the standard limited the application to seven neighbouring peers at any moment. The initialisation of communication in Bluetooth was costly and time consuming.

# 5  Performance Evaluation

As described earlier, the simulator was implemented as a layer above JiST. The simulator was required to take a collection of parameters which included: the experiment time (duration of simulation), the number of initial sellers, the number of buyers, the maximum radio range of the devices, the maximum speed, ratio of sharing peers (those peers willing to act as proxies), the waypoints and paths and finally the initial interval distance between peers (moving on a pipe). Peers were assumed to hold homogenous devices (i.e. the same device with common capabilities). All distances and speeds were measured in meters. World fault injection was possible. The ether of the environment and a message's traveling abilities could therefore be limited or influenced. Each experiment was repeatedly executed a multiple of five times to verify the data results produced.

Experiments were divided into *generational* and *real* mobility sets. The generational mobility tests sought to test and measure the fundamentals of the MCP in providing a means by which content could be efficiently and effectively shared, whilst peers also exploited their social network. In this context, the social network was defined for two variations: a) a *contact social network*, a social network created through repeated contact of peers and b) *choice social network*, a social network where peers chose their friends or common group. To experiment with a choice network a subset of peers were made available. The real mobility sought to measure the effectiveness and behavior of the MCP in an environment with added reality, where more realistic mobility had been recorded. It was sought that real mobility sets act as a confirmation of the Market Contact Protocol's abilities to contagiously spread messages when tested with generated mobility data.

The body of 5.2 considers Simple MCP. 5.3 considers Simple and Extended MCP *message generation*. A comparison of *infection* data produced for Simple and Extended MCP is exhibited within the appendices of this report in 9.1.

## 5.1  Preliminary Tests

Before implementing the MCP fully a play version of the MCP simulator was used to flesh out the feasibility of writing the simulator in JiST. Initially the MCP simulator was simple, using Euclidian co-ordinates and a very primitive two-dimensional visualization system, so as to understand the basic interactions of peers. The behavior was initially random in nature.

Initially, a collection of peers existed within a length by length square field, common with the ideas of SWANS. Figure 34 describes this approach. SWANS used a field to describe a test area. Tests were run for increasingly large numbers of peers. Peers were placed at random locations prior to the start of an experiment. Each peer was assigned a random vector direction translating into a bearing and velocity.

When a peer neared the boundary of the test field, they would simply bounce off that boundary at the reflex angle. Peers were assumed to all share content and use an altruistic sharing policy. Each peer had a set period for messages broadcast. The behavior exhibited was very similar to that of a perfectly spreading infectious virus, where the virus had no self destructive characteristics.

The self destructive characteristic was later implemented as the time-to-live for a message within the MCP simulator. In all cases measurements concluded that given unlimited time a message would infect all peers. The simple simulator proved the effectiveness of infectious methods to spread message content, just by the application of short distance contact. The speed of infection was also noted as significantly rapid.

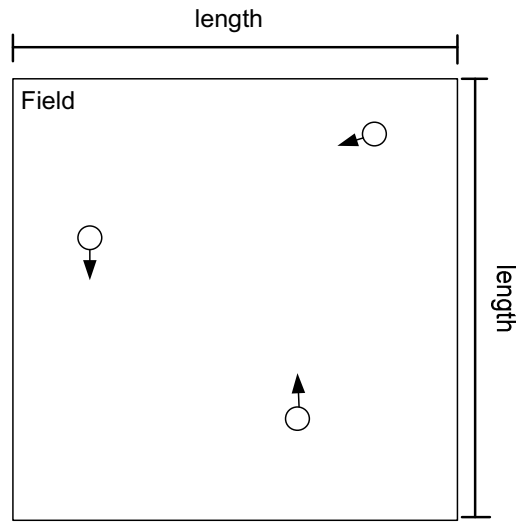

**Figure 34. The limited square field as used in SWANS. Implemented in the preliminary simulator with random walking peers.**

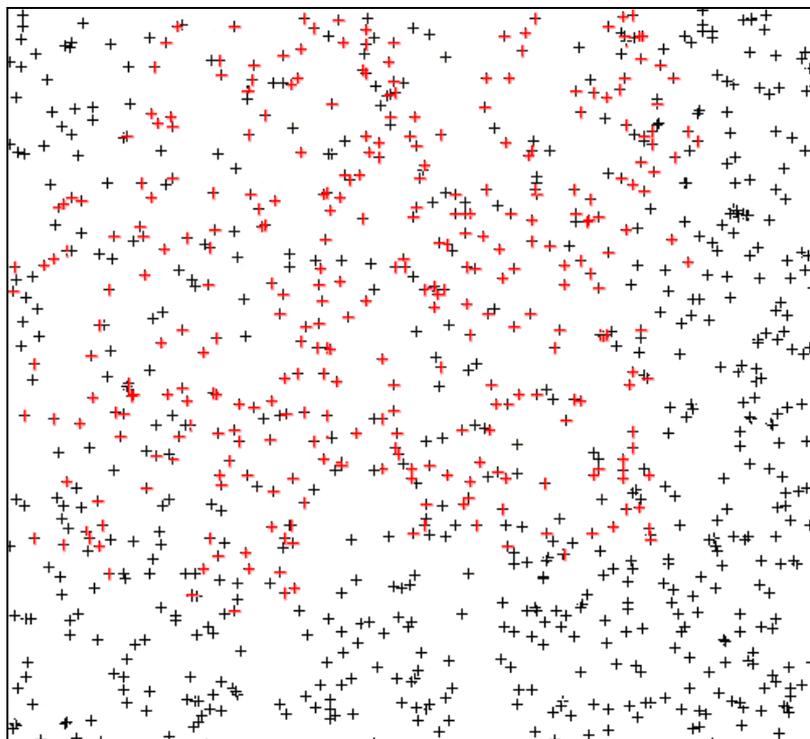

**Figure 35. Actual simulation of 1000 peers, each peer had a maximum communication range of 10 meters. Infected peers illustrated in red, while non-infected peers are represented in black.**

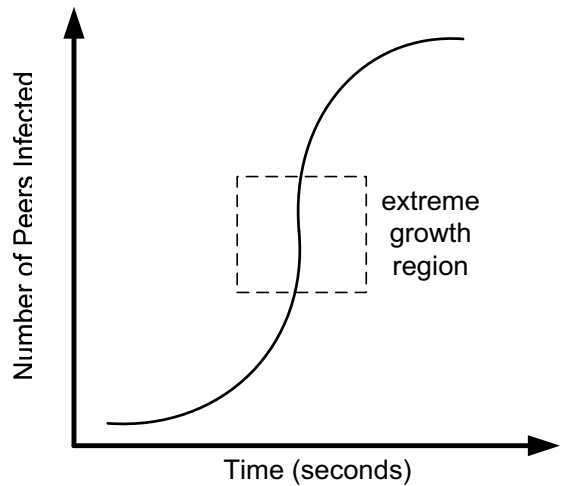

**Figure 36. Expected infection growth in dense systems of peers (sigmoid function). This growth is common to the spread of a worm and is defined by Verhulst's equation [78].**

From the analysis of this simple MCP simulation system it was found that dense systems follow the Verhulst [78] equation (5.1), noted sometimes as the logistic function model. The model was initially used by Verhulst to describe population growth, where $P$ represents the population, $t$ the time, $r$ the growth rate and $K$ represents the carrying capacity of the system. The function (5.1) states that: (a) the rate of reproduction in a system is proportional to the existing population, all else being equal and (b) that the rate of reproduction within the system is proportional to available resources. The availability of resources describes the limit to population growth.

$$P(t) = \frac{KP_0 e^{rt}}{K + P_0(e^{rt} - 1)} \quad \text{where} \quad \lim_{t \to \infty} P(t) = K \ \dots (5.1)$$

$$p_{n+1} = rp_n(1 - p_n) \ \dots (5.2)$$

Verhulst's equation can be described discretely using the logistic map (5.2). The logistic map is extended by May [50] as a basis on which simple non-linear dynamical equations can exhibit chaotic behavior, where $p_n$ exists between 0 and 1, and defines the population at year $n$. $p_0$ is the initial population. $r$ is positive and defines a combined description of population "reproduction and starvation".

Hence from this analysis and conclusion it subsequent experiments of the MCP aimed to find similar behavior in such systems defined by unidirectional, bidirectional and grid structures.

## 5.2 Generational Mobility

For the generational mobility experiments the paths generated and used for testing were within a region near to Imperial College London centered on the Queensgate Road (Figure 37). In this respect the movement of people within the area was first observed and then applied to the model. The benefit of this move to real geographic data was that real movements and real data could later be applied or compared with the MCP simulator model.

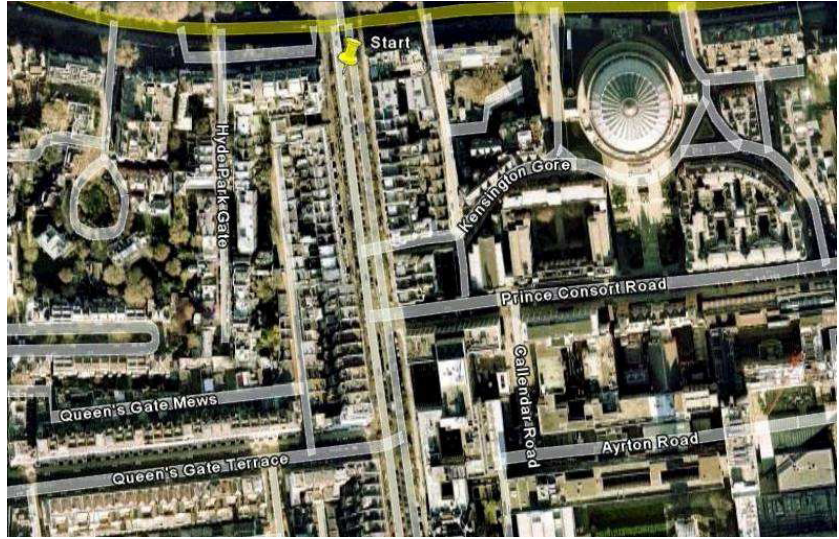

**Figure 37. The geographic test area on which unidirectional, bidirectional and grid structures were overlaid. The original starting location is visible to the north of the picture.**

### 5.2.1 Unidirectional Mobility

Unidirectional movement tests began at a waypoint of *(-0.180414, 51.501427)*, while the ending was the waypoint *(-0.178214, 51.492243)*, the distance was 1.01 kilometres. Peers traveled south from the original waypoint. The experiments considered the movements of peers in a single direction traveling at both *regular (uniform)* and *irregular (random)* pace (a constant average speed and a start stop, random speed) of mobility (Figure 38).

For each experiment a benchmark was measured, so that successive results could be easily compared. In a perfect system the world would consistently be updated with messages from peers. Peers would be altruistic in their sharing. Where a regular pace was used, the content sharing peer was changed between the first, middle and final peer.

The performance of the protocol was measured over a single hour (3600 seconds). All peers began their operation with an initial maximum range of 10 meters to add some realism to the radio communication techniques of present day (e.g. Bluetooth and WiFi). Peers initially shared all messages pushed to them and had a message broadcast period of 60 seconds.

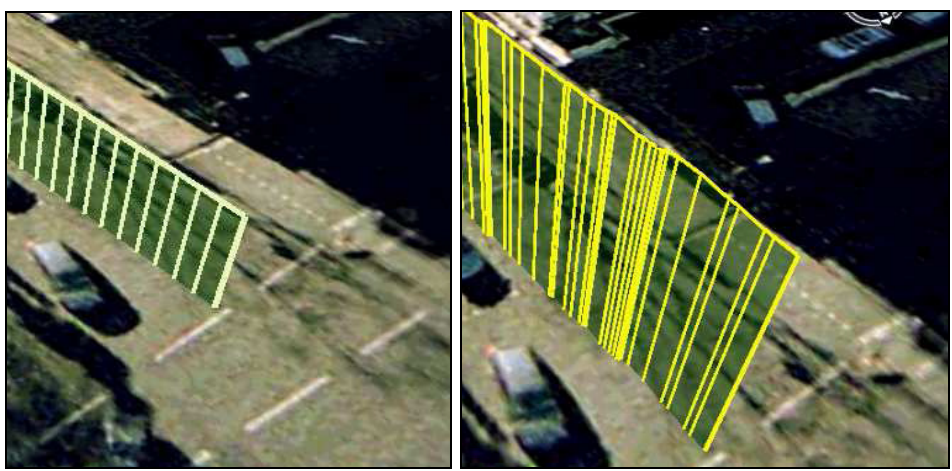

**Figure 38. Paths and points generated using KML represent the *regular* mobility of peers (left). *Irregular* mobility of peers (right). Each time step is represented by a vertical line.**

The default maximum speed of a peer was 1.0 meters per second (walking pace is roughly 1.1 meters per second) or 60 meters per minute (3.6 kilometers per hour). The default interval between nodes was 8 meters (within the range of the peer devices). For each experiment a change was made to one or more of these parameters and the results interpreted. Stated more formally, where we have a set of southbound peers and a single originator peer ($0_S$) sharing its content:

```
Peers_southbound = {0_S, 1_S, 2_S … X_S}, where X-many peers exist in the set
Originator peers = {0_S}
```

Through optimising the message broadcast *period* it was expected that benefits to the power usage of mobile devices would be exhibited. As previously considered, a high rate of broadcasts might be advantageous for sharing information, but it may be unfavorable for the overall battery life of the broadcasting mobile device and for the ether in which that device resides, adding noise to the band through flooding messages. Extended MCP makes provision for intelligent controlled broadcasting. The experiment in effect tested the ability of the MPC in its simplified form to communicate a message to all peers over the period of one hour. If given unlimited time the protocol would without doubt spread the initial message to all peers. Increasing the period improves the messages transport speed. There are two processes which are observed which cause a message to travel more quickly through the dynamic system. In the first case if peers broadcast a message more often and in the second case where more than a single peer receives a message broadcast. Where the period is low (e.g. 1 second) a large number of messages are sent within the space of a single hour, the likelihood then is that more peers collect the content.

We can conclude that the flooding broadcast approach is particularly successful in reducing the time for messages to reach all peers within a straight line piped system, traveling at constant speed with a set interval distance between peers. The disadvantage is the redundancy of messages i.e. peers may have already received the message and broadcasts are a waste of power. Figure 40 confirms the observation of improving a message's travel time within the system, where the interval distance between peers was less than half that of the broadcasting range we observe a second peer collecting a single broadcast (hopping). For future reference we refer to this as the *multiplicity* of the message at each time step. The message has the opportunity of spreading faster where there are more hosts to help the message spread. We can observe that there is a significant effect of changing the *period* between 10 and 20 seconds for intervals greater than 2 meters. The rate of change in subsequent periods after 20 seconds is observable. We can conclude that the speed at which messages travel through the system is greatly reduced if changing the period within the context of a unidirectional path.

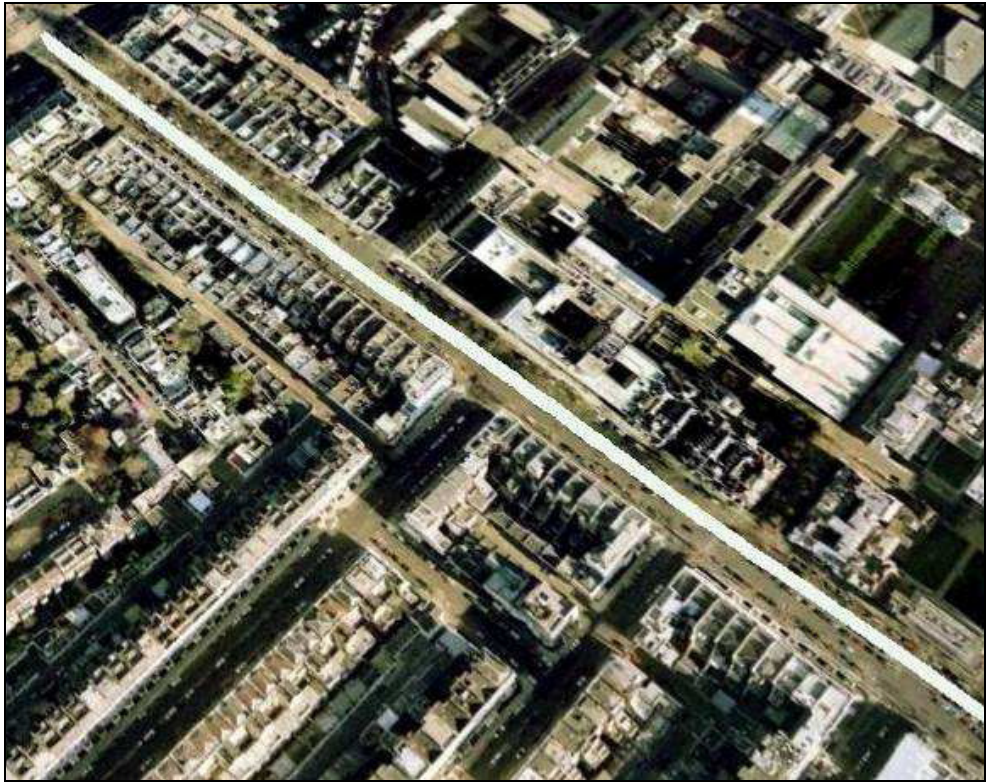

**Figure 39. Line path on which peers moved.**

Where a break is found within the unidirectional pipe due to *sharing* (Figure 41) or velocities being in excess of one another it may become impossible for a peer to share its messages with the network. If the interval between peers is too extensive the same problem occurs. In the case where the interval was too large messages were confined to a portion of the path of peers. In effect one creates an island of message content without the likelihood of that content traveling any further within the system. The messages cannot continue to spread along a pipe of peers. Hence, the availability of sharing by specific peers produced a detrimental effect on the pipe where the interval between peers was greater than the range of peers to broadcast. In cases where the pipe was not broken and sharing was altruistic for all peers, all peers received the original message within less than 1 hour of simulation time. The cost however of receiving a single message is large as the broadcast nature of the MCP leads to thousands of messages being sent.

Changing the range of peer contact improved message communication and message speed message speed by increasing the rate of message infection (multiplicity of seeding messages). Changing the speed of peers had no change to the system, such that the peers all had a common speed in a singular direction.

Irregular or random mobility had unexpected effects. Irregular mobility proved to have a higher rate of message infection than regular (uniform) mobility. Both infection rates observed grew linearly due to the density of peers being a line. A changing period had randomising effect on the outcome of messages reaching receiving peers. In essence it was difficult and unclear to determine a clear optimised point for the broadcasting of messages. The outcome for message infection was however more positive. Results suggested that message infection in an irregular mobility scenario was faster than that of uniform mobility.

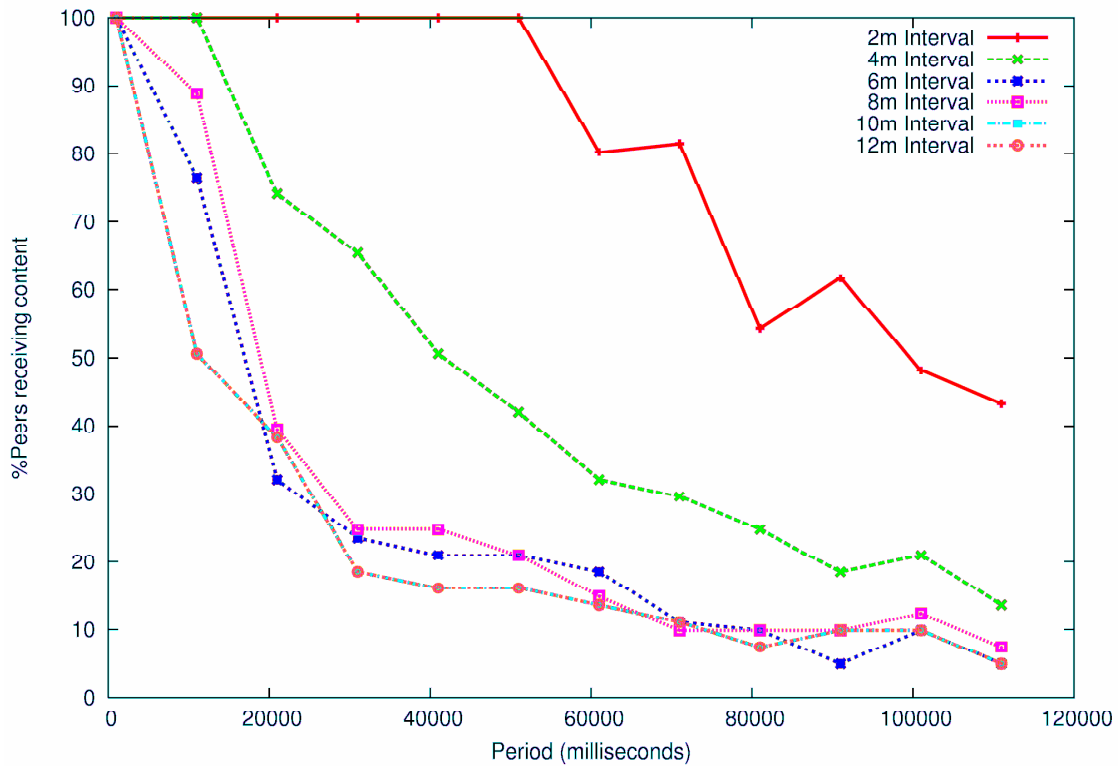

**Figure 40. Uniform unidirectional mobility: Percentage of peers who successfully received content broadcast by the original seller peer for a series of intervals between 2 and 12 meters and changing period. 10 and 12 meter intervals share the same results.**

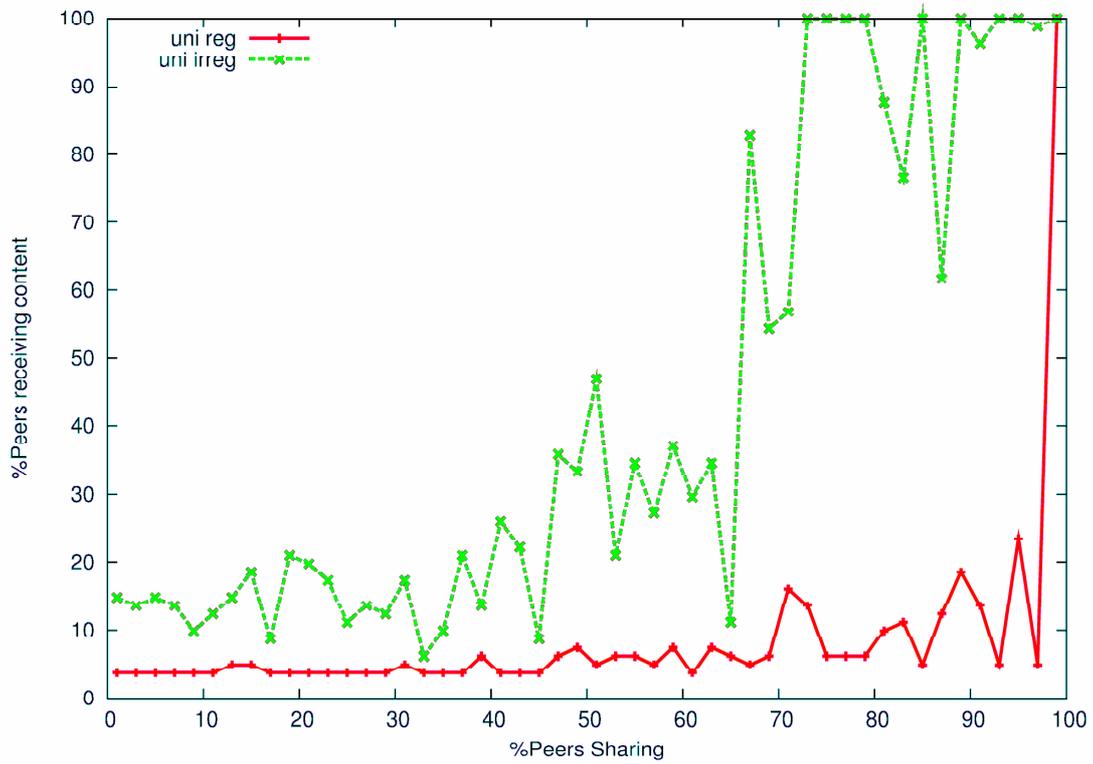

**Figure 41. Unidirectional regular and irregular mobility with changing sharing.**

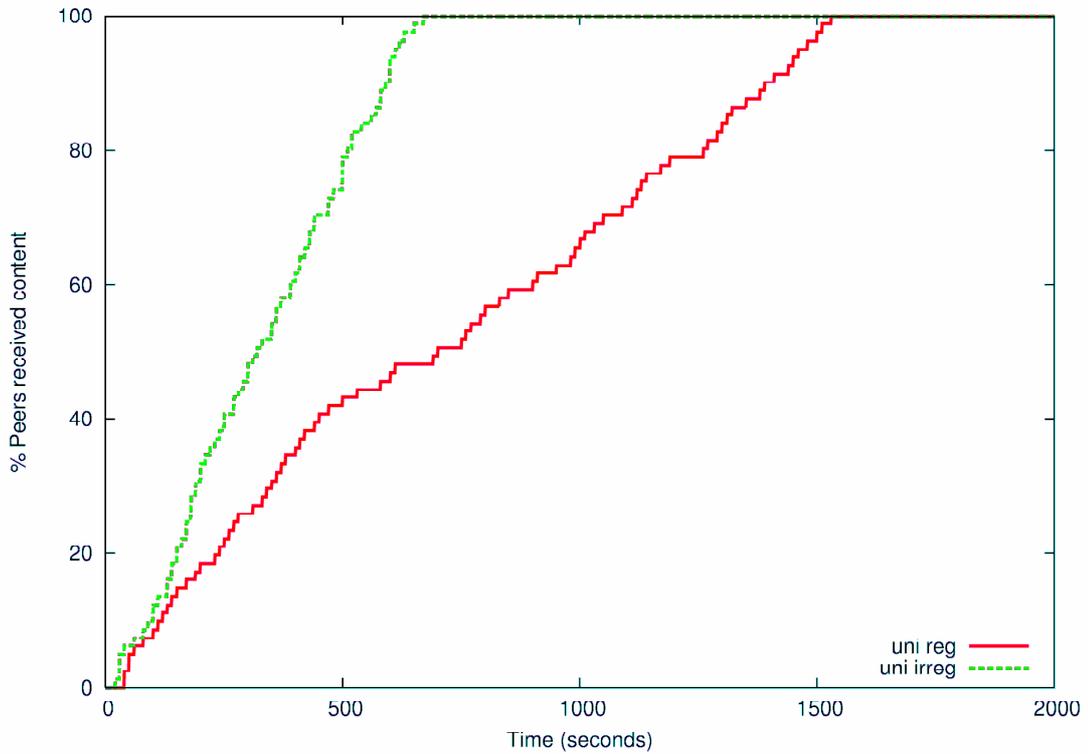

**Figure 42. Message infection given a unidirectional track with regular and irregular mobility.**

## 5.2.2 Bidirectional Mobility

Within bidirectional movement we used the repeated approach as with unidirectional movement to simulate the mobility of peers in their movement down and up a road i.e. two path tracks existed for peers to move along in opposite directions. A single peer would act as a message originator and seed its message at a point in the system. The results were impressive because of the movement of peers in opposite directions improved the contact of peers. We could divide the peers into two sets – namely *northbound* and *southbound* peers, where we had *X* many southbound peers and *Y* many northbound peers. The originator was placed as either a north or south bound peer to seed its content.

$$\text{Peers}_{southbound} = \{0_S, 1_S, 2_S \dots X_S\}$$
$$\text{Peers}_{northbound} = \{0_N, 1_N, 2_N \dots Y_N\} \qquad \text{Originator peers} = \{0_S\}$$

When considering *period*, most cases of varying interval and period were of little consequence. Messages were able to spread to all peers within a period of a single hour, no matter the changes made to period. The results pointed to the conclusion that in a dense environment of inter-relating sets of peers or one in which all peers come into contact with all other opposite sets we can assume that message infection would reach all peers.

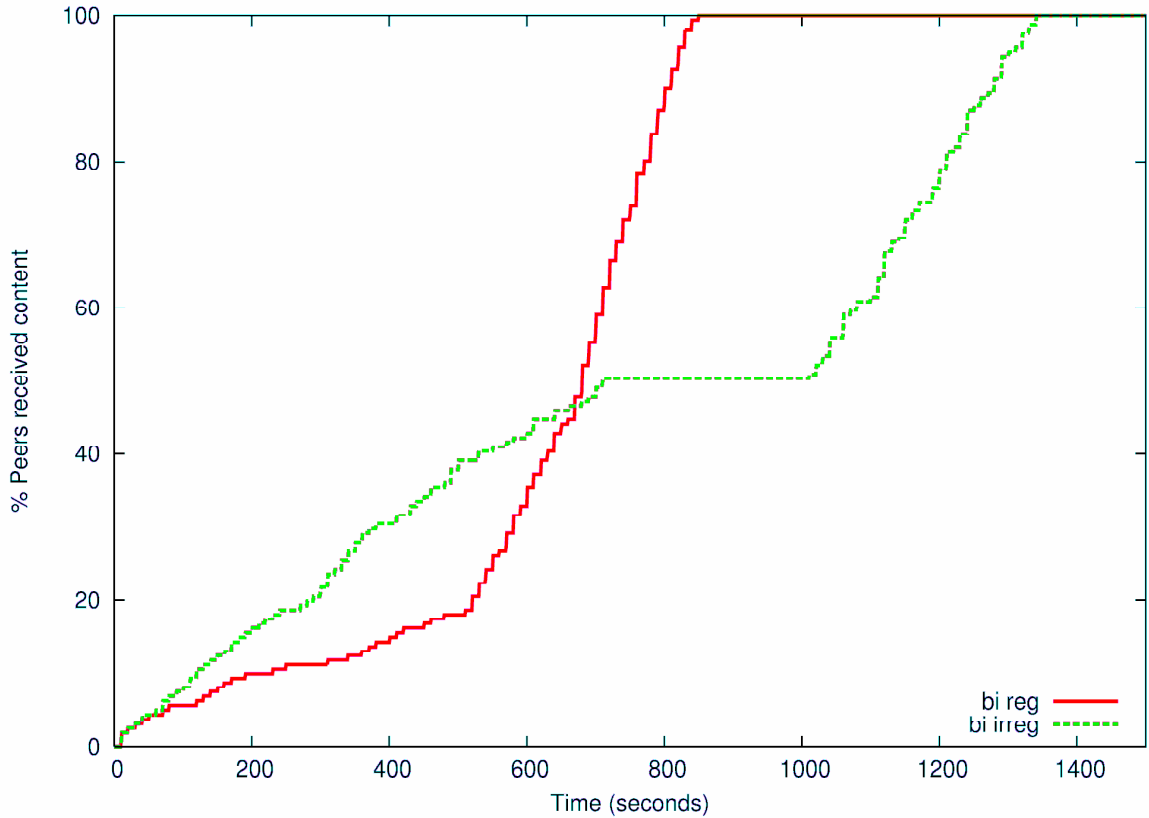

**Figure 43. Message infection given a bidirectional track with regular and irregular mobility.**

As with unidirectional mobility changing speed had effect only if peers became out of reach of one another's range. Speed in effect is a problem relative to a specific node. Peers moving too fast reduced their chances of receiving messages broadcast. Where speeding peers do help a system is where they come into contact with other groupings of density. A speeding peer helps to seed a message over larger distance, but also increases its chances of missing or dropping messages broadcast within specific locations. Where systems exist where there is large redundancy, we could expect all peers to eventually receive a message along a particular bidirectional track. Changing range in bidirectional movement was detrimental if the path of another set of peers was not covered. In effect if peers traveled along one side of a road and another seeding peer traveled along an opposite path, on the opposite side of a road, range needed to encompass the other peer to receive message content. Once a message was received within a specific set of co-interacting peers the effective spread of messages was improved linear growth. Results for variations in sharing produced an inverse effect to that of unidirectional mobility. Peers traveling a regular mobility pattern produced increased sharing characteristics.

Figure 43 illustrates the set interactions. We experience an improved rate of *sharing* between 500 and 600 seconds where the mobility is regular. Similarly we see this where the sharing is stunted in irregular mobility between 700 and 800 seconds. The collection of peers who receive content remains constant. This is due to the entire southbound set having been saturated, but not within the range of the bidirectional regular range of the opposing set. Once both sets meet message sharing continues at an increased rate, due to the immediacy of a more dense collection of peers.

### 5.2.3 Grid Mobility

Within the context of a grid we assume that a grid is comprised of various bidirectional movements of peers. Once again, regular and irregular mobility are considered. A grid consists of a single peer deciding to share content.

Eight sets of peers exist to fill the grid - two sets of bidirectional peers for each grid line. Each path was between 1.0 and 1.03 kilometers in length. If a peer neared a junction of a path, the inherent range which it could communicate along was affected. In this way, even if a peer had a maximum broadcasting range that range was hampered by the structures of the urban environment to add realism to the simulation. In most cases a radio communication range of 10 meters was not impeded, due to its minimal distance.

Within uniform mobility grid patterns, changing the period of peers within the system had an opposite effect as exhibited in linear uniform mobility. The larger the interval between peers the better the performance, where the period was degraded to 60 seconds. In effect the larger the interval between peers the longer a peer could increase its period the threshold of communication became erratic. Following the points of greatest coverage, the numbers of peers receiving an original message dropped steeply. While, changing the period for irregular mobility resulted in results which seemed random.

Sharing content followed a trend considered comparable both with regular and irregular mobility. The number of peers receiving an originally seeded message by a peer followed a gradual increase.

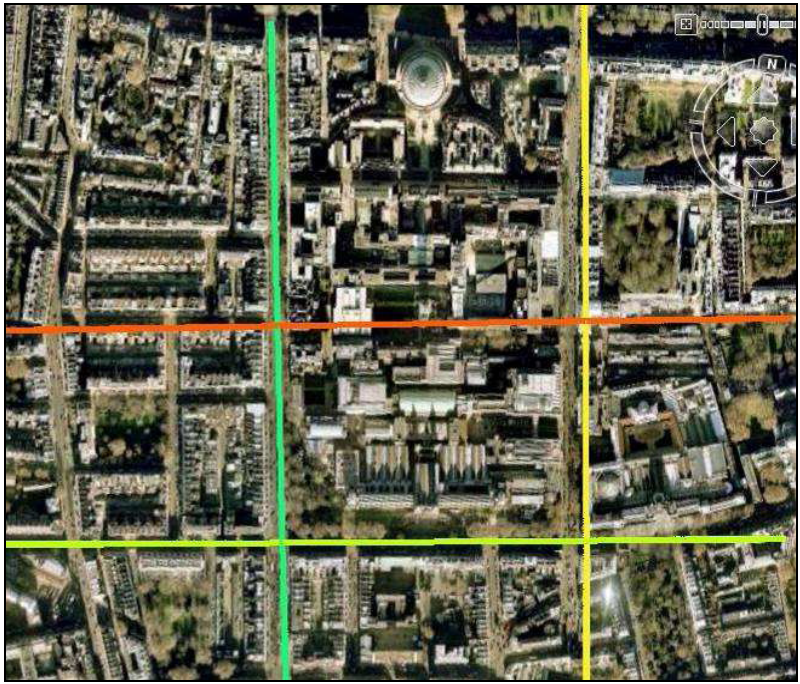

**Figure 44. Grid on which regular and irregular mobility peers traveled. Lines are the actually paths peers traveled.**

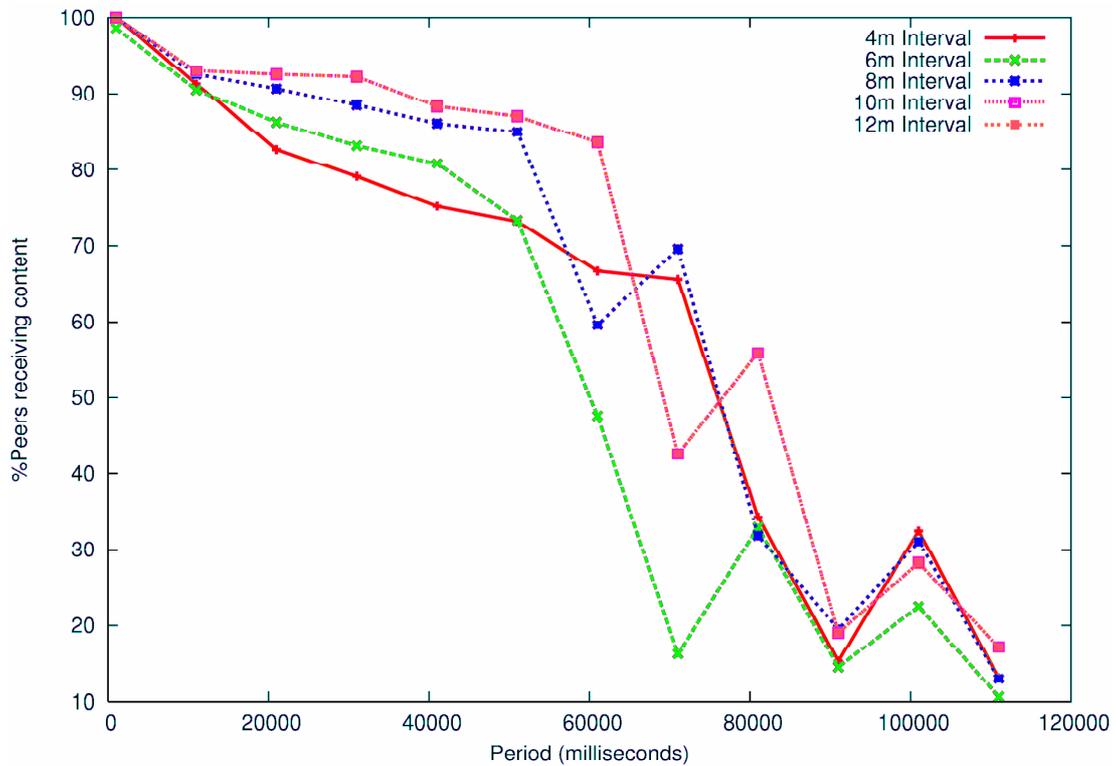

**Figure 45. Regular grid mobility: Percentage of peers who successfully received content broadcast by the original seller peer for a series of intervals between 4 and 12 meters and changing period. 10 and 12 meter intervals share the same results.**

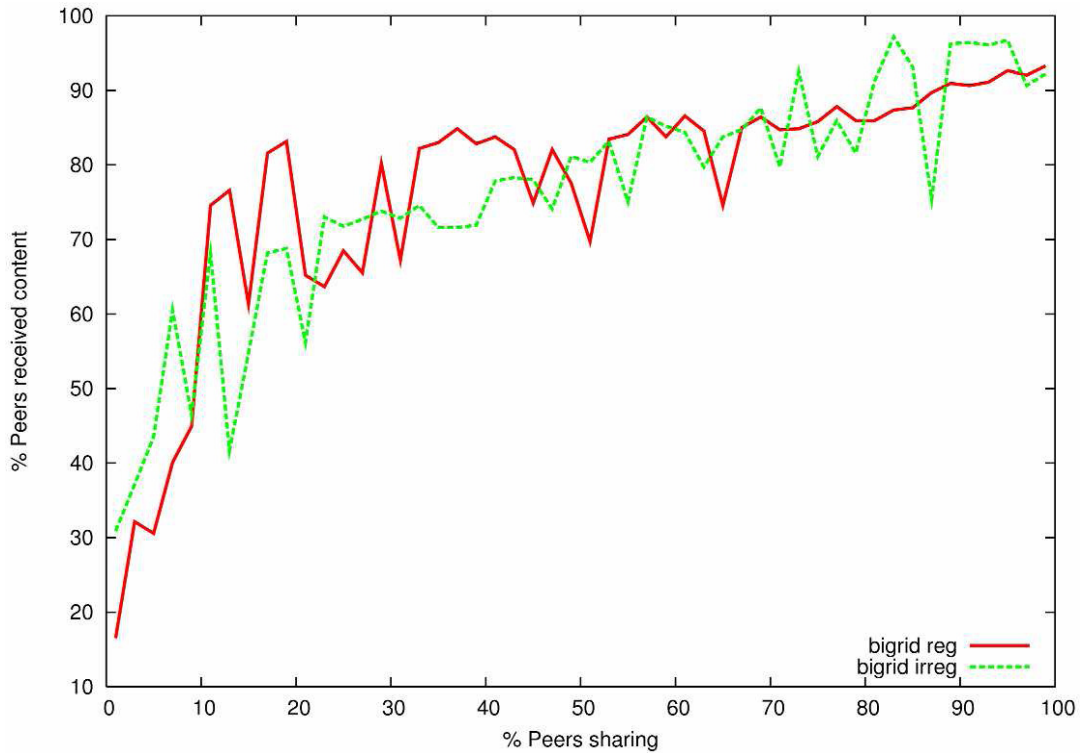

**Figure 46. Bidirectional Grid, percentage of peers receiving original message for regular and irregular mobility where there is a changing set percentage of sharing peers. Growth consistent with carrying capacity.**

The maximum shared outcome was not achieved within the experiment simulation bounds of 3600 seconds. We should be reminded that sharing considers the number of peers willing to act as proxies for the sharing of messages. In this regard the MCP was impressive in its spread of content, where 30% of peers shared; the outcome was that 70% of peer received the content. The number of messages within the system was extremely large.

## 5.3 Message Growth, Context-awareness and Infection

The message growth inside a broadcasting protocol like the MCP is considered a major disadvantage. The approach causes the MCP system to contain many redundant messages and increased noise. Spectrum is wasted with repeated messages and radio pollution. Due to this consideration, the Extended MCP sought to improve or reduce the inefficiencies of Simple MCP, finding a *threshold* point to achieve the same outcome as Simple MCP using more deliberate and efficient context-awareness. We defined the *message count* as the total number of messages sent by all peers within the MCP system.

Threshold is a count of the number of neighbouring peers required before an MCP Seller peer will broadcast its message content. We considered the number of messages and infection when investigating threshold.

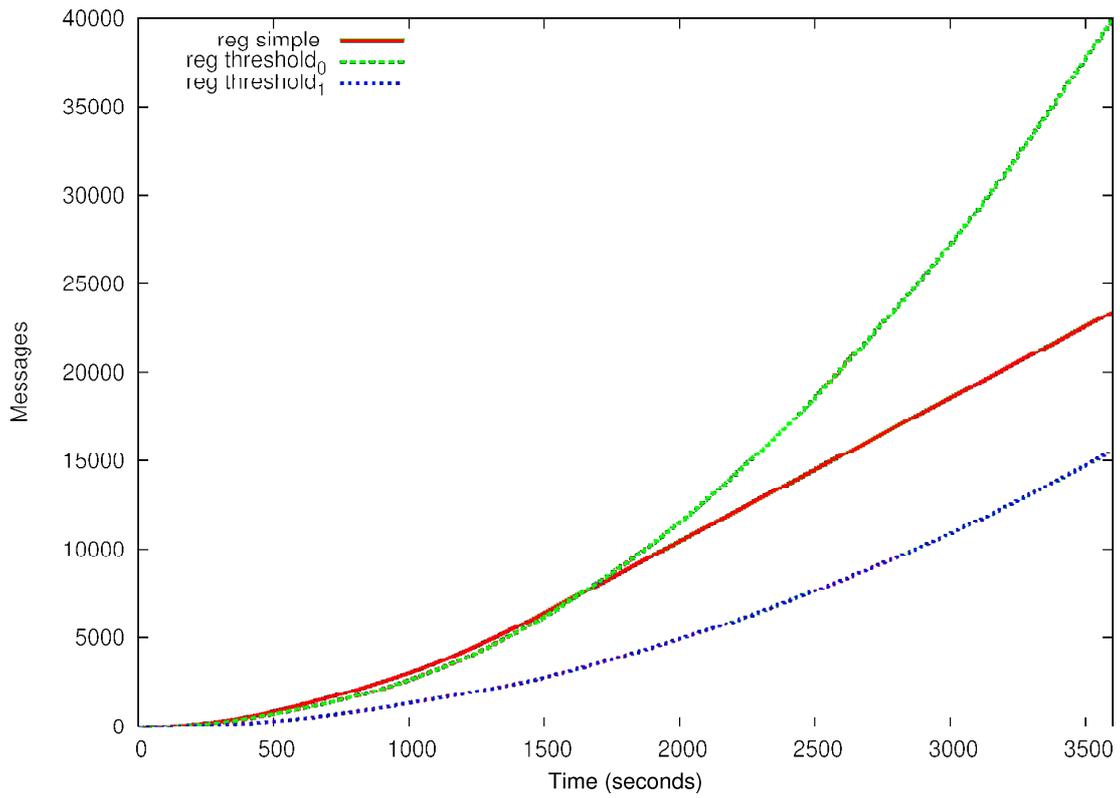

**Figure 47. Unidirectional Path (regular mobility): Simple and Extended (threshold) message numbers.**

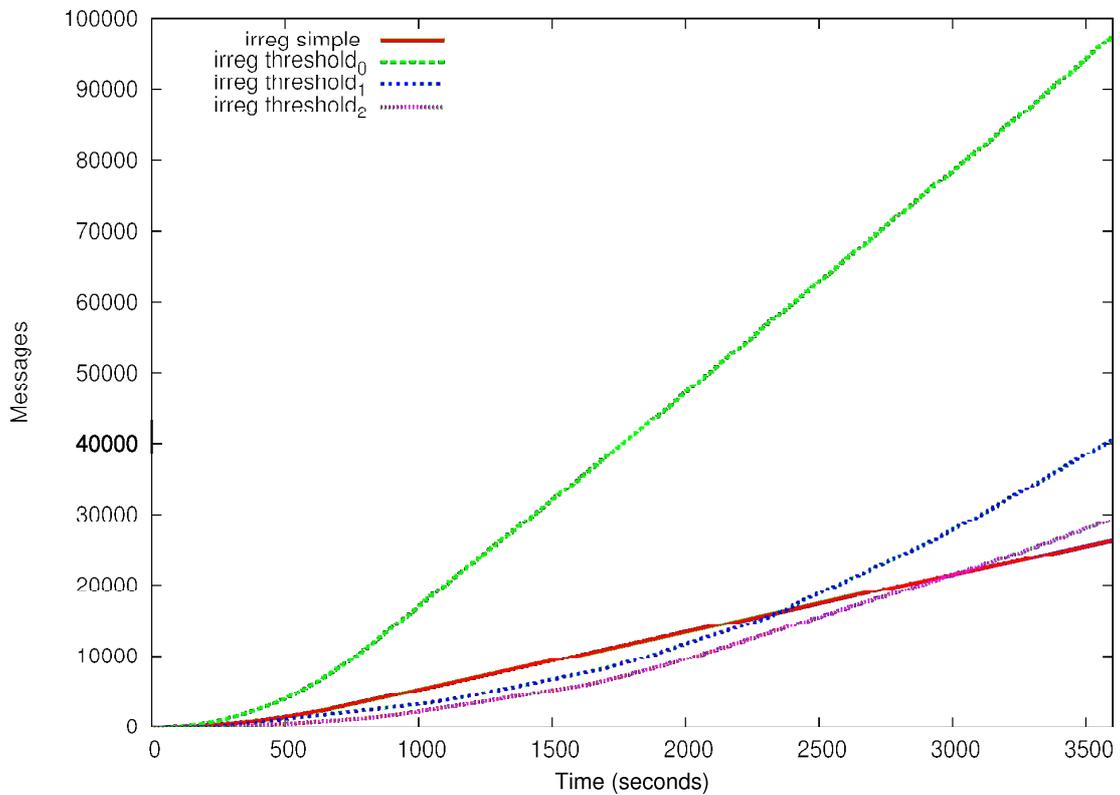

**Figure 48. Unidirectional Path (irregular mobility): Simple and Extended (threshold) message numbers.**

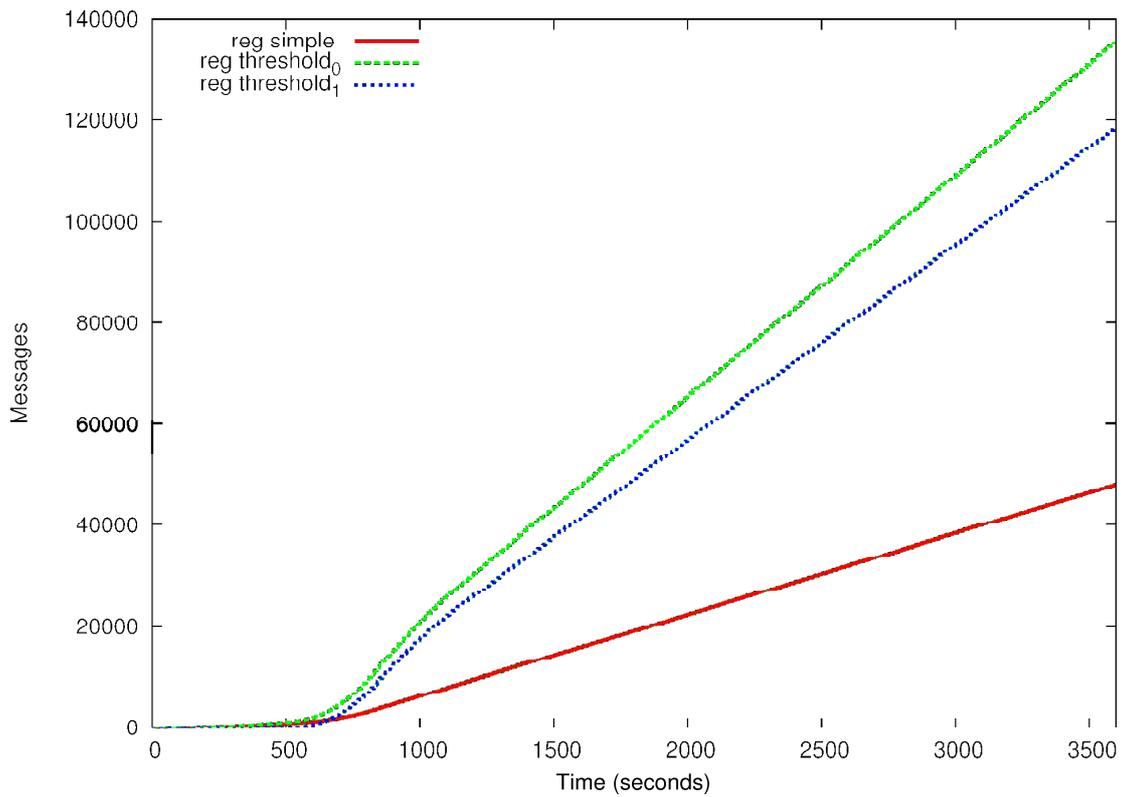

**Figure 49. Bidirectional Path (regular mobility): Simple and Extended (threshold) message numbers.**

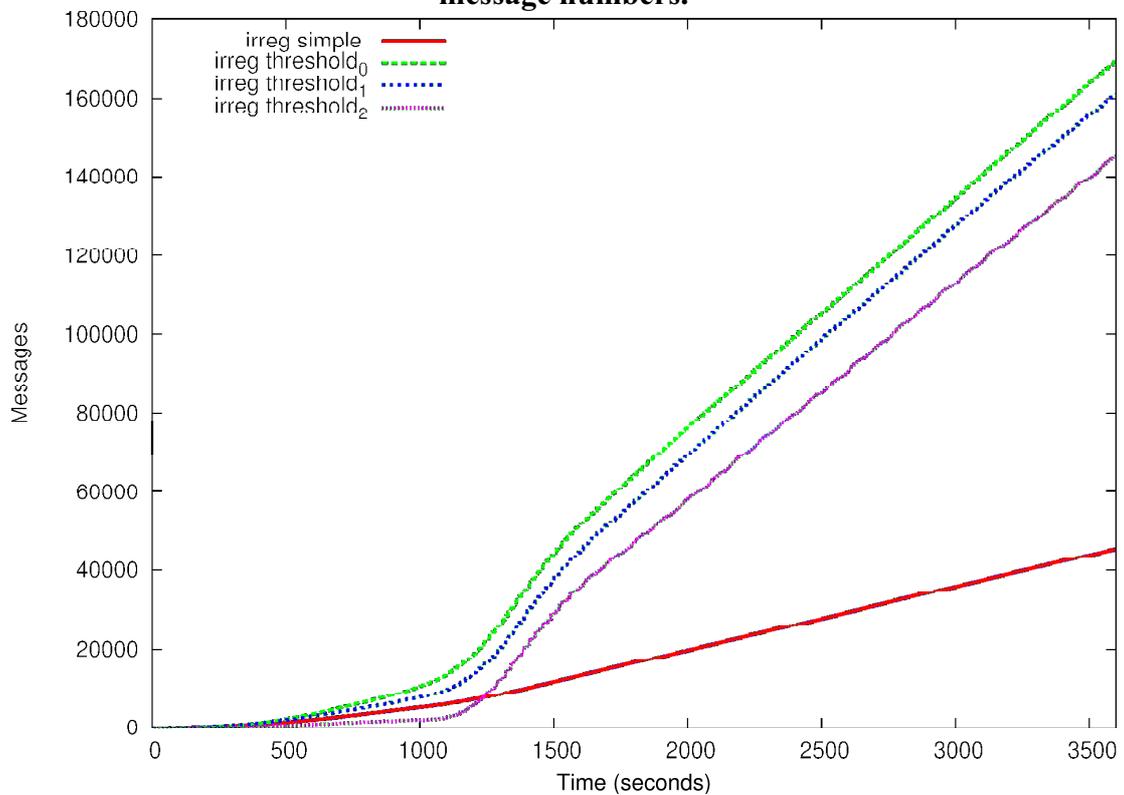

**Figure 50. Bidirectional Path (irregular mobility): Simple and Extended (threshold) message numbers.**

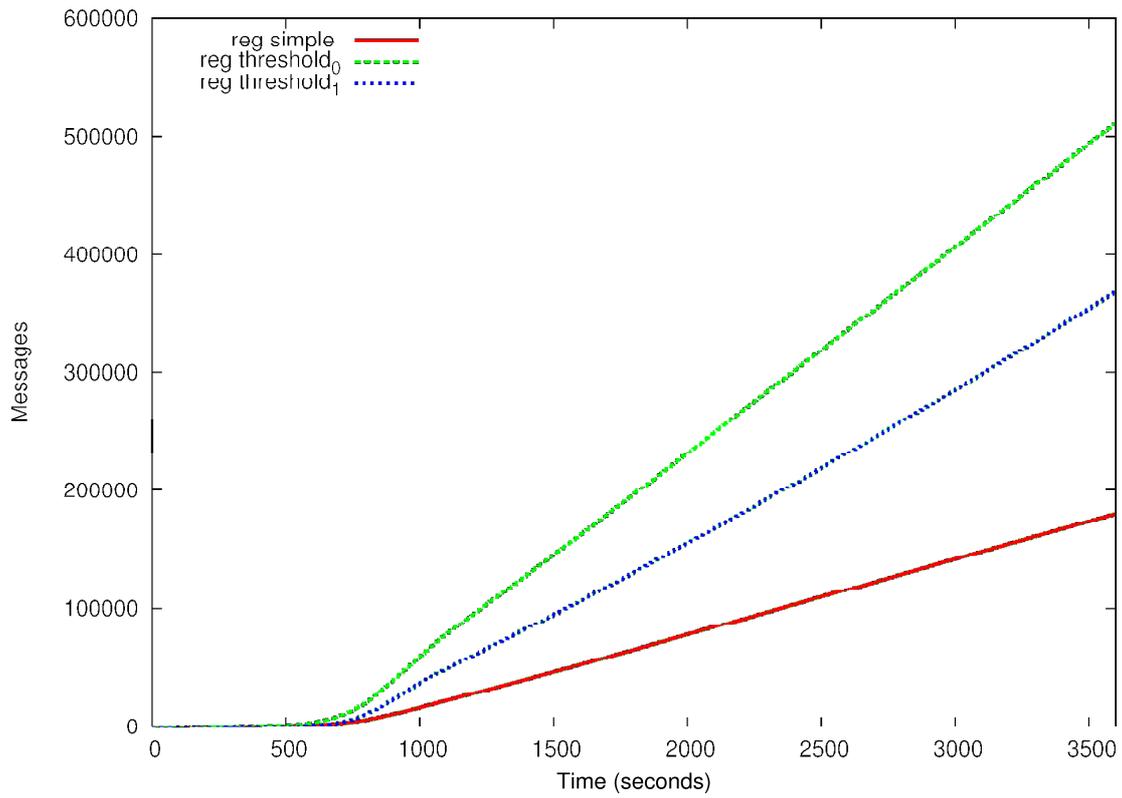

**Figure 51. Bidirectional Grid (regular mobility): Simple and Extended (threshold) message numbers.**

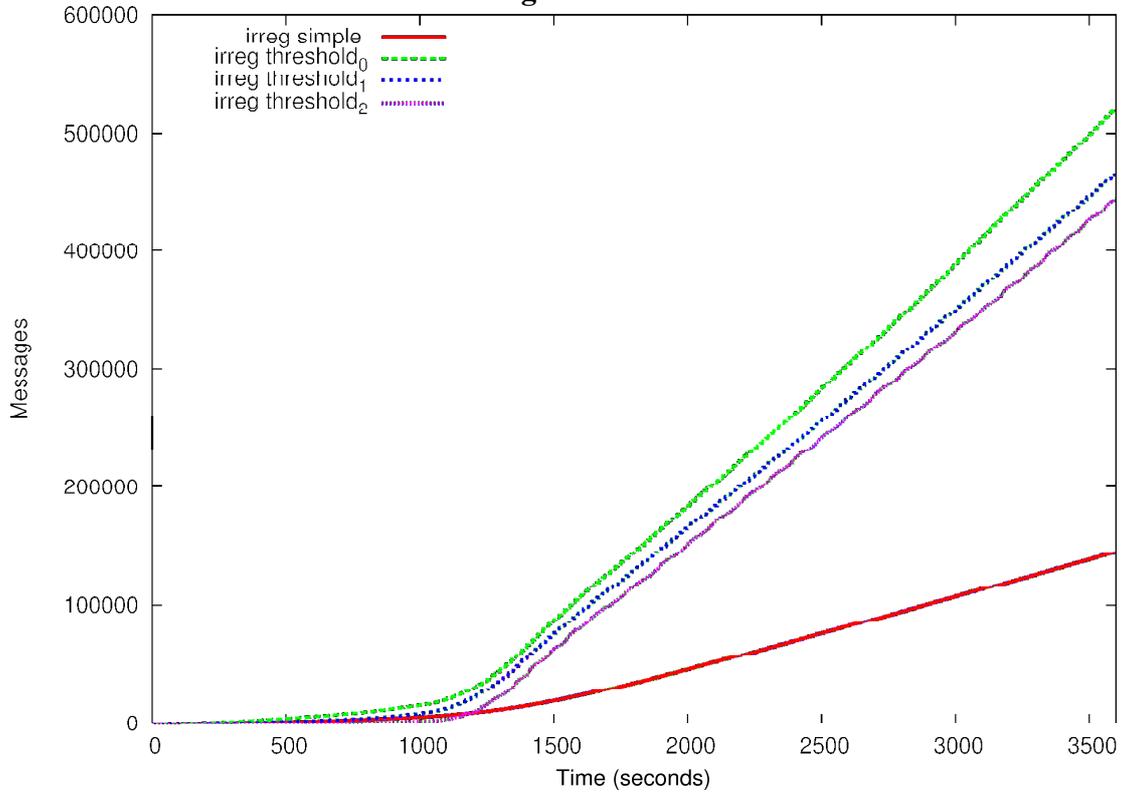

**Figure 52. Bidirectional Grid (irregular mobility): Simple and Extended (threshold) message numbers.**

Threshold is only applied within the Extended MCP. Message size is largely dependent on the type of information being communicated. We discuss the benefits of Simple and Extended MCP towards the end of this section.

In all cases of unidirectional mobility (regular or irregular), the infection performance of Simple MCP was better than that of all thresholds held in Extended MCP (Appendix 9.1). Increasing thresholds reduced both the number of messages within the system and the infection of peers where Extended MCP was applied. In this regard threshold was found to stunt the infection of neighbouring peers in return for the saving of power resources of a peer. The peer might save power by only sampling (short bursts of energy). The cost for sampling was however diverted to neighbouring peers. Any Seller MCP peer which sampled, injected its "sampling" message to all neighbouring peers within range. The received sampling request spawned a n-many neighbour reply. The acknowledgements required increased logic on the part of the Seller MCP peer and counting, which required time and power. Extended MCP could however not reach the level of efficiency that Simple MCP achieved given the circumstances.

Hence, the effect of sampling in a sparse system of peers resulted in wastage of power and contact time. Context-awareness only became a cost effective where the factor of threshold (neighbouring peers) exceeded a base factor computed by the density of peers, a consideration of broadcasting period (large period) and the mobility scenario. The performance of the MCP in Extended MCP was reduced in view of increased Messages.

In the scenario where MCP payload data was substantial, the sampling approach of Extended MCP provided savings in bandwidth. The sampling messages were small messages simply used for contact counting and neighbouring peer context. Using smaller messages to check the viability for the transmission of larger messages resulted in major gains as the size of the Message payload increased. The message infection rate was however slowed and required that Buyer MCP peers maintain availability for an extended period of time.

A method to reduce the number of messages within the system was applied. It sought to kill messages after they have been received and forwarded a limited number of times or after a limited time period. Within the MCP this approach did reduce message numbers significantly however it limited the spread of messages to peers available at a specific period of time. New peers entering the system would have no knowledge of historical messages if they did not travel within the system when a message was alive. The only hope of receiving a historical message was for the originator to rebroadcast the same message, once again producing a cascading effect of messages traveling within the system.

## 5.4 Summary

The MCP in its simplified form is fundamentally interested in the sharing of content messages using means of incentive (altruism, monetary, etc). The results evaluated described the effect of changing parameters on the system, where those parameters included the: *period*, *sharing ratio*, *interval* and *speed* of peers. Most parameters exhibited tradeoffs in their application. The results evaluated associated a relationship between the number of messages within the MCP system and the infection rate.

Increasing the *period* reduced the message speed of sharing (infection). It reduced the chances that a particular peer would share its messages with other peers. In each scenario simulated, period possessed a threshold. Increasing multiplicity exhibited an improvement in message sharing. A reduced period improved the message contact of the system however the tradeoff was the increased numbers of messages within the system and a greater use of limited resources.

Changes in the *sharing ratio (incentive)* limited the spread of messages within the system to islands in a unidirectional system. The sharing was improved if peers traveled in a more indeterminist fashion (variable *speed*); however this result was due to minor islands of content eventually traveling within range of other sets of peers.

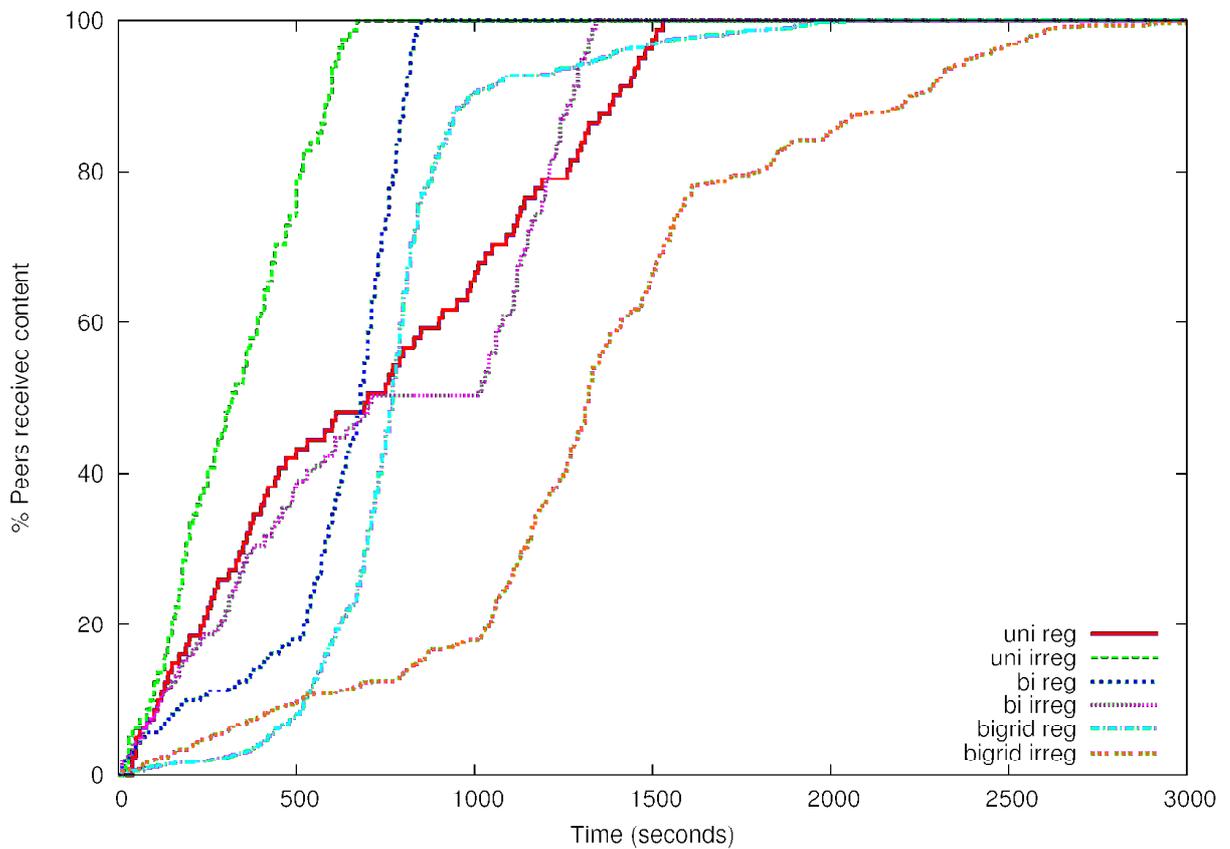

**Figure 53. Simple MCP: A comparison of infection recorded against time for regular and irregular mobility. The outcome is a greatest rate of growth at the midpoint in time whilst sharing.**

The *interval* between peers had an effect on the spread of messages and the optimisation of the period of peers. Interval is inter-related to and distinguishable with the density of peers. Hence density should be considered a determining factor of the message infection rate. Described alternatively, the interval affected period parameters and hence density. The more dense a portion of the system, the more quickly messages could travel due to multiplicity, the more message pollution generated.

The *speed* of peers affected content sharing if the peers traveled faster than the range of receiving peers. If a speeding peer was an originator or a proxy it could seed a message at greater distances in a shorter space of time. This could be an enviable position in which peers hover around specific location. However a speeding peer, if a buyer was unlikely to receive many messages, unless they were clearly and instantaneously at a specific location along the speeding peers path.

The *infection rate* of peers was compared for all scenarios (unidirectional, bidirectional and grid paths) and combined with regular and irregular mobility. The results confirmed the sigmoid increase of infection over time common with Verhulst's population growth equation [78]. While unidirectional and bidirectional infection was linear, grids were not. This confirmed the initial expectation about the outcome of message sharing between peers which was first considered, prior to experimentation (Figure 53). When presented with more complex systems like the bidirectional grid we observed that at the point where the rate of growth (derivative of the sigmoid function) was at its maximum for Simple and threshold infection, the functions were found to match one another within margin for infection. Threshold was found to limit the initial and end growth rates of a system, but not the infection of greatest rate. This infection rate was also found to occur at reasonably the same time.

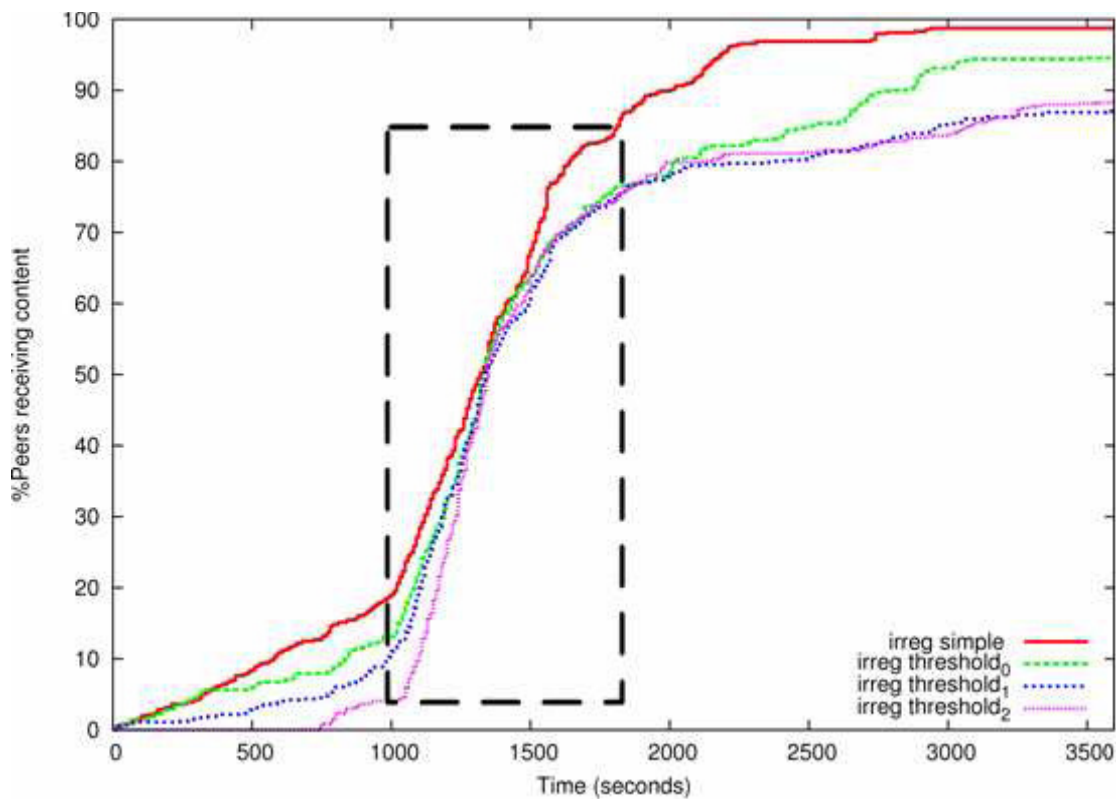

**Figure 54. Bidirectional Grid with irregular mobility, describing Simple and Extended infection (thresholds), where the change in threshold did not affect the rate of infection within the mid range of the function but did affect the beginning and final infection outcomes.**

It was found that by introducing Beacon peers (non-moving ordinary peers) to strategic locations in a system the message share capability of the system could be improved. Beacon peers act as standing peers which lack mobility. They act altruistically in the storage and broadcast of messages. With the MCP simulator they were analogous to libraries of content. Peers passing the Beacon peers uploaded their messages to it and it broadcast large collections of messages. In this way Beacons provided a half way point for altruistic application of the protocol by providing a geographic infrastructure for the retention of messages. As stated, placed at strategic points the outcome other placement was improved message speed. They acted as a centralised message storage point available to all peers.

The results evaluated considered the message efficiency of both Simple and Extended MCP, where Extended MCP utilised *context-awareness* (sampling) to reduce the number of messages within the system. Reducing the number of messages within the system aimed to reduce the power cost and the pollutant effect of message broadcasting. It was found that the complexity of the sampling approach, while being useful in unidirectional scenarios, was negative in complex systems of peers if the Message size was small. Sampling produced a cascade of extra information and redundant data where peers would require improved logic abilities to sort and compute future decisions on actions to take. Too many messages could overwhelm a peer. As the *threshold* factor was increased it did however reduce the number of messages within the system.

If we measured the *bandwidth costs* of messages in the MCP then Extended MCP was a better approach than Simple MCP. Extended MCP employed smaller messages for sampling and these smaller messages were used to decide on the viability of large message exchange. In this case Extended MCP was more advantageous to content sharing than Simple MCP. In an environment where the average data content size was large Extended MCP outperformed Simple MCP.

A method to kill messages after a specific amount of time was successful in reducing the number of messages within the system significantly. Messages traveled in wave like forms across the system for periods of time as they traveled from node to node. The disadvantage was such that that these messages disappeared from the system, unless they were rebroadcast a short period later.

# 6 Realistic Data Sets

This chapter describes the use of contact trace datasets provided by CRAWDAD [18] within a modified version of the MCP simulator. Contact data was used as an alternative approach to measure the effectiveness of the MCP and confirm the assumptions and findings made by previous MCP simulations. It provides a half way point in the substantiation of the feasibility of the MCP. The MCP simulator was extended into an alternative version (the MarketContact CRAWDAD simulator) to deal with specific contact data, which in most cases was represented in differing and complicated formats.

We should take note that much of this data is uninterested in geographic mobility and in most cases does not present effective resolution into the durations of connectivity between traveling peers. Instead instants of peer density have been sampled and collected over time. With this is mind the datasets were applied to the MCP with various assumptions and limitations. Project time constraints limited the application of further contact datasets. The predominant collection of data was collected using small mobile wireless boxes called iMotes [39] (BTnodes are a similar alternative device [9]). iMotes were powered by an ARM processor, had a Bluetooth capability with a 30 meter maximum radio range, comprised a local SRAM and Flash memory and supported TinyOS. An assumption for contact was made, that contacting peers could only communicate at the instant in "iMote time" (units) where a Buyer peer could see the Seller peer.

## 6.1 Cambridge/Haggle Datasets

The iMote Bluetooth experiments used were collected samples from the Cambridge/Haggle [31] project. They consisted of three sets of data utilising *density* and *time* information to produce a model for message content sharing (*infection*).

The first and second ($CAM_1$ and $CAM_2$) sets provided a collection of traces of contact between groups of users over the period of a number of days within an office environment at the University of Cambridge Computing Department. These first trace datasets constituted limited numbers of peers interacting. Two sets of devices existed i.e. internal and external devices where internal devices comprised iMotes and external devices comprised all other devices. This information becomes more relevant later when evaluating the results produced by the MarketContact-CRAWDAD simulator. The subsequent sets of data followed similar assumptions to the first dataset. The third data set ($CAM_3$) provided a collection of traces for groups of users carrying iMotes at the IEEE Infocom Conference held in Grand Hyatt Miami. The duration of the experiment was four days. Within the context of the CRAWDAD MCP simulator, iMote devices were able to share data with one another. In many cases devices could be seen but they had no ability to achieve mobility. These devices existed within the external devices definition. The CRAWDAD MCP simulator used contact between peers at specific moments in iMote or device recorded time (minutes) to opportunistically share content.

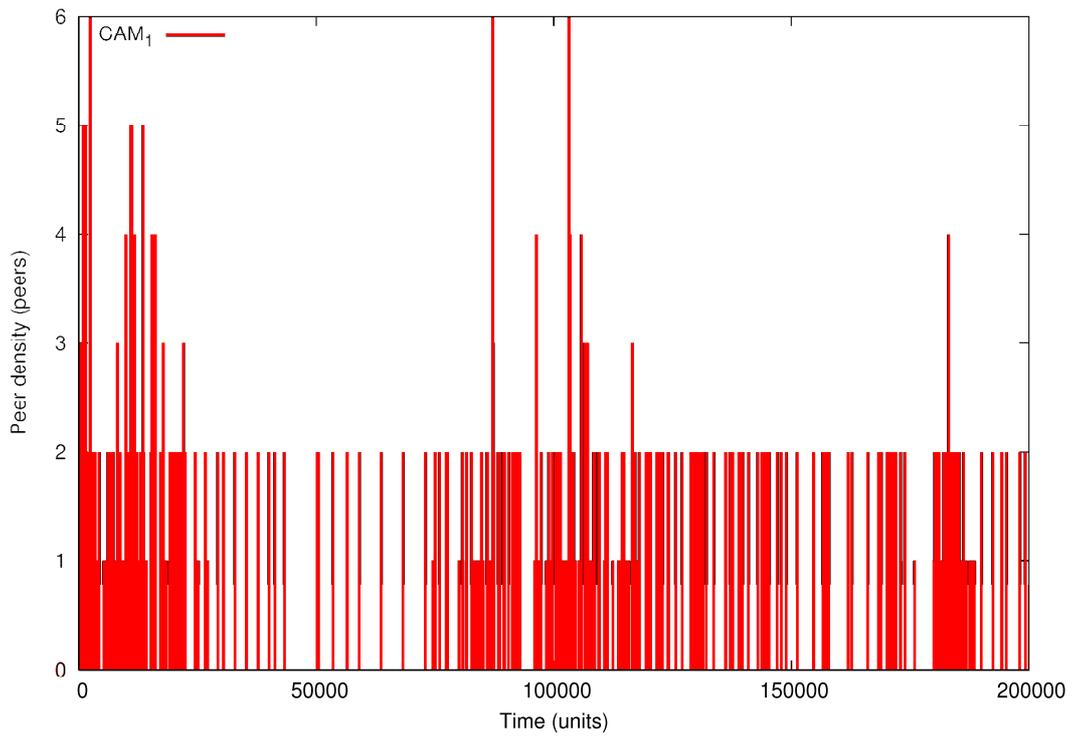

**Figure 55. Peer density viewed from a single Seller peer for the CAM₁ Dataset over a period of time.**

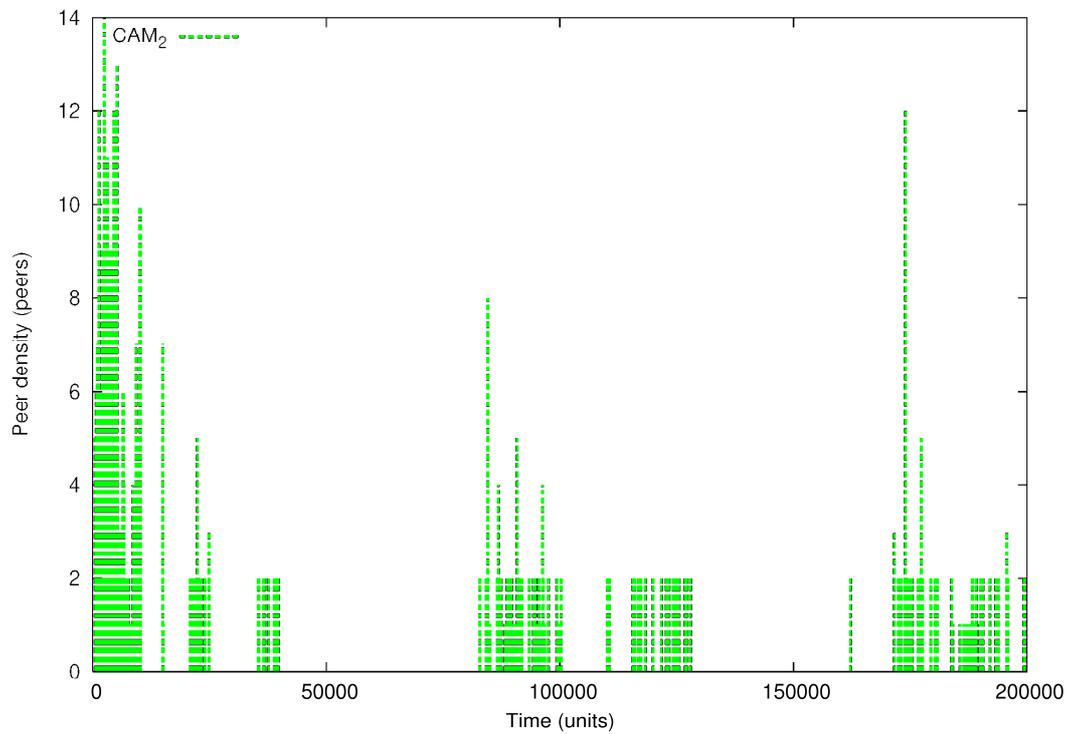

**Figure 56. Peer density viewed from a single Seller peer for the CAM₂ Dataset over a period of time.**

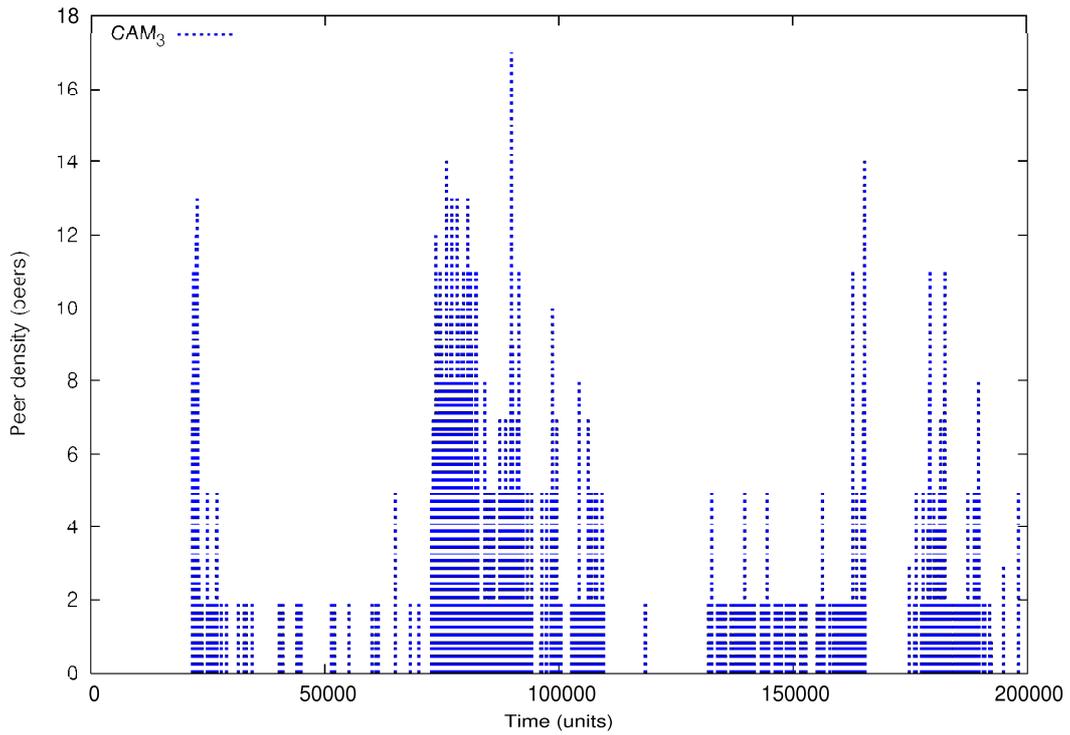

**Figure 57. Peer density viewed from a single Seller peer for the CAM₃ Dataset over a period of time.**

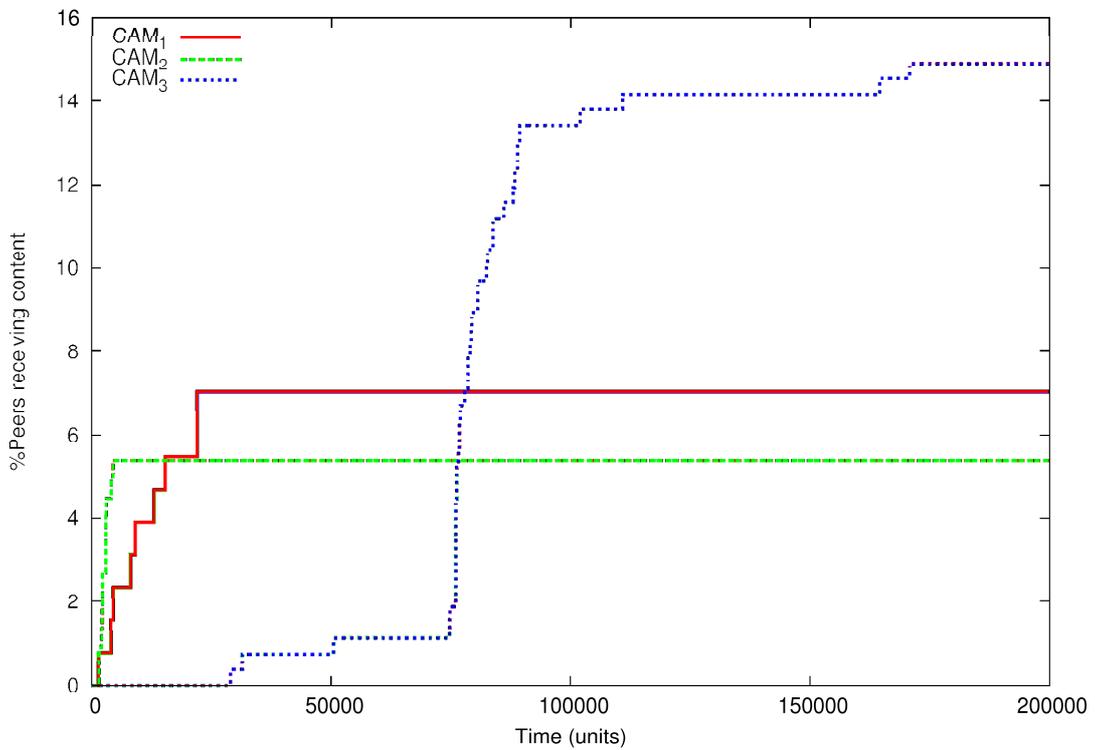

**Figure 58. Percentage of peers receiving the initial message sent by a single seeding peer using CAM₁, CAM₂ and CAM₃ contact datasets.**

Given the results evaluated we shall first describe the points of greatest contact, those points in time where peers were in more commonly dense locations. A peer was chosen as an initial MCP Seller peer. The experiments yielded similar results where the initial seeding peer was changed. All peers received the content initially shared given enough time. Over the period of days, peers would come into contact with another on numerous occasions. The experiments considered 0 to 200000 time units. We assumed that the message or content being shared survived as long as the duration of each experiment. Within the results illustrated, a period of 10 units was used between broadcasts for Seller peers.

The evaluation of limited peers yielded mixed results. The intervals of growth were seen to take place at the points where peers were surrounded by a specific density of neighbouring peers. When considering the sparse and small collections of peers the MCP achieved linear growth (CAM$_1$ and CAM$_2$). The common logistic function could be seen in dense systems of peers, such as those in CAM$_3$ (Figure 58). The logistic functions were common with the growth illustrated in the MCP simulator using generational paths. We should note that density readings used Extended MCP sampling prior to the sending of messages. Indeed if we consider this major and then successively smaller growth we found that all Cambridge/Haggle datasets exhibited this common infection behavior.

The changing density can be seen to have an initial effect. This is because the density was measured from the viewpoint of a single peer, in this case the initial seeder. Changing the period produced a similar result to that found with the mobility MCP Simulator.

## 6.2 Alternative Traces

Alternative datasets were considered for use with the MCP CRAWDAD simulator. These included the MIT Reality Mining and the National University of Singapore (NUS) datasets. The MIT Reality Mining dataset was incorporated into the MCP CRAWDAD simulator however the results were disappointing due to the complexity of the datasets provided. Experimenting with the MIT Reality Mining data set traces yielded no useful results. The NUS data was a collection of inferred contact traces taken from 22341 students' class schedules and class rosters for a semester in 2006. The benefit of this data was its increased number of peers. A disadvantage was that the data was again difficult to adapt to the MCP simulator. Further traces pose a future goal to add further validity to the MCP as it is executed within a more realistic simulation hybrid environment.

## 6.3 Summary

The results from the Cambridge/Haggle [31] datasets provided a confirmation that peer density was fundamental to the infection of peers over a specific interval of time. The growth of infection over extended intervals of time (days) was initially great followed by successively smaller growth stages. Described alternatively, the rate of message infection provided diminishing rates of infection for each successive unit of time beyond the immediate rapid infection point. Explained from work in Chapter 5, we consider this the carrying capacity of resources available to Seller peers and their proxies. We saw that carrying capacity was limited largely to internal devices (iMotes). The Cambridge/Haggle density data suggested that growth was greatest at those points where density of peers was largest, however successive contacts of smaller density helped to keep message infection rising in large sets of social groups. Contact density was argued to be intermittent, where infected peers might follow a set pattern of social contact during a day's mobility. Hence we experienced intervals of density peaks for traveling peers.

The application of trace data also confirmed the quality of the MCP simulator, the MCP itself and the ability for real data to support those fundamental assumptions on which the sharing of content occurred in social networks of mobile peers.

# 7  Conclusions and Future Work

This chapter highlights the motivation, ideas used, achievements and knowledge gained from in the generation and implementation of the MCP. We describe the achievements of the MCP simulator and the MCP with emphasis on the issues of incentive, contact counting, peer value, sampling, period, policy and peer density. We start by describing the simulator's achievements, following with advantages and disadvantages of MCP finally describing future areas of work.

## 7.1  Achievements

This project has presented a novel decentralised approach to content sharing for mobile devices, based on the utilisation of the Market Contact Protocol (MCP). The MCP is founded on the basis of *opportunistic networking*, *incentive, contact counting* and *policy conditions*. In both its Simple and Extended forms, the MCP provides a method which is adaptable to generic, but gradual message communication and content sharing. The protocol was extended to allow for its improvement through contact counting, peer value and intelligent context-awareness, with the major aim of improving bandwidth efficiency based on the environment in which a peer found itself.

An adaptable collection of simulators were developed to simulate the MCP. The MCP simulator presented a collection of generational mobility models which could be manipulated and amended easily to suit the conditions of the intended simulation system. The input paths require no instruction in programming to test the protocol's effectiveness. The simulator itself, developed on top of JiST (Java in Simulation Time), is capable of simulating extremely large systems of 125 000 peers within a more realistic setting comprising real geographic coordinates and parameters. Experiments were superimposed on a real world coordinate system. While mobility models could be considered limited they provide a benchmark of possible best case outcomes for the MCP protocol. Due to the more realistic simulation parameters and the capability to superimpose results on the real world, the results and paths of simulation could be verified against future data results collected from other sources such as real contact data collections.

The MCP protocol itself came in two flexible and adjustable versions (Simple and Extended), with both approaches firmly based on the ideas of opportunistic networking, incentive and market-based systems. Peers acted as buyers and sellers analogous to a real market. The peer Seller components used policies to conclude whether to announce or broadcast messages held, while the Buyer components were ever listening for possibly interesting messages as described by the user. Based on the Buyer policies (strategies to facilitate content sharing), intercepted messages were replied to, stored or broadcast onward within the system.

Simple MCP sought to provide a basic rapid approach to content sharing using broadcasting Seller peers. Sellers actively broadcast their intention to share or market their data to receiving or listening Buyer peers. Each message presented to a peer was comprised of four fields describing the origin of the message, the message payload, incentive to carry the message and information on when and how a message should be destroyed. The MCP's message incentive approach is new and incentive was considered vital with devices that had limited computational, storage, communication and power resources. When carrying a message on behalf of an original Seller peer, intermediary peers were considered Proxy peers. Proxies would act on the behalf of sellers, broadcasting the Seller's messages and acting as agents for the sellers they served. The process was recursive in nature.

Extended MCP sought to improve the protocol by providing peers the opportunity to improve message communication through context-awareness, contact counting, peer value and intelligence, providing sampling as a preemptive step to message broadcasting and improved logic to handle optimal environmental conditions (density fluctuations in peer population). Extended MCP while more expensive in its generation of messages, was cost effective for the sharing of large content collections where there was motivation to reduce the bandwidth of data messages sent. Sampling incurred a smaller cost to resources than that of resending a long message data stream, which could fail in its communication.

Results produced by the MCP simulators exposed a relationship between message sharing and Verhulst's population growth function. The MCP experiments showed that the protocol performed extremely well in dense settings but that it had limited effectiveness in sparse and minimally sharing scenarios. The effectiveness of both Simple and Extended MCP was largely dependent on the environment in which a peer found itself. Simple MCP while wasteful was less wasteful to radio spectrum than Extended MCP in sparse simulation settings. The Extended MCP became a useful alternative in massively dense communication settings where the cost of sending data messages was high. The complexity of the Extended MCP sampling process added extra message overhead to the system. It was found that in effect Simple MCP was less pollutant and quicker to spread messages than Extended MCP in most scenarios. Extended MCP provided a method by which policies could be applied more easily within the MCP based most importantly on peer value and incentive. The Extended MCP threshold approach skewed the Verhulst growth of infection, by manipulating the number of messages present within the system. In complex and dense systems of varying peer mobility, it was found that Verhulst's growth function always existed. This function provides a basis on which message infection can be forecast.

The protocol itself was wasteful to the limited resources of mobile devices and the ether in which the peers existed unless messages were actively destroyed. The more peers receiving a message, the more messages are spawned due to the multiplicity of neighbouring peers. Thousands of messages were redundantly sent within the system, not all would be received and some would be repeatedly received. Where MCP messages were given limited lifetime there was a significant reduction in the number of messages held within the system. Message killing (forwarding-and-destroy strategy) was largely beneficial, where a time was allocated to the life of a message from each successive seller or proxy. The disadvantage was that the same message if rebroadcast needed to travel throughout the entire system once more to be received by new peers joining the system. Given this, the MCP was shown to be extremely useful for the rapid sharing of content in dense zones of activity. The addition of strategically placed standing Beacon peers was seen to improve the system even more acting as reservoirs or geographical information libraries for passing peers.

While the MCP is still as of yet unproven in the real world, the J2ME implementation and the results produced from simulation and modeling, provide a compelling basis on which to motivate the development of systems which could use the MCP in the sharing of content or the generic provision of communication. Before the MCP can be applied, some technical achievements still need be met, most especially the communication techniques presently available to mobile devices. WiFi and Bluetooth are lacking, but should not be entirely dismissed; they are still the only standards available to test prototype systems.

The MCP proves itself as an alternative for data sharing in environments where umbrella networks are not available or cannot be used and where users wish to become more social in their market interactions. Its utilization of incentive, contact counting and sampling makes this possible. With this in mind, the MCP also provides a decentralised alternative which could reduce centralised load on a content sharing location. Within future content generation likely to grow the MCP is a reasonable alternative to more complicated approaches for content sharing. The MCP makes communication more geographic and hence more socially interactive. With the addition of context-awareness, message destruction and improved logic the MCP has been demonstrated in simulation to provide a rapid and flexible means by which to share content in dense communities where social contact is common.

## 7.2 Future Work

Future work still exists within the context of real world testing, intelligent broadcasting using context-awareness, investigating the most beneficial approaches to incentive, privacy and security concerns, GPS mobility traces and the social implications of such a system.

The real world implementation of the MCP protocol is hampered by the present communication techniques available and the mobile devices themselves. Bluetooth and WiFi are not technically optimal platforms on which to test or implement the protocol. Neither Bluetooth nor WiFi provides a low energy cost capability for the broadcast of messages. They rather require the connection of devices before communication can take place. While a J2ME implementation of Simple MCP was developed it lacked the ability to act pervasively within its environment. User interaction and coaxing was required.

While some logic has been considered in this project, improving the intelligence of Extended MCP would help to reduce the cost and pollution of communication using the MCP. To achieve this would require a better understanding of peer mobility within specific environments and a way for devices to intelligently recognize these mobility scenarios. At present the MCP floods its environment with messages announcing its content. Improved logic or intelligence for spreading content based on parameters and context-awareness would make the MCP content sharing approach more efficient. Many devices will increasingly have GPS (Global Positioning Systems) on board them. These GPS parameters could help to improve the intelligence of peers broadcasting, determining speed, direction and the statistical change of correct message broadcasting; however this would require increased computation.

Incentive is fundamental to the action of the MCP. As with the nature of P2P systems incentive makes the system available for the communication of messages.

Without peers willing to act or become involved in the provisioning of the system the MCP is limited to pockets of users, where the full benefit of the approach of opportunistic networking cannot be full realised. Policies to improve incentive for proxies need to be explored. Investigating incentive and the groups who would make use of the system would improve the protocol's application. Improved algorithms for efficient and effective routing and message propagation need to be explored.

The immediacy, geographical nature and socialised basis of the MCP and the short contact range between peers in the MCP environment, makes the MCP a possible privacy hazard. Broadcasting can however be overheard, yet the same could be argued for all wireless communications. The act of broadcasting for communication between peers (one-way advertisement) has sought to reduce privacy concerns and it should be noted that in a system of communicating MCP peers privacy cannot be absolutely perfectly protected if peers wish to communicate with one another within short geographic sets. The need for privacy within the MCP depends largely on the type of information being shared, but the MCP is open enough to use encryption algorithms with its data messages. There is also the possibility for the MCP to be extended with present handshaking algorithms. Even with this consideration, the one way direction of broadcasting was motivated by privacy implications. Content is coupled with broadcasts and replying to messages is not expected. With this in consideration the MCP only requires an infrastructure where communication techniques provide for simple broadcasting. Various security protocols need to be considered where the data of peers may need to be protected or made confidential. This security is more applicable to the context in which the MCP is applied.

Approaches like CRAWDAD [18] and MIT's Reality Mining [22] have provided varying collections of mobility data. Most of this data is associated with contact points in networks. This has been somewhat useful in considering the spread of information in opportunistic networks. The data provided helps to investigate the grouping approach of peers and to verify the points at which content sharing could take place. Datasets could be easier to use if they provided a standard format, one where the ID of peers could be associated with one another at a specific time and a specific duration of contact could be easily recorded. The datasets presently available do not however look at the details of mobility, i.e. individual movements at points in time. There is the still the opportunity to collect large high resolution GPS based data sets with more importance laid on the positioning of devices in time. A higher resolution would help investigate the integration of real-time mobility with the application of content sharing protocols utilising the opportunistic networking paradigm. This mobility data (such as peer speed and location at specific times) could provide strategies on the best location and optimisation points to implement content sharing applications in the real world.

Another approach to contact information collection could be to use massively multiplayer game users, to provide information about their mobility. This could be achieved by providing an add-on to the gaming community and those members of it who would like to contribute to context-aware systems research. Approaches like this have been used by SETI@Home [6] to search radio signals received from space for extraterrestrial life. It is debatable as to how useful gaming mobility data could be, however it should not be discounted as a source for human behavioral information.

It has been suggested that the geographical proximity of peers and the decentralised market based environment which the MCP provides could be used to create underground social networks. Such networks could include messaging and content sharing marketing criminal behavior or terrorist activity. The decentralised nature of the ether makes the usage of the MCP difficult to check. This is both a blessing and a curse. A solution to this problem is not clear.

## 7.3 Discussion

At present, technical restrictions have made the MCP difficult to implement within real world devices, however this will change over time. The ad-hoc environment in which the MCP needs to perform is fundamentally complex, unreliable and erratic, where varying factors affect the outcome of simple communications. The MCP requires a reasonably reliable communication approach and more device density before context-awareness can make significant changes to content sharing. This report has not delved heavily into the urban environment and its problems, which are considerable, given that wireless mobile devices have limited battery and communication capacities.

However, with this in mind, exciting initiatives like TerraNet [75] seek to provide wireless peer-to-peer infrastructures on which the MCP could be applied very soon. While the problems inherent in these systems are still considerable, they put forward the basis for cost effective, decentralised systems where there is no need for a centralised umbrella network for information communication. These systems are more likely to be challenged not by their working ideas, but perhaps the barriers to entry made by physical constraints to spectrum availability and economic motivation. The social implications of decentralised control, where information has total freedom to travel, could also make the systems both a powerful tool and a problem to the society that uses them. This outcome depends largely on how information sharing is used and to what end.

# 9 Appendix

This appendix holds results and items not included in the body of the main report, but extra detail which provides a more global view of the results produced by the MCP simulator and the MCP in its application, optimisation and behavior.

## 9.1 Further Generational Mobility Results

The extra graphs below consider the effect of limiting messages within the MCP system using both Simple and Extended MCP approaches. The Simple MCP approach simply broadcast content packages within message announcements, while the Extended MCP sought context-awareness to reduce the bandwidth cost of peer. Extended MCP was successful in reducing the cost of communication; however its added complexity made it inefficient in sparse settings. Its added complexity also increased the number of messages it produced within the system making it more expensive in message count than Simple MCP.

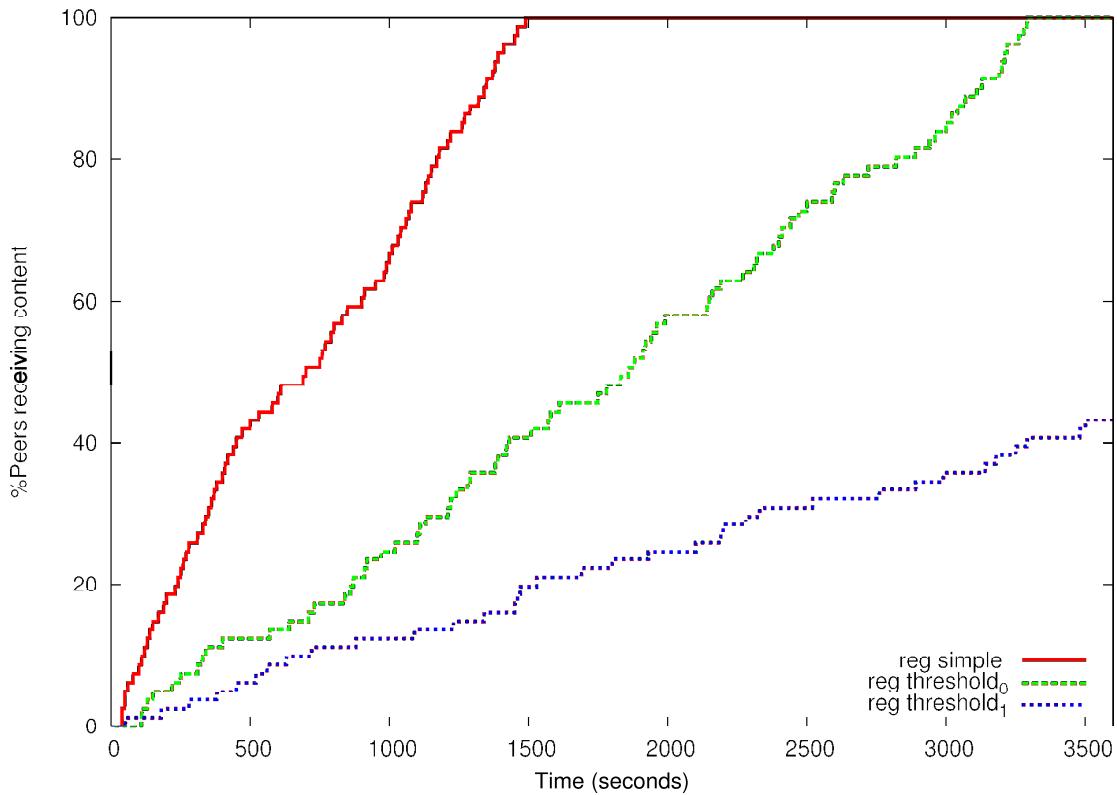

**Figure 59. Unidirectional Path (regular mobility): Simple and Extended (threshold) infection between 0 to 1 neighbours.**

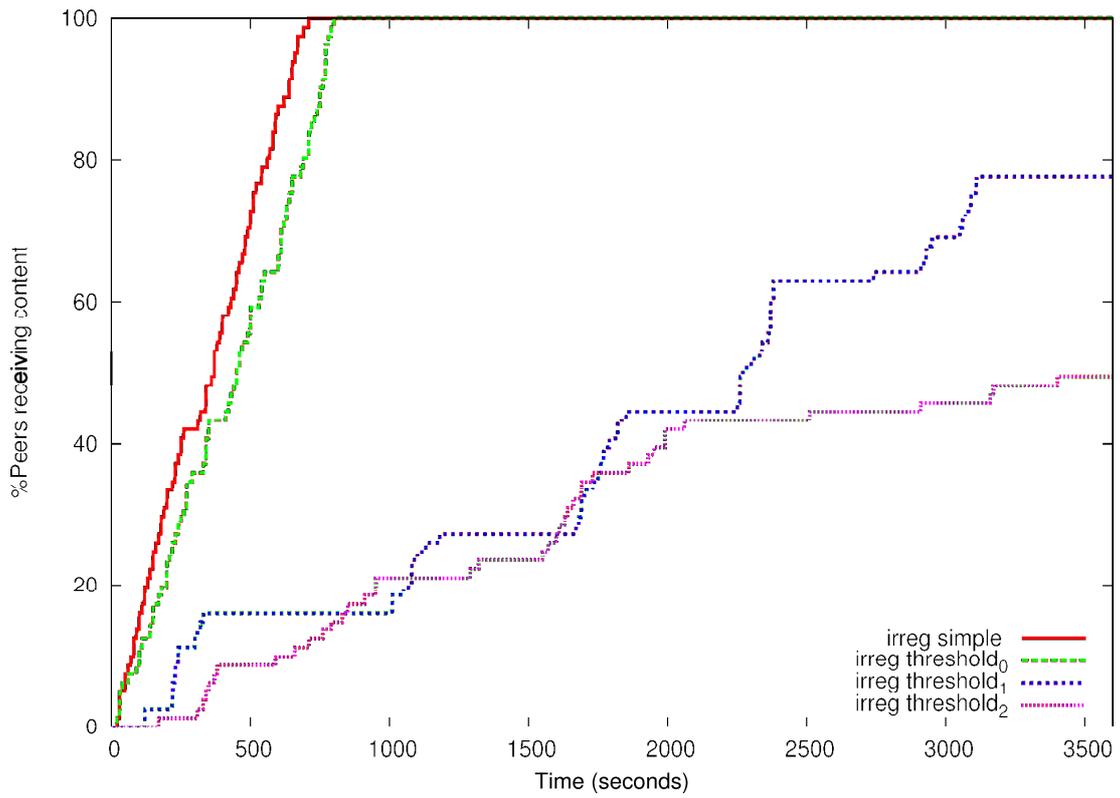

**Figure 60. Unidirectional Path (irregular mobility): Simple and Extended (threshold) infection between 0 to 2 neighbours.**

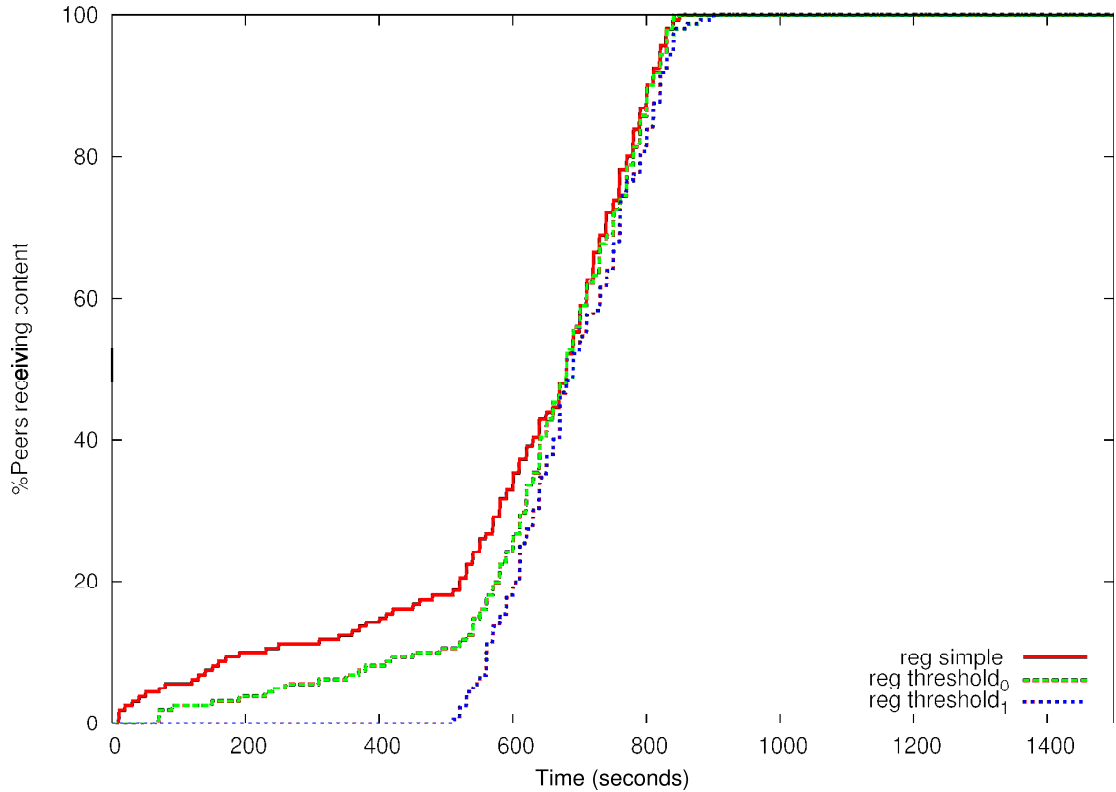

**Figure 61. Bidirectional Path (regular mobility): Simple and Extended (threshold) infection between 0 to 1 neighbours.**

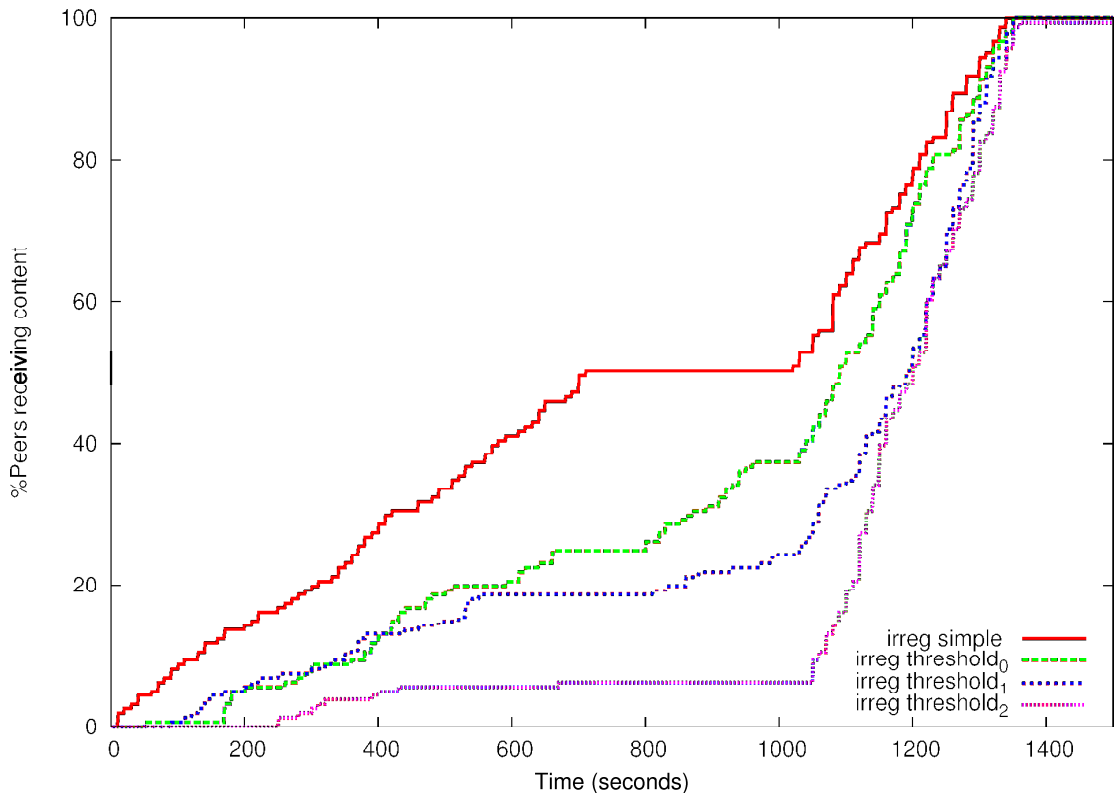

**Figure 62. Bidirectional Path (irregular mobility): Simple and Extended (threshold) infection between 0 to 2 neighbours.**

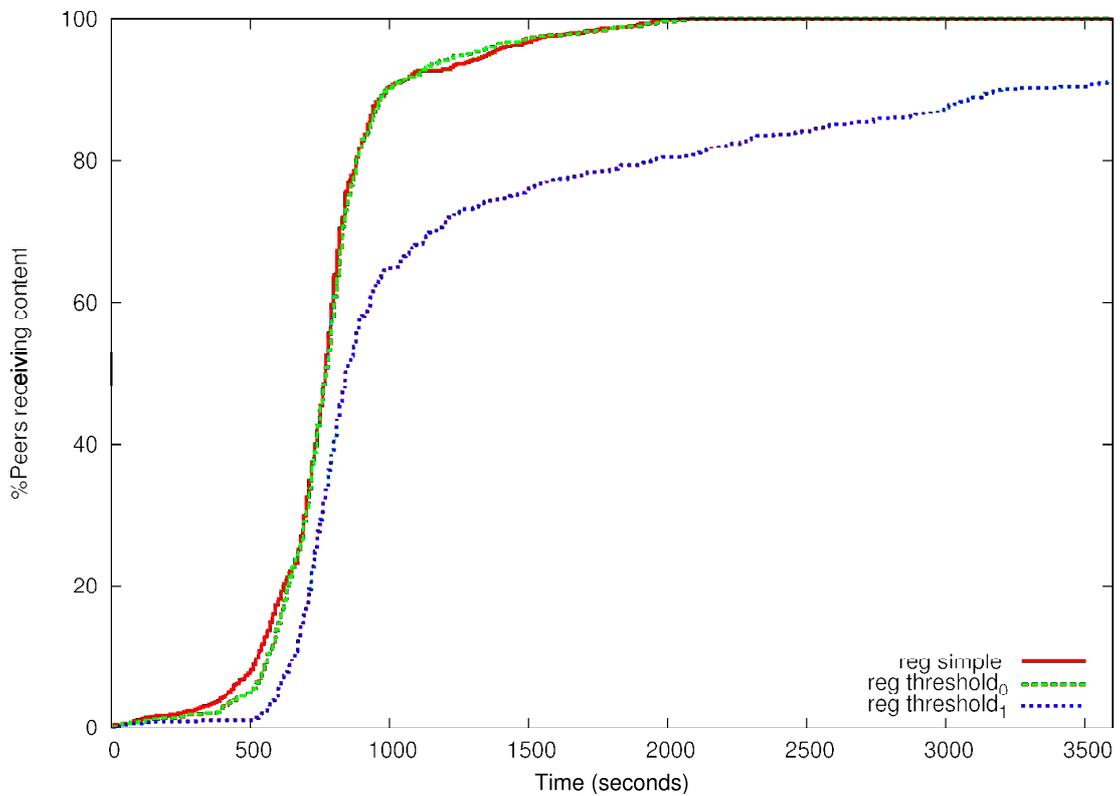

**Figure 63. Bidirectional Grid (regular mobility): Simple and Extended (threshold) infection between 0 to 1 neighbours.**

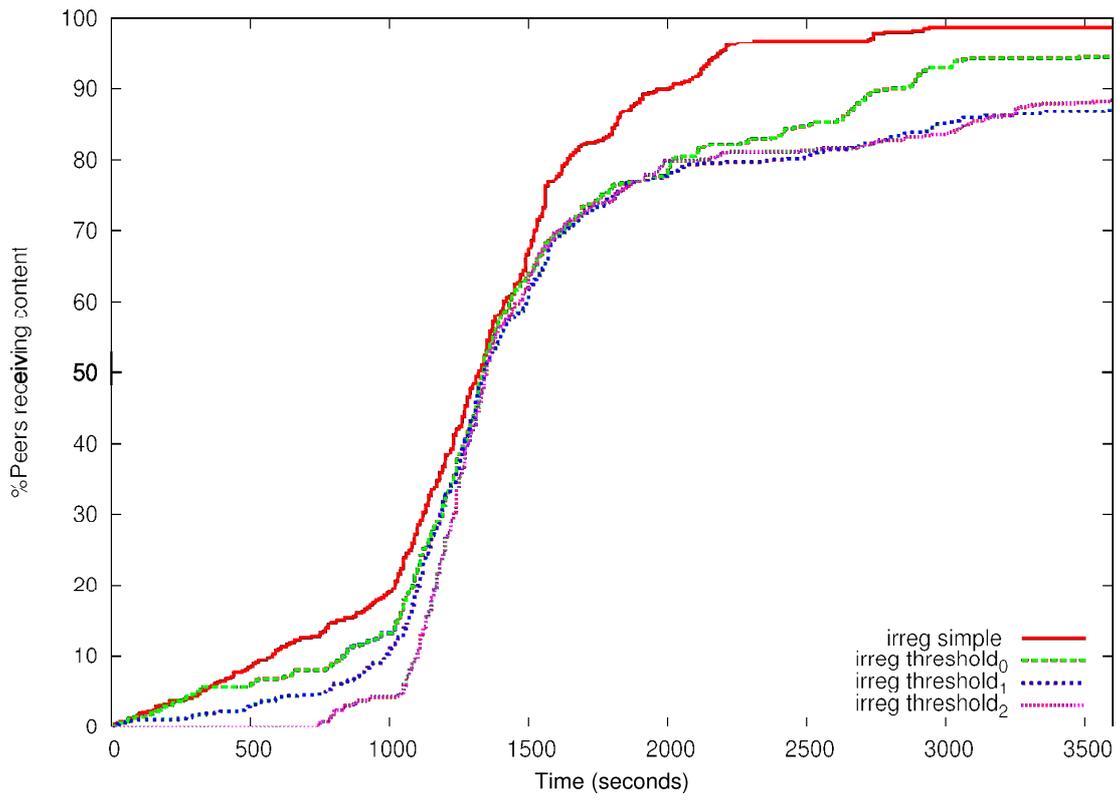

**Figure 64. Bidirectional Grid (irregular mobility): Simple and Extended (threshold) infection between 0 to 2 neighbours.**

## 9.2 Sample Runs

**PRELIMINARY SIMULATOR**

```
STATS
Test Started:   Tue Jul 17 13:47:25 BST 2007
Duration JiST Time:   3600 seconds
SIMULATION PARAMETERS
Number of Nodes:101
Range:       10 meters
Period:      1 seconds
----------------------------------------------------
Sharing Probability   50
Nodes who got the content   99%
Last Message Sent     3599000
Last Message Received 1894000
Total Messages Sent   145324
Total Messages Received       130
----------------------------------------------------
System Usage:
freemem:  1991376
maxmem:   66650112
totalmem: 4198400
used:     2207024
```

**JIST.MARKETCONTACT-BASIC**

```
Inputs
SimTime  = 3600
Range    = 10
ProbShare= 100
Period   = 10000
MaxSpeed = 1.0
Interval = 8
File     = uniline.map
Set Stats Completed...
Reading In File...
1 U (51.501427,-0.180414) | (51.492243,-0.178214) S 1 C 80 |
ADDED SUCCESSFULLY  1.0324581511821542 km
Starting Paths...
Uniform Movement Path Seen...
Sharing  = [65, 26, 48, 57, 70, 42, 74, 11, 58, 10, 71, 36, 60,
55, 33, 31, 78, 76, 52, 51, 45, 6, 8, 12, 61, 46, 54, 0, 75, 67,
53, 63, 29, 39, 19, 27, 62, 9, 35, 28, 79, 23, 72, 40, 15, 30,
49, 64, 1, 38, 18, 50, 43, 7, 34, 24, 25, 32, 16, 13, 17, 3, 41,
77, 47, 44, 2, 56, 59, 68, 20, 22, 14, 4, 37, 5, 66, 69, 73, 21]
NonSharing = []
```